\documentclass{jpp}

\usepackage{graphicx}
\usepackage[makeroom]{cancel}

\usepackage{natbib}
\usepackage{amsmath}
\usepackage{amssymb}
\usepackage{color}
\usepackage{bm}
\usepackage{authblk}

\ifCUPmtlplainloaded \else
  \checkfont{eurm10}
  \iffontfound
    \IfFileExists{upmath.sty}
      {\typeout{^^JFound AMS Euler Roman fonts on the system,
                   using the 'upmath' package.^^J}%                                                                                                 
       \usepackage{upmath}}
      {\typeout{^^JFound AMS Euler Roman fonts on the system, but you
                   dont seem to have the}%                                                                                                          
       \typeout{'upmath' package installed. JPP.cls can take advantage
                 of these fonts, if you use 'upmath' package.^^J}%                                                                                  
      }
  \else
  \fi
\fi

\ifCUPmtlplainloaded \else
  \checkfont{msam10}
  \iffontfound
    \IfFileExists{amssymb.sty}
      {\typeout{^^JFound AMS Symbol fonts on the system, using the
                'amssymb' package.^^J}%           
       \usepackage{amssymb}%           
       \let\le=\leqslant  \let\leq=\leqslant
         \let\geq=\geqslant
      }{}
  \fi
\fi

\ifCUPmtlplainloaded \else
  \checkfont{msam10}
  \iffontfound
    \IfFileExists{amssymb.sty}
      {\typeout{^^JFound AMS Symbol fonts on the system, using the
                'amssymb' package.^^J}%          
       \usepackage{amssymb}%                
       \let\le=\leqslant  \let\leq=\leqslant
         \let\geq=\geqslant
      }{}
  \fi
\fi

\ifCUPmtlplainloaded \else
\IfFileExists{amsbsy.sty}
             {\typeout{^^JFound the 'amsbsy' package on the system, using it.^^J}%                                                                  
     \usepackage{amsbsy}}
    {}
\fi

\newsavebox{\astrutbox}
\sbox{\astrutbox}{\rule[-5pt]{0pt}{20pt}}

\newcommand{\Alfven}{Alfv\'{e}n}
\newcommand{\Alfvenic}{Alfv\'{e}nic}
\newcommand{\V}[1]{\mathbf{#1}} 
\newcommand{\T}[1]{{\tt #1}}

\title[Characterizing the Signature of Electron Landau Damping]{Characterizing the Velocity-Space Signature of Electron Landau Damping}

\author{Sarah~A.~Conley\thanks{Email address
for correspondence: sarah-horvath@uiowa.edu}$^{1}$, Gregory~G.~Howes$^{1}$, Andrew J. McCubbin$^{2}$}
  
\affiliation{$^1$Department of Physics and Astronomy, University of Iowa, Iowa City IA 52242, USA\\
[\affilskip]
$^2$ John Hopkins Applied Physics Laboratory, Laurel MD 20723, USA
}

\pubyear{2023}
\volume{?}
\pagerange{?}
\begin{document}
\maketitle

\begin{abstract}
Plasma turbulence plays a critical role in the transport of energy from large-scale magnetic fields and plasma flows to small scales, where the dissipated turbulent energy ultimately leads to heating of the plasma species. A major goal of the broader heliophysics community is to identify the physical mechanisms responsible for the dissipation of the turbulence and to quantify the consequent rate of plasma heating. One of the mechanisms proposed to damp turbulent fluctuations in weakly collisional space and astrophysical plasmas is electron Landau damping. The velocity-space signature of electron energization by Landau damping can be identified using the recently developed field-particle correlation technique. Here, we perform a suite of gyrokinetic turbulence simulations with ion plasma beta values $\beta_i = 0.01, 0.1, 1,$ and $10$ and use the field-particle correlation technique to characterize the features of the velocity-space signatures of electron Landau damping in turbulent plasma conditions consistent with those observed in the solar wind and planetary magnetospheres. We identify the key features of the velocity-space signatures of electron Landau damping as a function of varying plasma $\beta_i$ to provide a critical framework for interpreting the results of field-particle correlation analysis of \emph{in situ} spacecraft observations of plasma turbulence.
\end{abstract}

%\begin{PACS}
%\end{PACS}

\section{Introduction}
\label{sec:Intro}
Plasma turbulence plays an important role in the transport of energy throughout the heliosphere, governing the conversion of the energy of large-scale magnetic fields and plasma flows into heat of the plasma species. Under the hot and diffuse conditions typical of heliospheric plasmas, collisionless interactions between the electromagnetic fields and the individual plasma particles govern the rate of damping of the turbulent fluctuations and the resulting particle energization. Landau damping \citep{Landau:1946} is one of several collisionless mechanisms that may transfer electromagnetic energy to particle energy in a weakly collisional plasma, and it has been proposed to be of considerable importance in dissipating the turbulent energy in the solar wind 
\citep{Leamon:1998a,Leamon:1999,Quataert:1998,Howes:2008b,Schekochihin:2009,Howes:2011b,Howes:2011c,TenBarge:2013a,Howes:2015,TCLi:2016}. Recent analyses of observations in the Earth's turbulent magnetosheath plasma from the \emph{Magnetospheric Multiscale} (\emph{MMS}) mission \citep{Burch:2016} have provided the first \emph{in situ} evidence of electron Landau damping acting to dissipate turbulence in any space plasma \citep{Chen:2019, Afshari:2021}. Understanding how energy is transferred to plasma particles in turbulent dissipation is a major goal in heliophysics, needed to explain the long-standing problem of how the solar corona is heated to temperatures above a million Kelvin \citep{Edlen:1943,Edlen:1945} in which turbulence is thought to play a crucial role \citep{Withbroe:1977,Heyvaerts:1983,Parker:1988,Klimchuk:2006,Cranmer:2009b,Chandran:2010b}, or to predict the observed 
non-adiabatic temperature profile of the expanding solar wind \citep{Richardson:2003}. A variety of mechanisms are likely to be involved in the dissipation process; descriptions of these can be found in a recent literature review \citep{Verscharen:2019}. 

In the solar wind, observations show that the incompressible (\Alfvenic) component dominates turbulent fluctuations within the inertial range \citep{Bruno:2005,Alexandrova:2008}. As the turbulent cascade of \Alfvenic~fluctuations reaches the ion kinetic length scales, finite Larmor radius effects lead to the development of a nonzero component of the electric field parallel to the local mean magnetic field. Particles with a parallel velocity near the parallel phase velocity of an \Alfven~wave can resonantly interact with the parallel electric field of the wave and thereby gain or lose energy. This results in net energization of the particles if, in the region of the resonance, the initial slope of the distribution function is negative (positive) for particles with $v_\parallel > 0$ ($v_\parallel < 0$). This collisionless transfer of energy from the electric field to the particles is entropy conserving and thus reversible. However, in practice arbitrarily weak Coulomb collisions acting on small velocity scales in the particle distribution function serve to thermalize the energy, realizing thermodynamic heating of that plasma species \citep{Howes:2006,Howes:2008c,Schekochihin:2009}. Landau damping is capable of energizing ions at perpendicular scales near the ion Larmor radius, $k_\perp \rho_i \sim 1$, and electrons at smaller perpendicular scales $k_\perp \rho_i \gtrsim 1$  \citep{Leamon:1999,Quataert:1998,Howes:2008b,Schekochihin:2009,Howes:2011b,Howes:2011c,TenBarge:2013a,Howes:2015,Told:2015,Kiyani:2015,TCLi:2016}. 

Recent analyses of burst-mode measurements from the \emph{Magnetospheric Multiscale} (\emph{MMS}) mission \citep{Burch:2016} using the field-particle correlation technique \citep{Klein:2016,Howes:2017,Klein:2017} have provided the first direct observational evidence of electron Landau damping acting to damp turbulence in Earth's magnetosheath plasma \citep{Chen:2019}. Following this groundbreaking study, a survey of the field-particle correlation signatures in 20 \emph{MMS} intervals was conducted, revealing evidence for electron Landau damping in 19 of the intervals. In one-third of those intervals, the study indicated that this mechanism may be responsible for a dominant fraction of the damping required to terminate the solar wind turbulent cascade \citep{Afshari:2021}.

Motivated by these results, we previously used a high-resolution, gyrokinetic simulation to re-create the field-particle correlation signature of electron Landau damping in plasma conditions matching the first \emph{MMS} interval that yielded an \emph{in situ} observation of electron Landau damping \citep{Chen:2019}. The study demonstrated that electron Landau damping is observable in gyrokinetic turbulence simulations and confirmed the complex features of Landau damping in the regime of dispersive kinetic \Alfven~waves (KAWs) \citep{Horvath:2020}. In particular, we showed how the bipolar signatures of Landau damping of KAWs at different phase velocities may appear together in velocity-space to produce field-particle correlations with multiple bipolar signatures or with superimposed signatures that create a single, broadened structure. Furthermore, we demonstrated that the broadband frequency content in velocity-space fluctuations increases the difficulty of observing signatures in broadband turbulence due to the inability to choose the \emph{correlation interval} $\tau$ such that it averages over an integer multiple of all of the wave modes represented \citep{Klein:2016,Klein:2017,Howes:2017,Horvath:2020} 

In this paper, it is our goal to follow up on our previous work by applying the field-particle correlation technique to a suite of gyrokinetic simulations. Through this, we will construct a framework for identifying the presence of electron Landau damping \emph{in situ} throughout the inner heliosphere. Though the fiducial field-particle correlation signature of ion Landau damping is well understood \citep{Klein:2016,Klein:2017}, our first study showed that electron Landau damping produces more complicated signatures due to the dispersive nature of KAWs at spatial scales smaller than the ion Larmor radius, $k_\perp \rho_i \gtrsim 1$ \citep{Horvath:2020}. Further, the range of wavenumbers and frequencies over which  KAWs are dispersive changes with the ion plasma beta $\beta_i= 8 \pi n_i T_i/B^2$---the ratio of the ion thermal to magnetic pressure---and with the ion-to-electron temperature ratio $T_i/T_e$. Both of these quantities change throughout the heliosphere. In order to understand how the signature of electron Landau damping will appear at different locations in the inner heliosphere, therefore, we create four high-resolution gyrokinetic simulations at different ion plasma beta $\beta_i$ values and analyze these simulations with the field-particle correlation technique, described in Sec.~\ref{sec:fpc}. In addition, we create and analyze a set of single kinetic \Alfven~wave simulations to aid in interpreting the field-particle correlation signatures in turbulence. Both of these simulation types are described in more detail in Sec.~\ref{sec:agk}. The insights gained from the single-wave runs are discussed in Sec.~\ref{sec:lin}, and we present results from the turbulent simulations in Sec.~\ref{sec:turb}, and discuss the implications for observing these signatures \emph{in situ} in Sec.~\ref{sec:discussion}. 

%==========================================================================

\section{Methods}\label{sec:methods}

\subsection{The Field Particle Correlation Technique}\label{sec:fpc}

The field-particle correlation technique \citep{Klein:2016,Howes:2017, Klein:2017} uses single-point measurements of the velocity distribution function and electromagnetic fields to measure the rate of change of the particle \emph{phase-space energy density} $w_s ({\bf x},{\bf v},t)= \frac{1}{2}m_s {\bf v}^2 f_s({\bf x},{\bf v},t)$ and to identify particle energization mechanisms in weakly collisional plasmas. Energization mechanisms interact with particles in specific regions of velocity-space, thereby creating characteristic patterns as net energy is transferred between fields and particles. Correlating field and particle data together generates these characteristic velocity-space signatures and, in doing so, makes possible the identification of the physical mechanism at work and determines the net change in phase-space energy density. The field-particle correlation technique has been successfully applied to understand particle energization in a wide range of fundamental plasma physics processes: (i) collisionless damping of electrostatic \citep{Klein:2016,Howes:2017} and electromagnetic \citep{Klein:2017,Howes:2017b,Howes:2018} plasma waves; (ii) kinetic instabilities \citep{Klein:2017b}; (iii) damping of electromagnetic plasma turbulence through ion Landau damping \citep{Klein:2017,Howes:2018,Klein:2020}, ion transit-time damping \citep{Arzamasskiy:2019,Cerri:2021}, electron Landau damping \citep{Chen:2019,Horvath:2020,Afshari:2021,Horvath:2022}, ion cyclotron damping \citep{Klein:2020,Afshari:2023}, electron magnetic pumping \citep{Montag:2022}, ion stochastic heating \citep{Arzamasskiy:2019,Cerri:2021}; (iv) electron energization in collisionless magnetic reconnection \citep{McCubbin:2022}; and (v) ion and electron acceleration in collisionless shocks \citep{Juno:2021,Juno:2023,Brown:2023}. 

The field-particle correlation is derived by multiplying the Vlasov equation for species $s$ by $m_sv^2/2$ to obtain an equation for the rate of change of phase-space energy density $\partial w_s ({\bf x},{\bf v},t)/\partial t$.  Time-averaging the electric field term in the resulting equation over a correlation interval $\tau$ yields the form of the field-particle correlation. The contribution from the component of the parallel electric field yields the net energy transfer from Landau damping at a given spatial point $\V{r}_0$, and is called the \emph{parallel field-particle correlation} \citep{Howes:2017,Klein:2017}:
\begin{equation}
C_{E_\parallel}({\mathbf r}_0,{\mathbf v}, t; \tau) = \frac{1}{\tau} \int^{t+\tau/2}_{t-\tau/2} -q_s \frac{v_\parallel^2}{2} \frac{\partial f_s({\mathbf r}_0, {\mathbf v}, t')}{\partial v_\parallel}\  E_\parallel({\mathbf r}_0, t')\ dt'.
\label{eq:cepar}
\end{equation}
Taking a cylindrical coordinate system in velocity-space that is aligned with the direction of the local mean magnetic field, $(v_\parallel,v_\perp,\theta)$, we can integrate the parallel correlation over the gyrophase $\theta$ to obtain the parallel field-particle correlation in gyrotropic phase space, $C_{E_\parallel}(v_\parallel, v_\perp,t)$, with the correlation (time-averaging) interval $\tau$ centered at time $t$ and taken at a specified spatial location $\mathbf{r}_0$ (\emph{e.g.}, see contour plots in Fig.~\ref{fig:gyro}). Below we suppress the $\mathbf{r}_0$ argument, with the implication that each field-particle correlation is computed using data from a single spatial point. The resulting energy transfer between the parallel electric field and the particles includes two contributions in the case of Landau damping: (i) the conservative, oscillating energy transfer between the waves and the particles that is associated with undamped waves; and (ii) the secular energy transfer from the parallel electric field to the particles that is associated with collisionless damping via the Landau resonance. In the case of plasma turbulence, the oscillatory energy transfer is often larger in amplitude than the secular energy transfer. Applying a time average over a suitably chosen correlation interval cancels out the large-amplitude oscillatory energy transfer to reveal the smaller amplitude signature of the collisionless damping.  For turbulence simulations, one may choose a correlation interval $\tau$ that is longer than the period of waves at the simulation domain scale \citep{Klein:2016,Howes:2017, Klein:2017}. For spacecraft observations, in which there is a broadband frequency spectrum, including wave periods much longer than the \emph{MMS} burst-mode intervals used for the analysis, one may employ a high-pass filter to the electric field to eliminate the large-amplitude, low frequency contribution to undamped wave motion, isolating the wave periods at kinetic scales where collisionless damping is expected to arise \citep{Chen:2019,Afshari:2021}.

The velocity-space signatures of the particle energization can be presented in a variety of useful ways. To determine if the signature is coherent over time, $C_{E_\parallel}(v_\parallel, v_\perp, t)$ may be integrated over the perpendicular velocity $v_\perp$ to obtain the \emph{reduced parallel correlation} $C_{E_\parallel}(v_\parallel, t)$.  This rate of energization vs.~$v_\parallel$ can be plotted as a line plot at a single time $t$ (\emph{e.g.}, see lower panels of Fig.~\ref{fig:gyro}), or the $v_\parallel$-dependence at each time can be stacked to generate a \emph{timestack plot} of $C_{E_\parallel}(v_\parallel, t)$ (\emph{e.g.}, see Fig.~\ref{fig:tstacks}, central contour plots), revealing the persistence of the energization mechanism over time. To determine the net effect of the energization mechanism as a function of $v_\parallel$,  integrating the reduced parallel correlation over time yields the \emph{time-integrated, reduced parallel correlation} $C_{E_\parallel}(v_\parallel)$ (\emph{e.g.}, lower panels of Fig.~\ref{fig:tstacks}), providing the full time-averaged change in the phase-space energy density $w_s$ as a function of $v_\parallel$. Alternatively, one may instead integrate the reduced parallel correlation over $v_\parallel$ to obtain the total \emph{rate of change of spatial energy density} at position ${\mathbf r}_0$ due to the parallel electric field, $(\partial W_s/\partial t)_{E_\parallel}$ (\emph{e.g.}, left-hand panels of Fig.~\ref{fig:tstacks}). Note that $(\partial  W_s/\partial t)_{E_\parallel}=j_{e,\parallel} E_\parallel$ is simply the rate of electromagnetic work done by the parallel electric field on the electrons.

\subsection{\T{AstroGK} simulations} \label{sec:agk}

The Astrophysical Gyrokinetics Code, \T{AstoGK} \citep{Numata:2010}, was used to generate a suite of four plasma turbulence simulations spanning expected conditions in the solar wind of the inner heliosphere. These four simulations use a realistic mass ratio, $m_i/m_e = 1836$, in order to properly separate the ion and electron scales within the turbulent cascade, have a unity temperature ratio $T_i/T_e = 1$, and sweep through four different plasma beta values, $\beta_i={0.01,0.1,1,10}$. Each \T{AstroGK} simulation evolves the complementary, perturbed gyrokinetic distribution function, $g_s(v_\parallel,v_\perp, t)$, where $g_s(v_\parallel,v_\perp) = h_s(v_\parallel,v_\perp) - \frac{q_s F_{0s}}{T_{0s}} \langle \phi - \mathbf{v}_\perp \cdot \mathbf{A}_\perp \rangle_{\mathbf{R}_s}$, $h_s$ is the gyrokinetic distribution function, and $\langle \rangle_{\mathbf{R}_s}$ denotes a gyroaverage about a fixed gyrocenter coordinate $\mathbf{R}_s$ \citep{Numata:2010}. We choose $g_s(v_\parallel,v_\perp, t)$ at a fixed probe position $\mathbf{r}_0$ as the distribution function $f_s(\mathbf{r_0},\mathbf{v},t)$ in our computation of the parallel field-particle correlation by \eqref{eq:cepar}.

We choose the physical ($x,y,z$) and velocity ($\lambda, \varepsilon$) space resolution in each simulation to be $(n_x, n_y, n_z, n_\lambda, n_\varepsilon, n_s) = (64, 64, 32, 128, 32, 2)$, where $n_s$ is the number of plasma species and velocity-space is partitioned into a gyrotropic grid by pitch angle $\lambda=v_\perp^2/v^2$ and energy $\varepsilon = v^2$. Velocity $v$ is normalized to the thermal velocity, $v_{ts} = \sqrt{2T_s/m_s}$, for each species. Each simulation domain is an elongated 3D Eulerian slab with dimensions $L_\parallel \times L_\perp^2$, given by $L_\parallel = 2 \pi a_0$ and $L_\perp= \pi \rho_i$, where the ion thermal Larmor radius  is $\rho_i =v_{ti}/\Omega_i$, and the ion cyclotron frequency is given by $\Omega_i=q_iB_0/m_ic$. The elongation of the simulation domain is characterized by the small gyrokinetic expansion parameter, where $\epsilon \sim \rho_i/a_0\ll 1$ \citep{Howes:2006,Howes:2008,Howes:2011}. For these parameters, the fully resolved range of perpendicular scales is given by $2 \le  k_\perp \rho_i \le 42$, or $0.05 \lesssim  k_\perp \rho_e \lesssim 1$, capturing a broad range of dispersive kinetic \Alfven~wave frequencies (see Figure~\ref{fig:LGKDR}). Note that, for each timeslice, the fields are output at each point on the high-resolution spatial grid but the electron velocity distribution function is output only at $24$ individual ``probes" spread throughout the simulation box. Sixteen are in the mid-plane ($z=0$) and the remaining eight along the $z$-axis ($x=y=0$), as indicated in Figure~2 of \citet{Horvath:2020}.

\begin{figure}
    \begin{center}
        \includegraphics[width=0.65\textwidth]{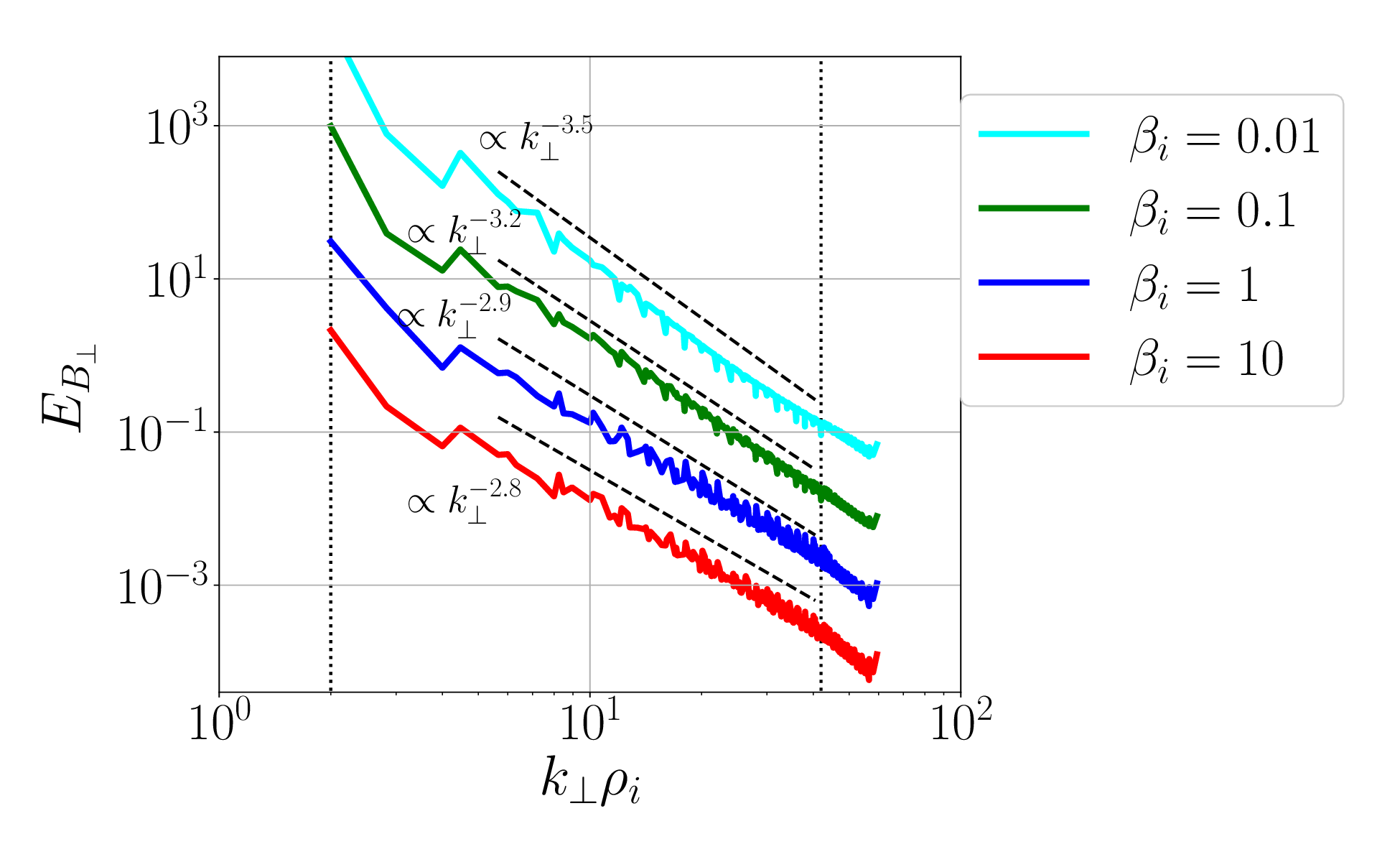}    % efpc2x_spectra_edited.eps
    \end{center}
\caption{Perpendicular magnetic energy spectra ($E_{B_\perp}$) of the four simulations, as a function of $k_\perp \rho_i$, time-averaged over a representative portion of the full simulations. The vertical dotted lines mark the driving scale ($k_\perp \rho_i = 2$) and the smallest fully resolved scale ($k_\perp \rho_i = 42$). By-eye power law fits are shown as references for each spectrum. }
\label{fig:Espectra}
\end{figure} 

An oscillating Langevin antenna \citep{TenBarge:2014} drives counter-propagating, perpendicularly polarized \Alfven~waves with $(k_x\rho_i, k_y\rho_i, k_z a_0) = (2, 0, \pm1)$ and $(0, 2, \pm1)$ to launch a turbulent cascade for each simulation. The driving frequencies $\omega_0$ and decorrelation rates $\gamma_0$ for each driven mode are listed in Table~\ref{tab:params}, where $\overline{\omega}\equiv \omega/k_{\parallel 0} v_A$ and $k_{\parallel 0}=2 \pi/L_\parallel$. The amplitude of the driving is chosen in accordance with critically balanced kinetic \Alfven~wave (KAW) turbulence \citep{Howes:2008b, Howes:2011b} in order to self-consistently produce a strong turbulent cascade that begins at the driving scale of $k_\perp \rho_i = 2$. The turbulent cascade transfers energy to ever smaller length scales, ultimately reaching $k_\perp \rho_i = 42$ where the resolved electron collisionless damping is sufficiently strong to terminate the cascade, enabling a turbulent steady state to be achieved. To prevent a build-up of fine-scale structure in velocity space that is unresolved by our finite grid, we set non-zero collisionalities for the ions and electrons. This allows the energy that is transferred to the particles via collisionless Landau damping to thermalize. These values may be changed somewhat during the course of a simulation to ensure good conservation of energy. In Table 1, we display the final values of the electron and ion collisionalities in normalized units ($\overline{\nu}_{e,f}\equiv \nu_{e,f}/k_{\parallel 0} v_A;\ \overline{\nu}_{i,f} \equiv \nu_{i,f}/k_{\parallel 0} v_A$) for each simulation. These values may be compared with the maximum damping rate ($-\overline{\gamma}_{max}$, not listed) and the driving-scale wave frequency ($\overline{\omega}_0$) to ensure that $\overline{\nu}_{s}$ is large enough to dissipate fine-scale structure in velocity space but small enough such that the driving scale remains collisionless. 
 
We present the average spectra of the perpendicular magnetic field energy for the four simulations in steady state in Fig.~\ref{fig:Espectra}. The spectra have been separated in amplitude in order to highlight the individual power law slopes, and by-eye fits (dashed black lines) are presented alongside the data. At $\beta_i = 10$ and $\beta_i = 1$, the slope of the perpendicular magnetic energy spectra agree well with dissipation-range spectra that have been observed in $\beta_i\gtrsim 1$ plasmas near 1 AU, $E_{B_\perp} \propto k_\perp^{-2.8}$ \citep{Sahraoui:2013b}. The simulation spectra steepen as beta decreases, up to $E_{B_\perp} \propto k_\perp^{-3.5}$ at $\beta_i = 0.01$. Such low beta plasmas are rarely observed near Earth \citep{Wilson:2018}, and dissipation-range simulations of turbulence at this beta are uncommon, though one recent work studying the low-$\beta$ limit of plasma turbulence also finds steep magnetic energy spectra \citep{Zhou:2023}. Additionally, our steepest spectra is within the range of early dissipation-range observations \citep{Leamon:1998a} and the trend we observe of the spectra steepening as beta decreases is consistent with the electrons transferring more energy from the turbulence within the dissipation range in the case of more significant damping \citep{Told:2015}.

\begin{table}
    \centering
    \begin{tabular}{c c c c c c c c c c c c c}
        $\beta_i$ & $T_i/T_e$ & $m_i/m_e$ & $-\overline{\gamma}_0/\overline{\omega}_0$ & $\overline{\omega}_0$ & $\overline{\omega}_{sd}$ & $\overline{\omega}_{max}$ & $\overline{T}_0$ & $\overline{T}_{min}$ & $\Delta t/T_0$ & {$\overline{\nu}_{e,f}$ }&  $\overline{\nu}_{i,f}$\\
        \hline
        0.01 & 1 & 1836 & 0.17 & 2.044 & - & 5.57 & 3.074 & 1.128 & 68.9 & 0.4 & 0.1\\
        0.1 & 1 & 1836 & 0.063 & 2.082 & 2.845 & 14.9 & 3.018 & 0.422 & 13.6 & 0.6 & 0.3\\
        1 & 1 & 1836 & 0.042 & 1.576 & 3.856 & 38.56 & 3.987 & 0.163 & 2.63 & 1.2 & 1.2\\
        10 & 1 & 1836 & 0.19 & 0.6832 & 0.989 & 13.55 & 9.197 & 0.464 & 1.01 & 1.0 & 2.0\\
    \end{tabular}
    \caption{Parameters of the four turbulent simulations.}
    \label{tab:params}
\end{table}

The linear phase velocity normalized to the electron thermal velocity, $\omega/k_\parallel v_{te}$, and the normalized damping rate $-\gamma/\omega$ for the KAWs over the parameter range of each of these four simulations are computed using the linear Vlasov-Maxwell dispersion relation solver \emph{PLUME} \citep{Klein:2015a} and are presented in Fig.~\ref{fig:LGKDR}. The normalized frequencies $\overline{\omega}_0 \equiv \omega_0/(k_{\parallel} v_A)$ of the domain scale KAWs with wavevector components $k_{\perp 0}\rho_i  = 2\pi \rho_i /L_\perp = 2$ and $k_{\parallel 0} a_0 = 2\pi a_0/L_\parallel =1 $ are listed in Table~\ref{tab:params}. Note that the \Alfven~wave and electron thermal velocity normalizations of the frequency are related by 
$\omega/k_\parallel v_{te} = \overline{\omega}\beta_i^{-1/2}(T_i/T_e)^{1/2} (m_e/m_i)^{1/2} $.
Also, the normalized \Alfven~wave frequencies have departed from the MHD limit of $\overline{\omega}=1$, since we are modeling kinetic \Alfven~waves with $k_\perp \rho_i > 1$ (Fig.~\ref{fig:LGKDR}(a)). The maximum frequency KAW found in each simulation can be determined from Fig.~\ref{fig:LGKDR} (listed in Table~\ref{tab:params} as $\overline{\omega}_{max}$), along with the frequency corresponding the onset of strong damping at $-\gamma/\omega \sim 0.1$ (Fig.~\ref{fig:LGKDR}(b)), which is listed as $\overline{\omega}_{sd}$ in Table~\ref{tab:params} if this frequency is above that of the domain scale.

\begin{figure}
    \begin{center}
        \includegraphics[width=0.49\textwidth]{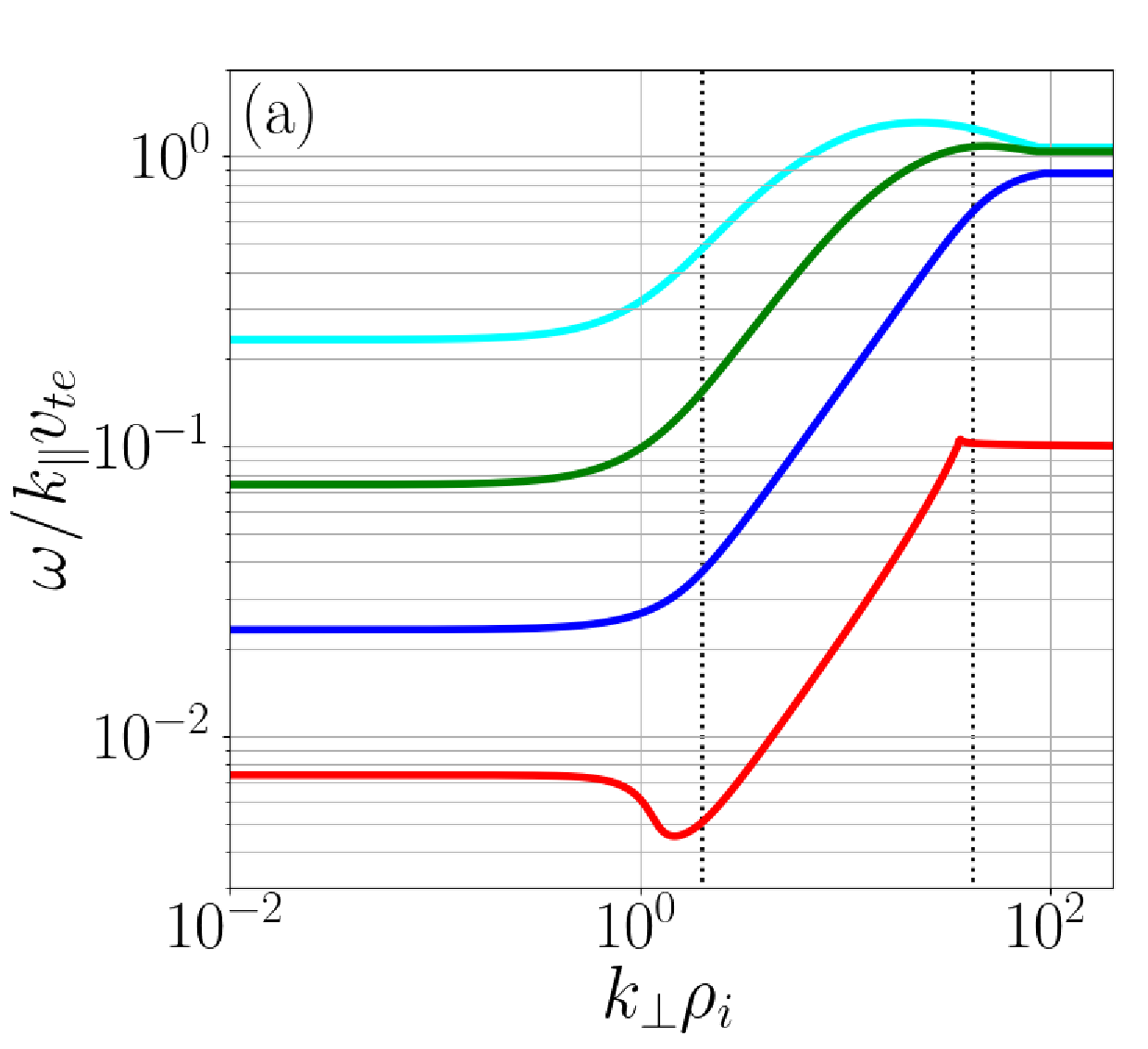}    % efpc2x_freq_vte_PLUME_v2.eps
        \includegraphics[width=0.49\textwidth]{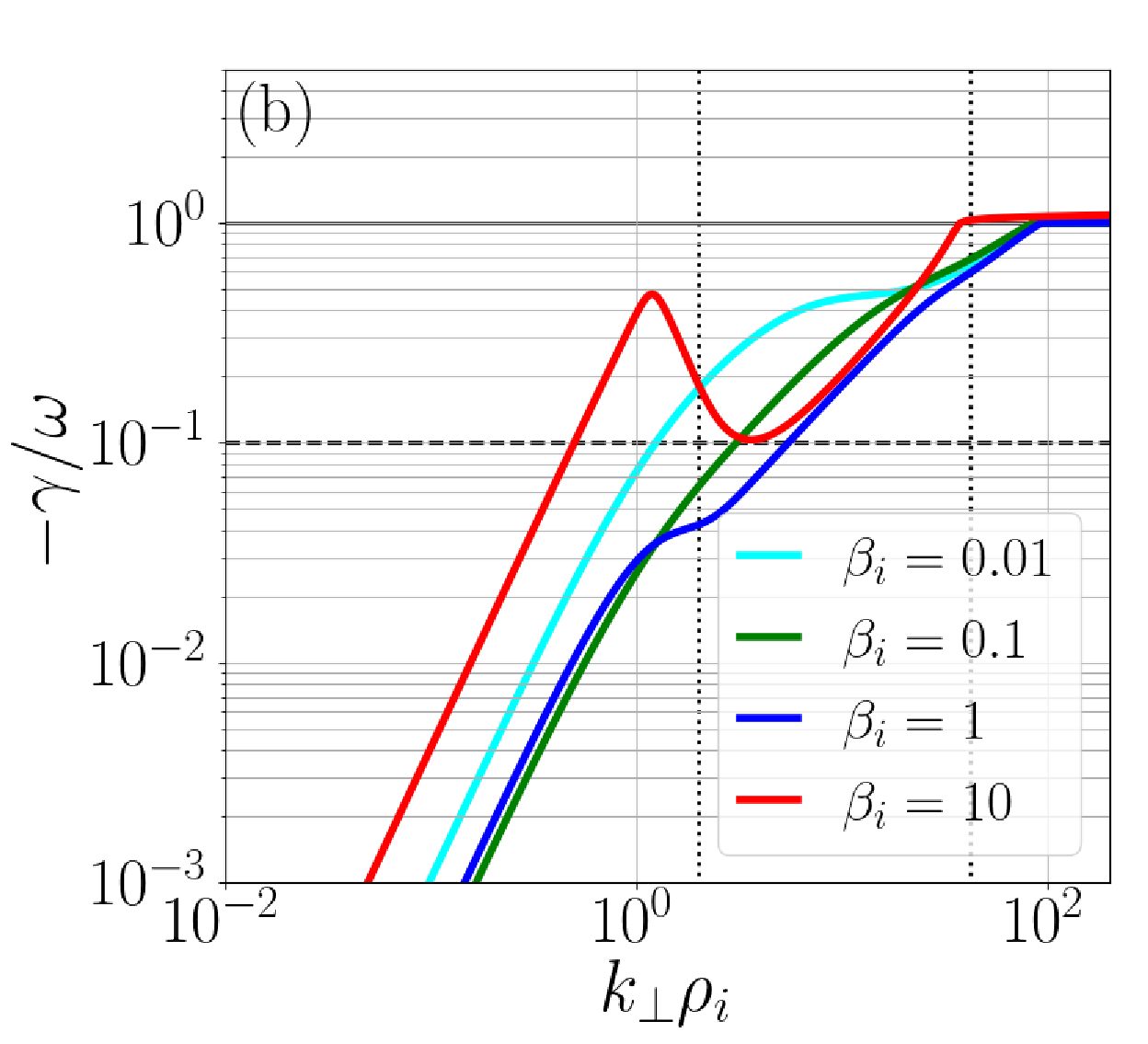}    % efpc2x_damping_PLUME_v2.eps
    \end{center}
 \vskip -1.75in
\hspace*{0.4in} (a)\hspace*{2.4in} (b)
\vskip +1.75in
    \caption{Linear Vlasov-Maxwell dispersion relation for plasmas with $T_i/T_e = 1$, $m_i/m_e = 1836$, and $\beta_i = 0.01$ (cyan), $0.1$ (green), $1$ (blue), and $10$ (red). (a): Parallel phase velocity normalized to the electron thermal velocity $\omega/k_\parallel v_{te}$. (b): Normalized damping rate $-\gamma/\omega$, with the onset of strong damping ($-\gamma/\omega \gtrsim 0.1$) marked by a horizontal dashed line.}
    \label{fig:LGKDR}
\end{figure}

The post-saturation length of each of the four simulations, $\Delta t/T_0$, is listed in Table~\ref{tab:params} in terms of time normalized to the period of the domain-scale kinetic \Alfven~waves, $T_0=2\pi /\omega_0$. Note that the timescale separation between the faster electron thermal velocity and the slower \Alfven~velocity, $v_{te}/v_A = \beta_i^{1/2}(T_e/T_i)^{1/2} (m_i/m_e)^{1/2}$, decreases for lower values of $\beta_i$. Since the maximum timestep in \T{AstroGK} is determined by the electron thermal velocity, lower $\beta_i$ simulations are substantially less computationally costly to run.

In addition to the nonlinear turbulence simulations described above, we also use \T{AstroGK} to model a total of $20$ individually-driven, damped kinetic \Alfven~waves at a single value of $k_\perp \rho_i$. For each value of $\beta_i = {0.01, 0.1, 1, 10}$ we use the oscillating Langevin antenna to separately drive linear modes at $k_\perp \rho_i = {2, 4, 8, 16, 32}$, where $(k_x \rho_i, k_y \rho_i, k_z a_0) = (k_\perp \rho_i, 0, 1)$ in each case. The parallel spatial grid and velocity-space resolutions for each wave are $(n_z, n_\lambda, n_\varepsilon, n_s) = (32, 128, 32, 2)$, where the variables and normalization are the same as described for the turbulence runs. Each driven, damped linear wave was evolved for a minimum of $10$ wave periods. 

%==========================================================================

\section{Velocity-Space Signatures of Single Kinetic \Alfven~Waves}\label{sec:lin}
To develop a framework for interpreting the velocity-space signatures of electron Landau damping in turbulence simulations, we first present the results of applying field-particle correlation analysis to the suite of single KAWs described in Sec.~\ref{sec:methods}. The analysis of single, damped waves eliminates the complications which arise from overlapping bipolar signatures of Landau damping in broadband turbulence, as illustrated in Figure~1 of  \citet{Horvath:2020}. Thus, we are able to more easily identify trends in the behavior of the field-particle correlation signatures of this mechanism as the ion plasma $\beta_i$ is changed. Later, these clearly identified features will aid our interpretation of signatures in the turbulence simulations, which we present in Sec.~\ref{sec:turb}.

\begin{figure}
    \begin{center}
        \includegraphics[width=0.48\textwidth]{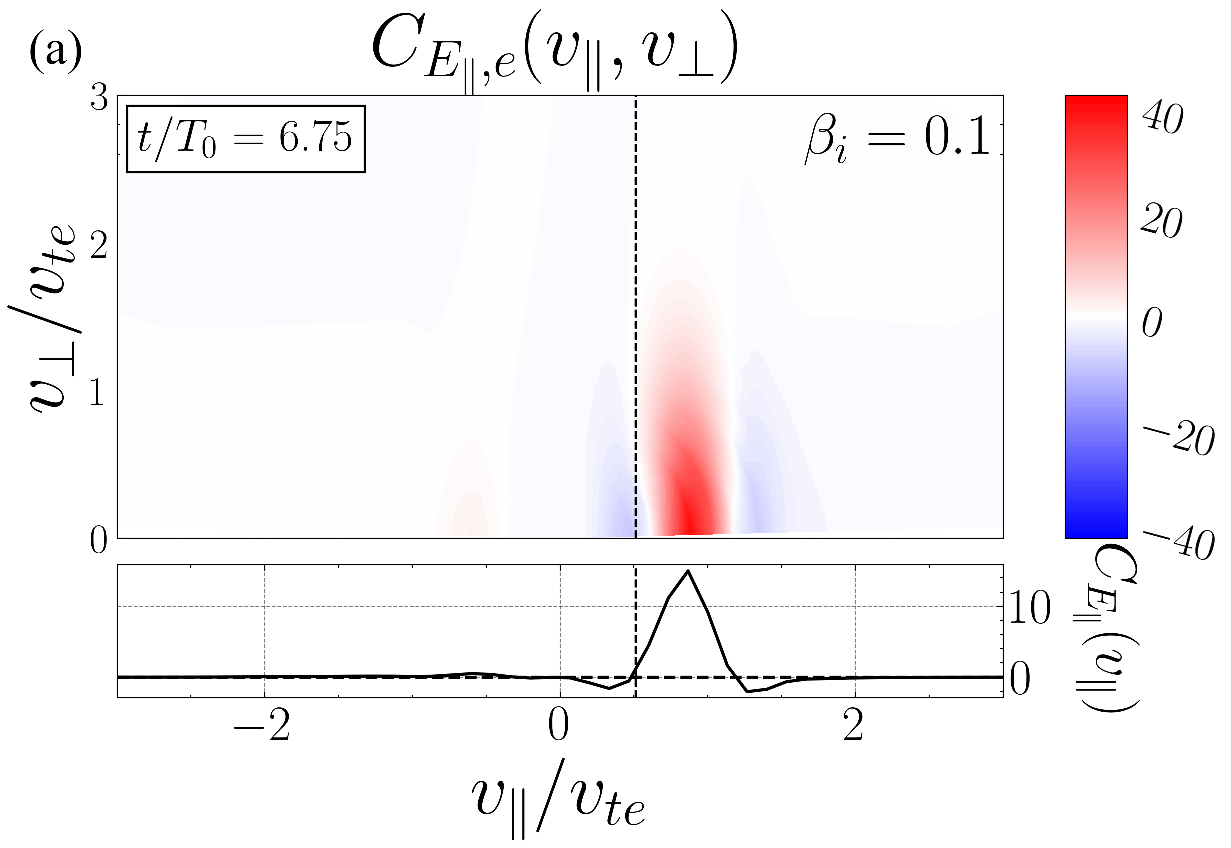}     % ekd_bp1_k8_probe001_s2_CE_nc064_t0432_contourf_a.eps
        \includegraphics[width=0.50\textwidth]{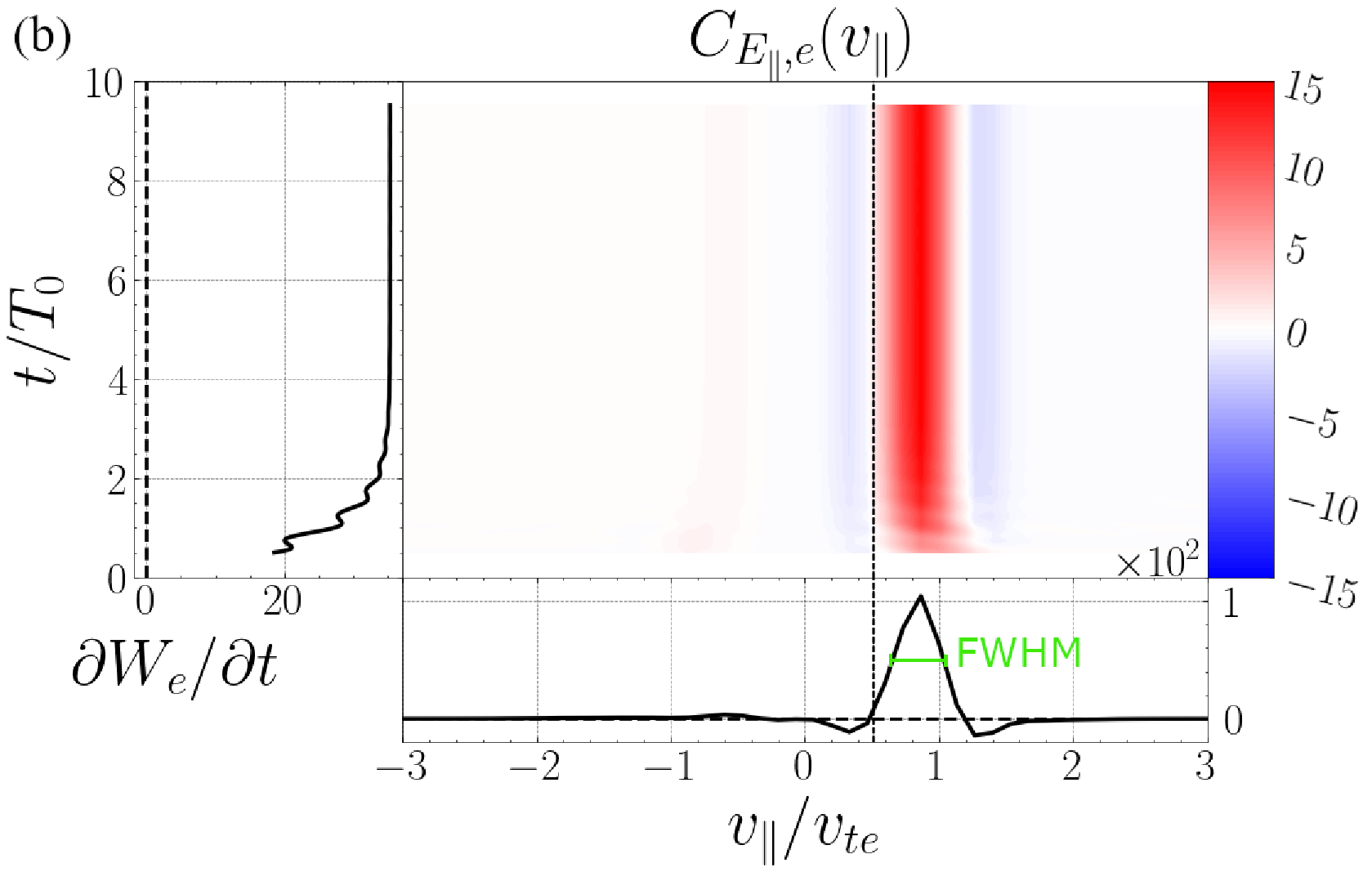}     % ekd_bp1_k8_probe001_s2_CER_nc064_contourf_gyroratio_annotated_b.eps
    \end{center}
    \caption{(a) Gyrotropic field-particle correlation signature $C_{E_\parallel}(v_\parallel, v_\perp)$ at $t/T_0 = 6.75$ and (b) timestack plot $C_{E_\parallel}(v_\parallel, t)$ of a driven, damped wave simulation with $\beta_i = 0.1$, $k_\perp \rho_i = 8$, {and a correlation interval of $\tau/T_0 = 1$.} In the time-integrated lower panel in (b), the full-width at half-maximum of the positive portion of the bipolar signature is indicated.}
    \label{fig:lingyro}
    \vspace{-0.1in}
\end{figure}

Fig.~\ref{fig:lingyro} shows (a) a gyrotropic signature $C_{E_\parallel}(v_\parallel, v_\perp, t)$  and (b) a timestack plot $C_{E_\parallel}(v_\parallel, t)$ of the field-particle correlation for an individual KAW simulation with $\beta_i = 0.1$ and $k_\perp \rho_i = 8$. 
In panel (a), the characteristic bipolar form of the field-particle correlation signature of Landau damping \citep{Klein:2016, Klein:2017, Howes:2017, Chen:2019, Horvath:2020} is clearly visible in both the colormap plot and in the line plot of the time-integrated, reduced parallel correlation $C_{E_\parallel}(v_\parallel)$ shown below. In this example, the correlation interval $\tau$ is equal to a full wave period ($\tau/T_0 = 1$) and the gyrotropic correlation in (a)  is shown centered at time $t/T_0 = 6.75$. The significant energy transfer shown in the bipolar velocity-space signature is localized in $v_\parallel$ around the parallel phase velocity of the damped wave (vertical dashed line, $v_{ph, \parallel}/v_{te} = \omega/k_\parallel v_{te} = 0.51$), with a loss of phase-space energy density (blue) 
at $v_\parallel < v_{ph, \parallel}$ and a gain of of phase-space energy density (red) at $v_\parallel > v_{ph, \parallel}$. The dominance of the positive region indicates a net transfer of energy from the parallel electric field to the resonant electrons. 
 As expected for Landau damping, the bipolar structure does not vary in the perpendicular direction apart from an exponential decrease at high perpendicular velocities due to the equilibrium velocity distribution. Therefore, we may integrate over all of $v_\perp$ to yield the total change in $w_s$ as a function of $v_\parallel$ at the spatial point being considered (lower panel). This lower panel is then calculated for all times and combined to create (b) a \emph{timestack} plot of the correlation $C_{E_\parallel}(v_\parallel, t)$. The panel beneath the timestack plot is the time-integrated, reduced parallel correlation $C_{E_\parallel, e}(v_\parallel)$. The panel to the left of the timestack is the rate of change of the electron spatial energy density due to the parallel electric field,  $(\partial W_e/\partial t)_{E_\parallel}$, which is calculated by integrating the correlation over $v_\parallel$ and is equal to
 $j_{e,\parallel} E_\parallel$.

\subsection{Velocity-Space Signature Location and Wave Phase Velocity}
\begin{figure}
    \begin{center}
        \includegraphics[width=0.49\textwidth]{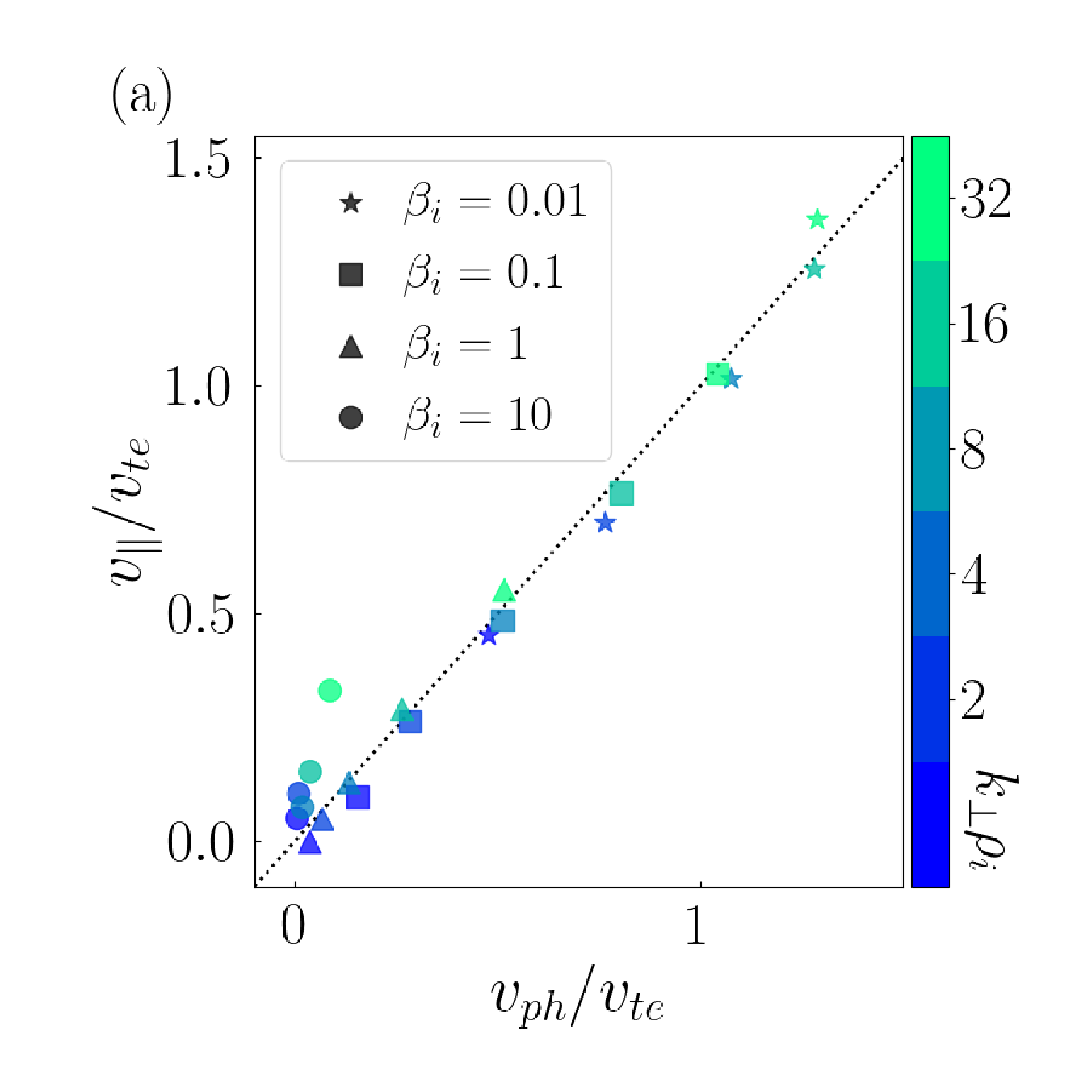}    % phase_test.eps
        \includegraphics[width=0.49\textwidth]{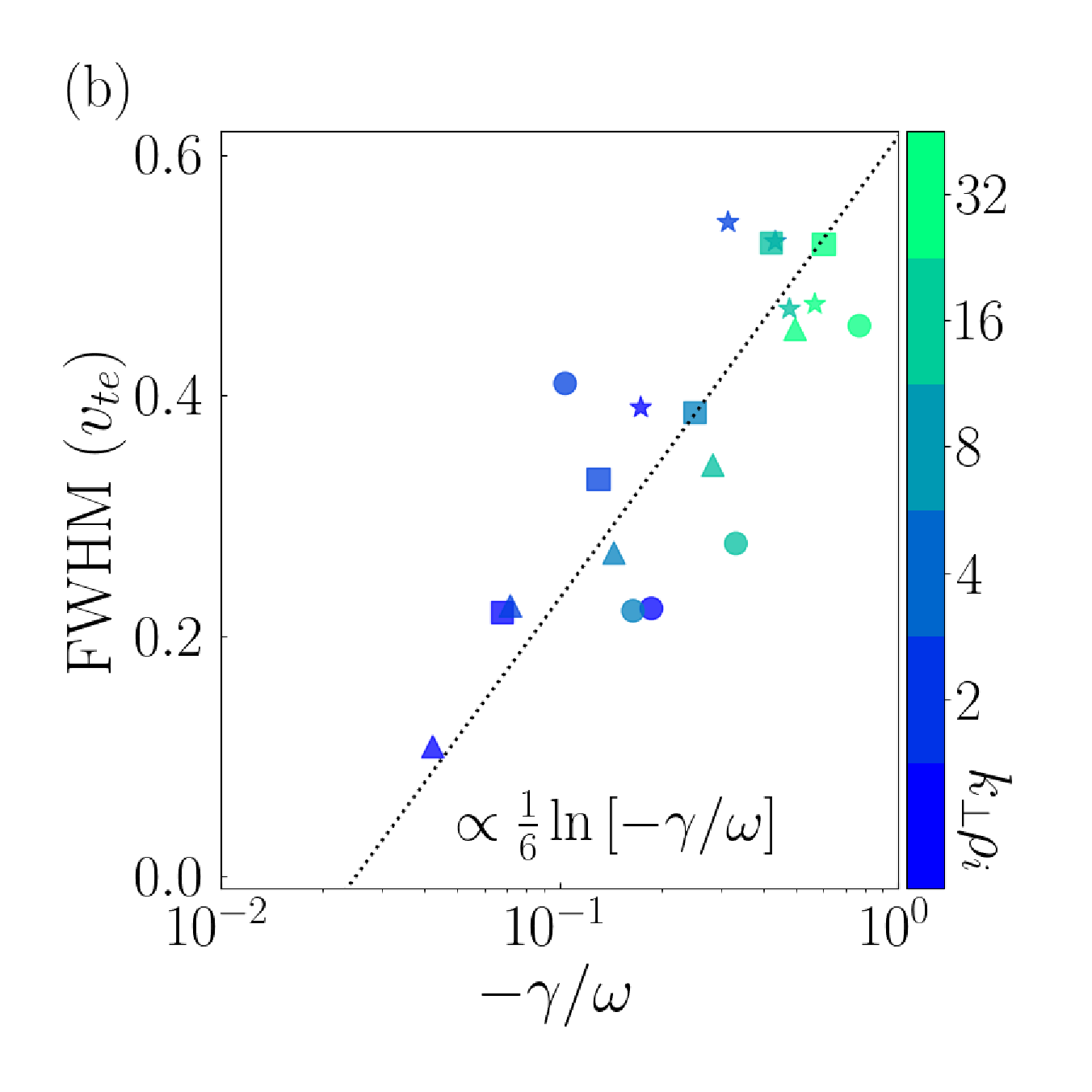}    % fwhm_test2.eps
    \end{center}
\caption{(a) Velocity-space signature location $v_\parallel/v_{te}$ vs.~parallel wave phase velocity $v_{ph,\parallel}/v_{te}$ for the linear runs. (b) Full-width, half maximum (FWHM) value of the bipolar signatures vs.~the damping rate $-\gamma/\omega$ of the wave. Plasma $\beta_i$ and $k_\perp \rho_i$ are differentiated using marker styles and color, respectively.}
\label{fig:vph_fwhm}
\end{figure}

A key feature of a bipolar signature of Landau damping is its concentration in velocity-space about the parallel phase velocity of the damped wave \citep{Klein:2016}. The form of the signature is consistent with flattening of the distribution function around the resonant velocity, as predicted by quasilinear theory of Landau damping, and indicates that a resonant mechanism is producing the energization \citep{Howes:2017}. Through the sweep of $\beta_i$ and $k_\perp \rho_i$ parameter space provided by our 20 single-wave simulations, we show here quantitatively that the location of the zero crossing in the bipolar signature of Landau damping (the velocity $v_{\parallel}/v_{te}$ where $C_{E_\parallel}(v_\parallel) = 0$) is indeed very well correlated with the parallel phase velocity of the damped wave. These two quantities are plotted against each other in Fig.~\ref{fig:vph_fwhm}(a), and the results lie almost entirely on the line of $v_\parallel = v_{ph}$. The linear Vlasov-Maxwell dispersion relation \citep{Stix:1992,Klein:2015a} was used to find $v_{ph}$ for each wavemode, and the zeros are found from gyrotropic plots with correlation interval $\tau/T_0 = 1$. Note that the exception to the close agreement between $v_\parallel$ and $v_{ph,\parallel}$ occurs for the signatures with $\beta_i = 10$ (circles). This error likely arises due to a combination of the small KAW phase velocity at high $\beta_i$ and the finite $v_\parallel$-resolution  $\Delta v_\parallel/v_{te} = 0.133$ of the simulations. The slight deviation of the other signatures from the $v_\parallel = v_{ph}$ line appears to fall within the error due to finite $v_\parallel$-resolution. 

\subsection{Velocity-Space Signature Width and Wave Damping Rate}
The second observation we are able to make using the single-wave simulations is that the width of the field-particle correlation signatures changes with both $\beta_i$ and $k_\perp \rho_i$. The width tends to vary directly with $k_\perp \rho_i$ (for instance, see the five single-wave signatures for $\beta_i=1$ shown in Appendix A) and indirectly with $\beta_i$. Similarly, the damping rate of the KAWs increases with $k_\perp \rho_i$ and decreases with $\beta_i$, suggesting a possible relationship. Note that this connection with $k_\perp \rho_i$ is true specifically for dispersive KAWs, where $v_{ph}$ generally increases monotonically with $k_\perp \rho_i$. Indeed, when the signature width---quantified by the full-width, half maximum (FWHM) of the positive portion of the bipolar structure (\emph{e.g.}, see the lower panel of Fig.~\ref{fig:lingyro})---is plotted against the corresponding normalized damping rate of the KAW in the simulation, $-\gamma/\omega$, a logarithmic dependence emerges, roughly described by $FWHM \propto \frac{1}{6} ln\left(-\gamma/\omega \right)$. This relationship is shown in Fig.~\ref{fig:vph_fwhm}(b). Though logarithmic dependencies are weak, broadening of the signatures due to increased damping rates has plausible theoretical support. Specifically, it may be analogous to the frequency-space broadening of the Fourier transform of a damped sine wave. A pure sine wave yields a Dirac delta function in frequency space under a Fourier transform, while a damped wave yields a peak of finite width due to its decrease in amplitude over time. 

\subsection{The Effect of Weighting by $v_\parallel^2$ in the Parallel Field-Particle Correlation}

\begin{figure}
    \begin{center}   
        \includegraphics[width=0.49\textwidth]{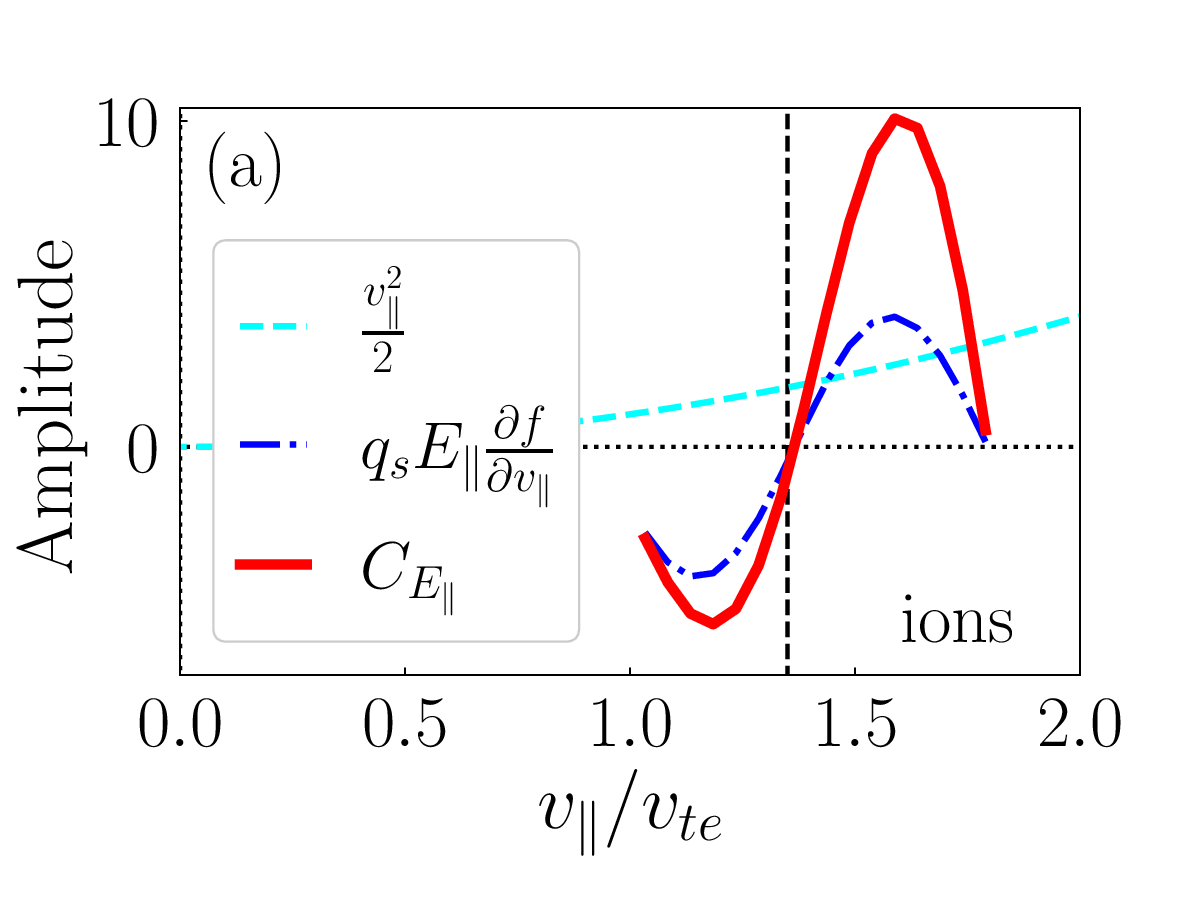}   % model_corr_ion.eps
        \includegraphics[width=0.49\textwidth]{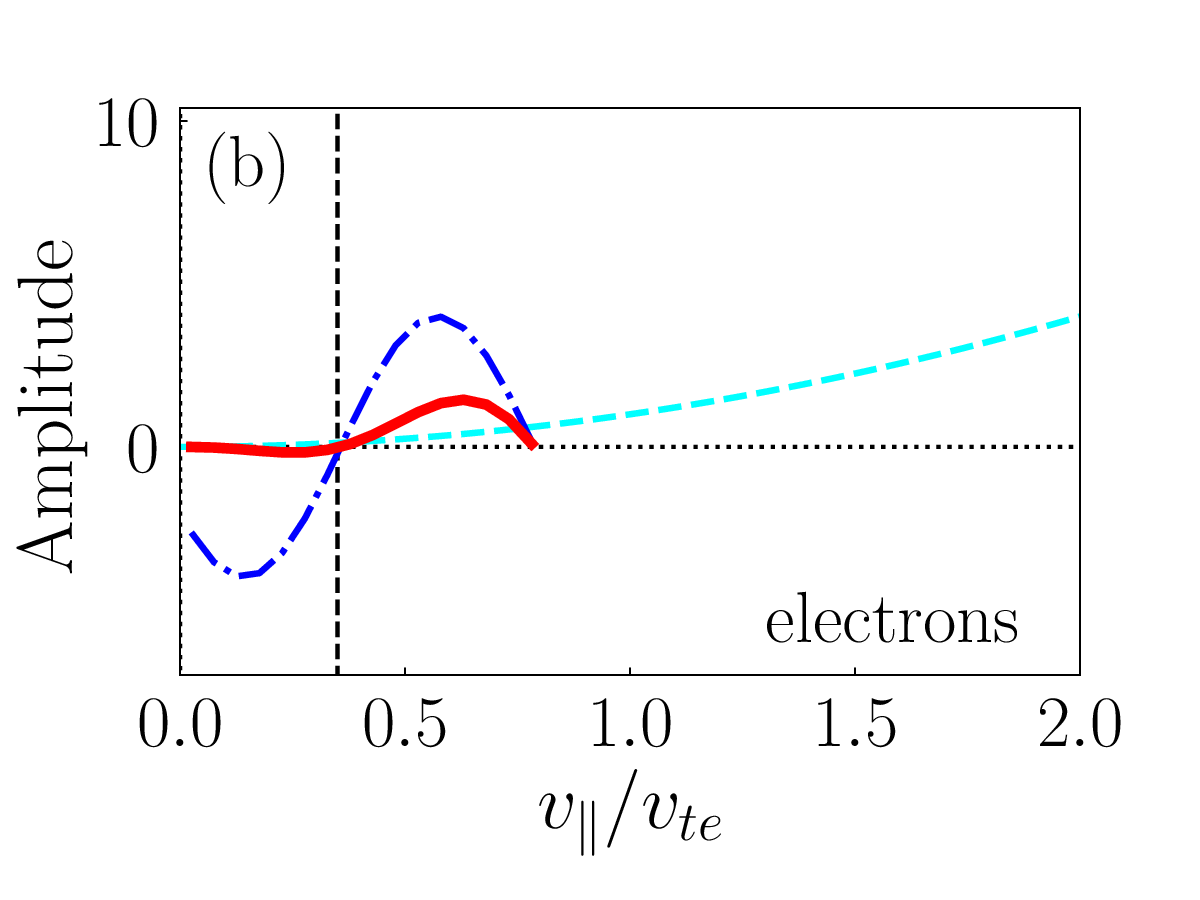}   % model_corr_elec.eps
    \end{center}
    \caption{Diagram of the effect of $v_\parallel^2$-weighting on bipolar signatures of Landau damping.}
    \label{fig:vparweight}
\end{figure}

Another feature that becomes clear through the suite of single-wave signatures is the effect of $v_\parallel^2$ in the definition of the field-particle correlation, Eq.~\eqref{eq:cepar}. Since particle velocities are normalized by the species thermal velocity, this factor enhances the velocity-space signature for $v_\parallel/v_{te}>1$ and suppresses the signature for $v_\parallel/v_{te}<1$.\footnote{ Note that the inflection point of $v_{te}=1$ is peculiar to our normalization choice. However, a similar effect would be observed for any choice of normalization.} This behavior is due to the mathematical form of the field-particle correlation, which arises from the derivation of the rate of change of phase-space energy density. Therefore, the enhancement or suppression of the energy transfer rate due to $v_\parallel^2$ is a physical effect that can hinder the ability to recognize bipolar velocity-space signatures. Specifically, for damped waves that are resonant with particles at low parallel velocities with $v_\parallel/v_{te}<1$, the preferential suppression of signal at small $v_\parallel$ can mask the characteristic bipolar structure that identifies the energization as being due to Landau damping. Fig.~\ref{fig:vparweight} presents two sketches of this effect, showing the separate magnitudes of the weighting by $v_\parallel^2/2$ (cyan dashed) and the remaining parallel electric field correlation $q_s E_\parallel \partial f/\partial v_\parallel$ (blue dot-dashed) along with their product (red solid). The relative magnitudes of the negative and positive regions of the bipolar signature of $C_{E_\parallel}$ when (a) $v_{ph}/v_{te} > 1$ is contrasted with the relative magnitudes when (b) $v_{ph}/v_{te} < 1$. For the Landau damping of ions, the wave resonance is typically $v_{ph,\parallel}/v_{ti} \simeq 1$, as in Fig.~\ref{fig:vparweight}(a), where the bipolar structure of the signature of Landau damping is easily observed. For electron Landau damping, however, the resonances are typically less than the electron thermal velocity $v_{ph,\parallel}/v_{te} < 1$, and asymptotically reach $v_{ph}/v_{te} \rightarrow 1$ at the smallest spatial scales ($k_\perp \rho_e \rightarrow 1$), as seen in Fig.~\ref{fig:LGKDR}(a). Additionally, as $\beta_i$ increases, the parallel phase velocity $\omega/k_\parallel v_{te}$ across all dissipation range scales becomes an even smaller fraction of $v_{te}$. Therefore, for electron Landau damping signatures in high $\beta_i$ plasmas, the resonant velocities are often deep in the center of the electron distribution function, so the negative portion of the signature can be significantly suppressed and become nearly unobservable, as in Fig.~\ref{fig:vparweight}(b). For our twenty single-wave simulations binned in parallel velocity to a resolution of $\Delta v_\parallel = 0.133\ v_{te}$, the $v_\parallel^2$-weighting made the negative portion of the bipolar signatures difficult or impossible to observe for resonant waves with $v_{ph} \leq 0.3\ v_{te}$: the case for eleven out of the twenty signatures.\footnote{ { When the factor of $v_\parallel^2$ is removed from the correlation in such cases, the bipolar structure becomes clear.}}
\subsection{Non-resonant feature}

\begin{figure}
    \begin{center}
        \includegraphics[width=0.48\textwidth]{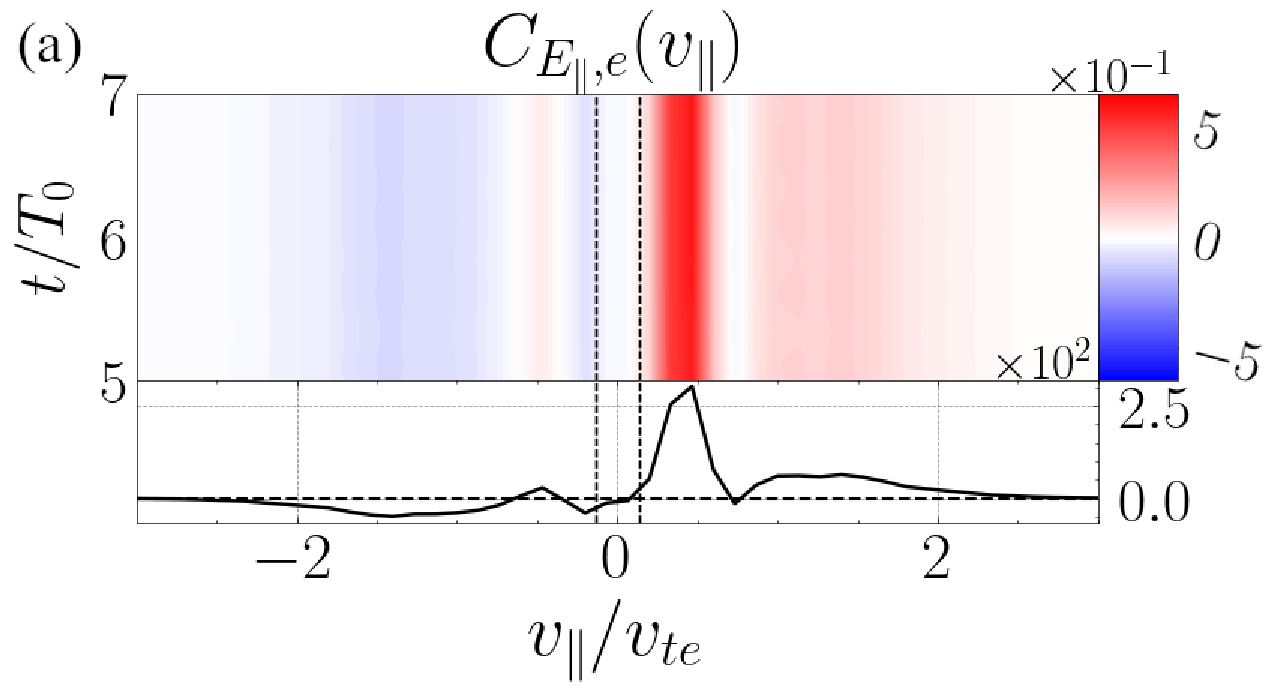}    % ekd_b1_k8_probe001_s2_CER_nc064_contourf_compressed_a.eps
        \includegraphics[width=0.48\textwidth]{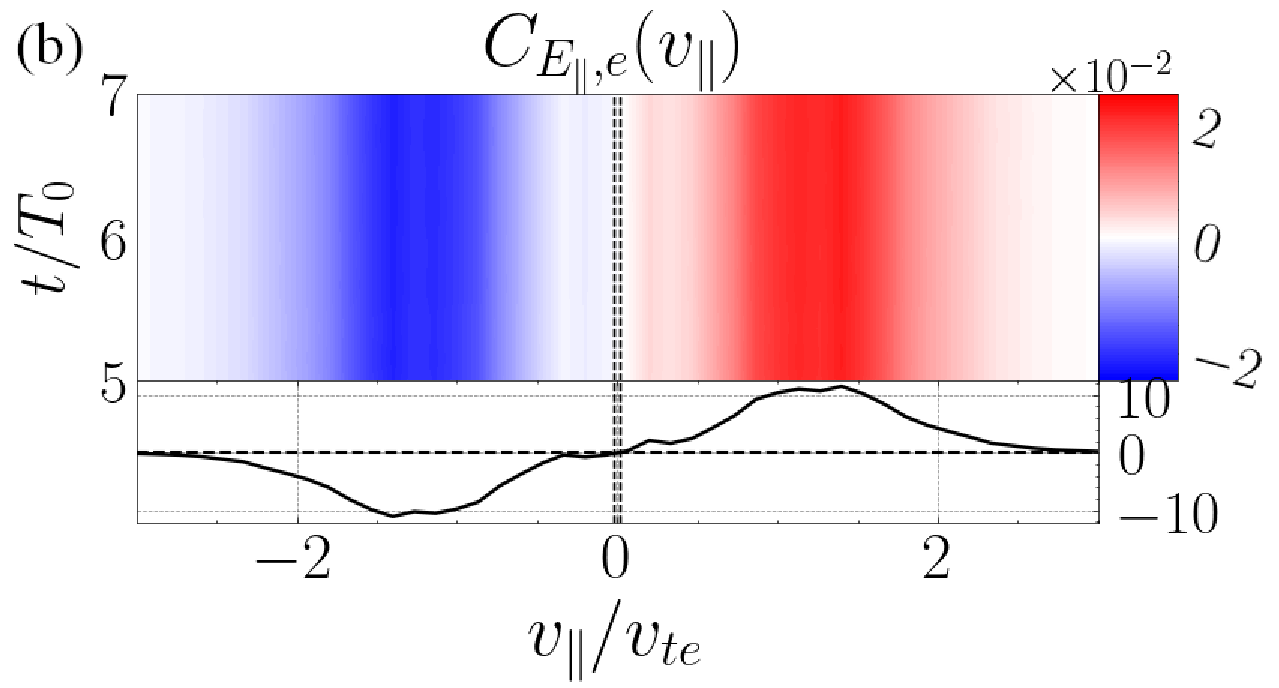}    % ekd_b10_k8_probe001_s2_CER_nc064_contourf_compressed_b.eps
    \end{center}
    \caption{Timestack field-particle correlation of driven, damped kinetic \Alfven~waves at $k_\perp \rho_i = 8$ with (a) $\beta_i = 1$ and (b) $\beta = 10$ and $\tau/T_0 = 1$ for both. The bipolar signature of Landau damping appears in (a) along with an antisymmetric, non-resonant feature. In (b), the non-resonant feature is dominant.}
    \label{fig:NRF}
\end{figure}

Finally, we use the single-wave simulations to address a prominent, non-resonant feature that begins to appear in the field-particle correlations for  $\beta_i \gtrsim 1$. Specifically, we see a large-amplitude feature across all parallel velocities that is antisymmetric about $v_\parallel = 0$. Fig.~\ref{fig:NRF} shows the non-resonant feature for two individual wave simulations at perpendicular wavenumber $k_\perp \rho_i = 8$ and (a) $\beta_i = 1$ and (b) $10$. At $\beta_i=1$, the non-resonant feature is visible but subdominant to the resonant, bipolar signature which is centered around $v_\parallel/v_{te} = 0.133$. At $\beta_i=10$, however, the amplitude of the non-resonant feature has completely eclipsed the amplitude of the resonant feature, which is centered close to zero at $v_\parallel/v_{te} = 0.018$. 

Our explanation of the source of this non-resonant feature are electron density oscillations, $\delta n_e$, that arise naturally in the turbulent plasma at $\beta_i \gtrsim 1$. A local spatial density perturbation has a Maxwellian $\delta f$, which is even in $v$ and therefore contributes an odd variation in $v_\parallel$ to the correlation due to the velocity derivative of the distribution function in Eq.~\eqref{eq:cepar}. The effect of a density perturbation on the parallel field-particle correlation is illustrated in a diagram shown in Figure~5(b) of \citet{McCubbin:2022}. \T{AstroGK} evolves the perturbed electron distribution function \citep{Numata:2010}, which in our kinetic \Alfven~wave simulations contains disturbances due to resonant field-particle interactions and due to non-resonant density oscillations. A density perturbation may appear in the field-particle correlation at high $\beta_i$ due to a combination of two factors. First, as $\beta_i$ increases, the relative magnitude of the compressive density perturbation to the resonant perturbations grows as the KAWs become more compressible \citep{TenBarge:2012b, Matteini:2020} and as the phase velocity of the damped KAW decreases and moves deeper into the core of the electron distribution function. Second, in order for the compressive signature to appear in the field-particle correlation, it must have an appropriate phase-offset from the parallel electric field fluctuation. At low $\beta_i$, $E_\parallel$ and $(\partial f_e/\partial v_\parallel)_{\delta n_e}$ are approximately out of phase, $\phi \simeq \pi/2$; as $\beta_i$ increases, the phase offset shifts and the product of $E_\parallel$ and $(\partial f_e/\partial v_\parallel)_{\delta n_e}$ contributes more significantly to the field-particle correlation. Appendix~\ref{sec:AppB} discusses the density perturbation and how it appears in the field-particle correlation in greater detail. In Section~\ref{sec:folding}, we present a technique for removing this feature from the data.

%==========================================================================

\section{Turbulence analysis}\label{sec:turb}

Next, we present the results of the field-particle correlation analysis of the four full turbulence simulations. The features of the bipolar signatures of electron Landau damping that were isolated in the suite of damped individual waves --- including variations in signature width, suppression of the negative portion of the bipolar signatures at low resonant velocity, and the presence of a non-resonant feature arising in $\beta_i = 1$ and $10$ --- are also present. To illustrate some of the commonly observed features, we present examples of the gyrotropic parallel field-particle correlation $C_{E_\parallel}(v_\parallel, v_\perp, t)$  in Fig.~\ref{fig:gyro} and of the timestack plot of $C_{E_\parallel}(v_\parallel, t)$ in Fig.~\ref{fig:tstacks}. 

\subsection{Gyrotropic Field-Particle Correlations}

\begin{figure}
    \begin{center}
        \includegraphics[width=0.49\textwidth]{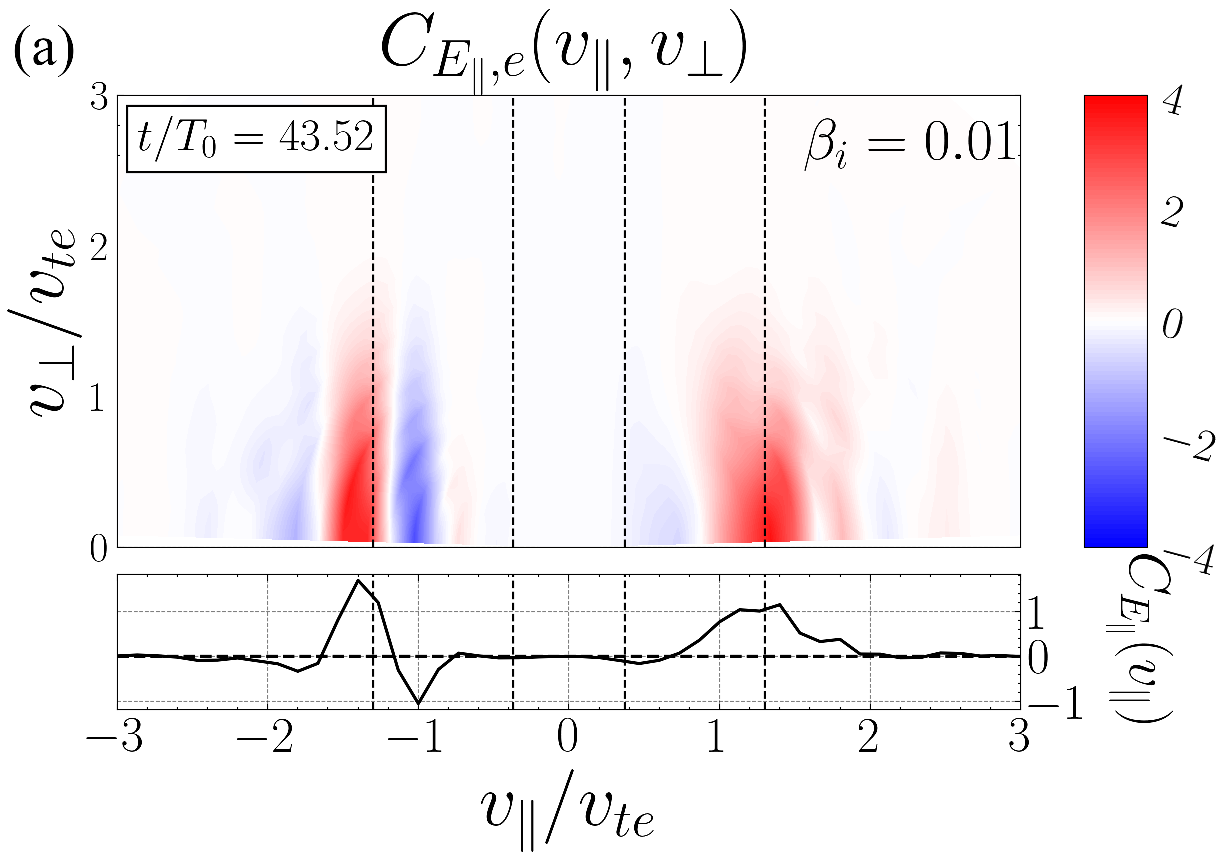}     % efpc21ao_probe013_s2_CE_nc1024_t2490_contourf_a.eps
        \includegraphics[width=0.49\textwidth]{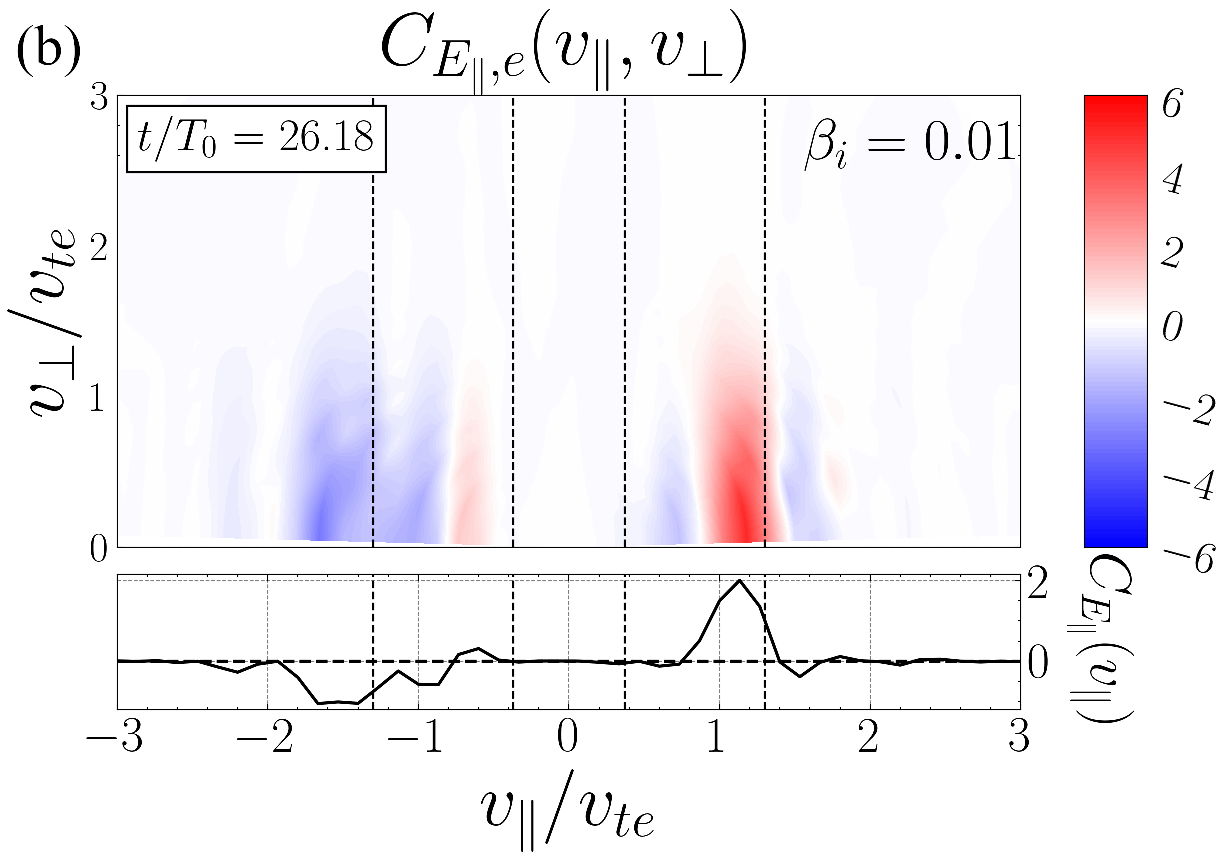}     % efpc21ao_probe020_s2_CE_nc1024_t1515_contourf_b.eps
        \includegraphics[width=0.49\textwidth]{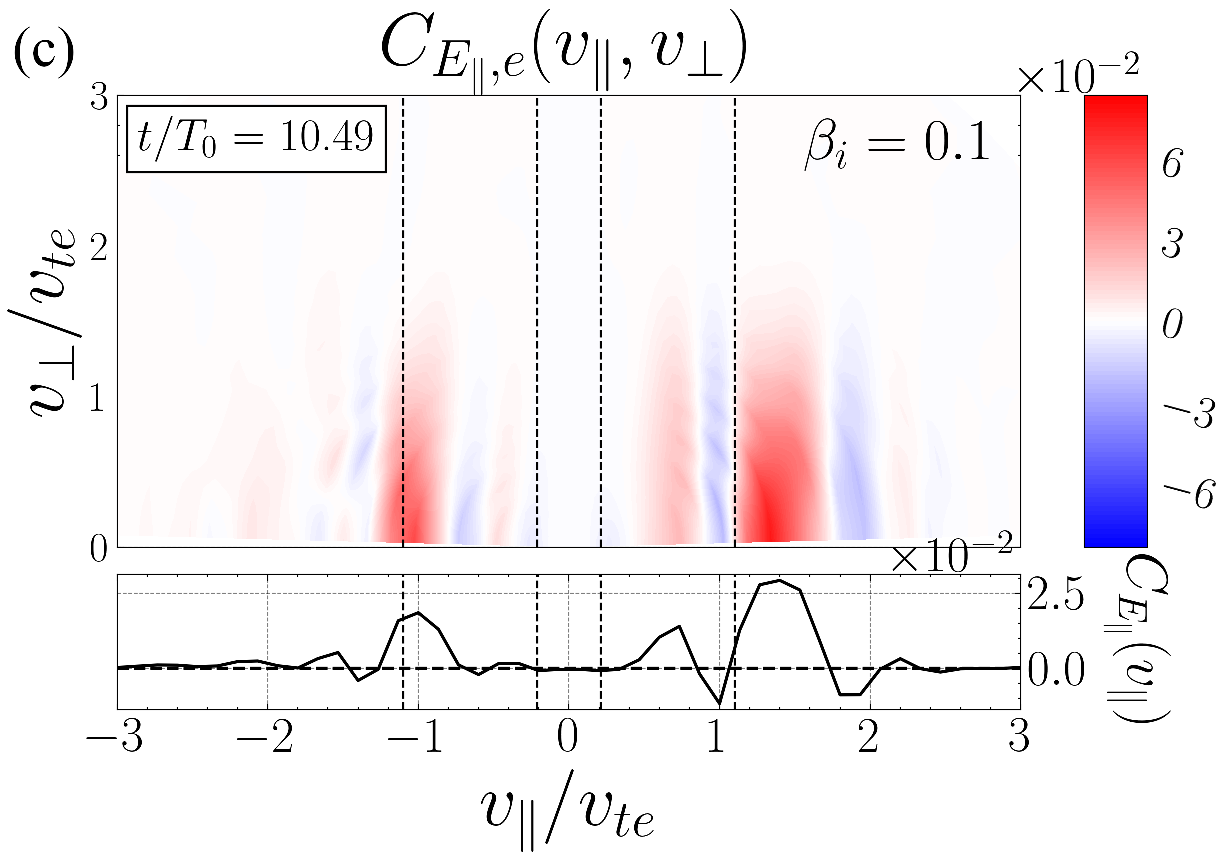}     % efpc22ak_probe021_s2_CE_nc512_t2057_contourf_c.eps
        \includegraphics[width=0.49\textwidth]{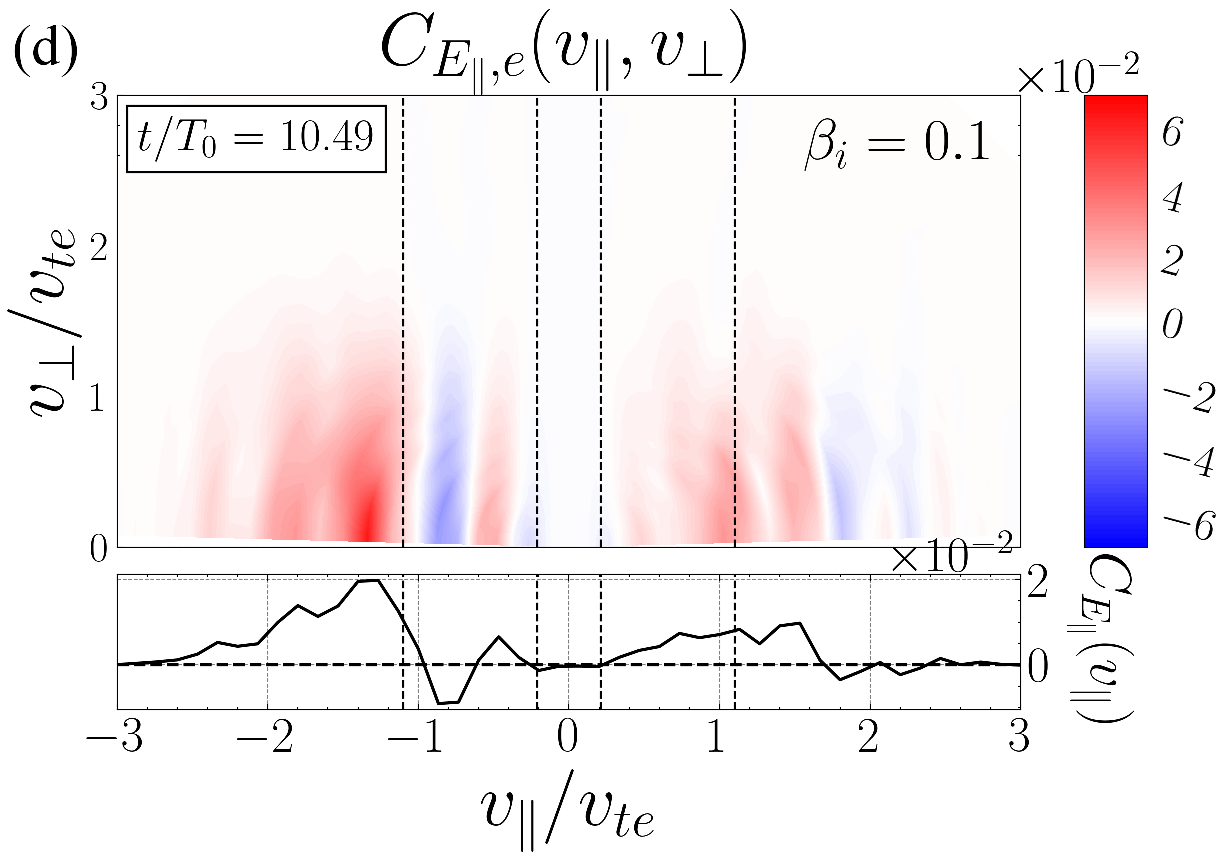}     % efpc22ak_probe024_s2_CE_nc512_t2058_contourf_d.eps
        \includegraphics[width=0.49\textwidth]{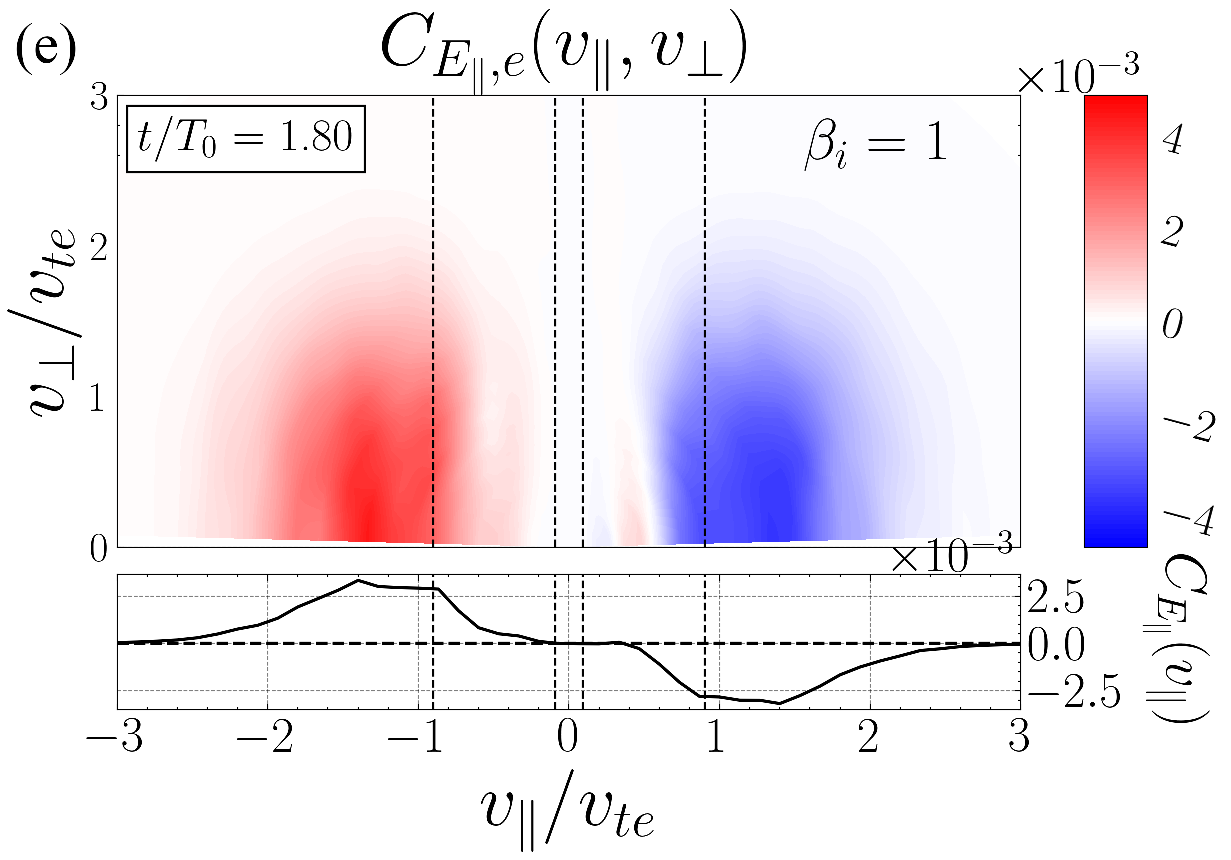}     % efpc23ai_probe013_s2_CE_nc256_t0629_contourf_v2_e.eps
        \includegraphics[width=0.49\textwidth]{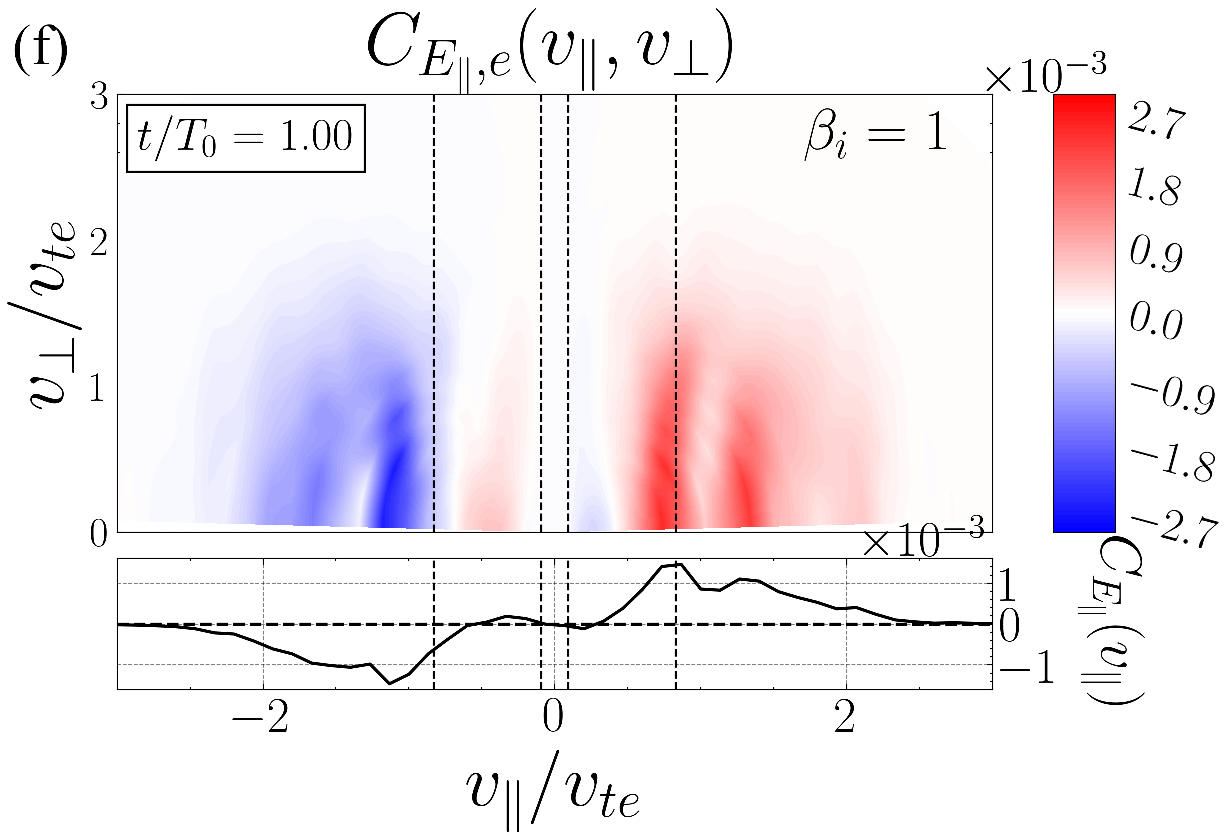}     % efpc23ai_probe015_s2_CE_nc256_t0230_contourf_f.eps
    \end{center}
    \caption{Gyrotropic parallel field-particle correlations $C_{E_\parallel, e}(v_\parallel, v_\perp)$ from different probes in the (a,b) $\beta_i = 0.01$ ($\tau/T_0 \simeq 17.2,\ \tau/T_{min} \simeq 47.0$), (c,d) $\beta_i = 0.1$ ($\tau/T_0 \simeq 2.47,\ \tau/T_{min} \simeq 17.7$) and (e,f) $\beta_i = 1$ ($\tau/T_0 \simeq 2.43,\ \tau/T_{min} \simeq 35.0$) simulations. The features of the velocity-space signatures are described in the text.}
    \label{fig:gyro}
\end{figure}

Fig.~\ref{fig:gyro} shows six example gyrotropic field-particle correlations, two from each of the turbulence simulations at $\beta_i = 0.01, 0.1$ and $1$. We exclude $\beta_i=10$, since signatures are not easily observed in this case, given that the $v_\parallel$-resolution of our simulations is binned at $\Delta v_\parallel/v_{te} = 0.133$ while the largest fully resolved wave mode in the $\beta_i=10$ simulation has a phase velocity around $v_{ph}/v_{te} \sim 0.1$. Though the positive portion of the damping signature can been seen in some cases, the negative portion is unresolved (and would be strongly suppressed due to the factor of $v_\parallel^2$). Further, the large relative magnitude of the non-resonant feature dominates the field-particle correlations at $\beta_i=10$.  

Each example plot in Fig.~\ref{fig:gyro} shows a contour map of $C_{E_\parallel, e}(v_\parallel, v_\perp)$, the gyrotropic correlation, that has been averaged over a correlation interval of length $\tau$, included in the caption in terms of both the largest ($T_0$) and smallest ($T_{min}$) KAW periods resolved within each simulation, and centered at some time $t/T_0$ (displayed in the upper left-hand corner). The range of phase velocities at which we expect to see signatures of Landau damping are marked with vertical dotted lines. The inner vertical lines correspond to the magnitude of the phase velocity of the lowest frequency waves in this range: having a period of either $T_0$ (the driving-scale wave) or $T_{max}$ (the wave at the onset of strong damping) for $\beta_i \geq 0.1$. This minimum damped frequency occurs when the damping rate reaches a threshold of $-\gamma/\omega = 0.1$, which we take to be the onset of strong damping. Fig.~\ref{fig:LGKDR}(b) shows that for $\beta_i = 0.1$ and $1$, this wave is above the driving-scale wave of $k_\perp \rho_i = 2$. For $\beta_i = 10$, the initial peak above $-\gamma/\omega = 0.1$ is due to ion Landau damping; therefore, we take the onset of strong electron damping to be around $k_\perp \rho_i \sim 3$. The vertical dotted lines at higher velocities correspond to the highest frequency wave that we expect to undergo Landau damping (with period $T_{min}$), found simply by the taking the peak of the curve in Figure~\ref{fig:LGKDR}(a). The phase velocities of these two limiting cases are marked for both upward ($v_\parallel/v_{te} > 0$) and downward ($v_\parallel/v_{te} < 0$) wave propagation. Each line plot in the lower panels of Fig.~\ref{fig:gyro} displays the reduced correlation integrated over all $v_\perp$, $C_{E_\parallel,e}(v_\parallel)$.

The top row of Fig~\ref{fig:gyro} shows gyrotropic correlations from the $\beta_i = 0.01$ turbulent simulation. Panel (a) is from a probe in the $z=0$ plane of the simulation box, calculated using a correlation interval of length $\tau \simeq 17.2\ T_0$ ($\tau \simeq 47.0\ T_{min}$) and shown centered at time $t/T_0 = 43.52$. This example contains a pair of bipolar signatures: one at positive and the other at negative parallel velocity, indicating damping of KAWs propagating in both directions relative to the background magnetic field. As shown in Fig.~\ref{fig:vph_fwhm}(a), the location of the zero crossing in the integrated correlation $C_{E_\parallel}(v_\parallel)$ in the lower panel corresponds to the phase velocity of the damped wave. Thus, the upward-propagating damped wave has a lower phase velocity than the downward propagating damped wave. However, the signature at $v_\parallel/v_{te} > 0$ is significantly wider than the one at $v_\parallel/v_{te} < 0$. This seems to contradict the conclusion of Fig.~\ref{fig:vph_fwhm}(b), which predicts a direct relationship between phase velocity and signature width for constant $\beta_i$. However, the $v_\parallel/v_{te} > 0$ signature has a subtle, divided positive peak. This may indicate that the signature is a superposition of the damping signatures of two waves closely spaced in parallel phase velocity such that their individual bipolar signatures cannot be resolved. Instead, the two signatures merge into a single, broadened peak \citep{Horvath:2020}. This interpretation is supported by gyrotropic plots centered at earlier times and by the timestack plot of the correlation at this probe point. The plot in panel (b) is an example of an imbalanced signature, which occurs when damping is present for only upward or downward propagating waves at a given time. In this case, an upward-propagating wave produces a bipolar signature but no clear resonant damping signature is visible for $v_\parallel/v_{te} < 0$. This correlation is averaged over the same interval $\tau$, but is taken at a probe point along the $z$-axis ($z \neq 0$) and is centered at an earlier time. 

The second row of Fig.~\ref{fig:gyro} contains correlations from the $\beta_i = 0.1$ simulation, taken from two different probe points along the $z$-axis ($z \neq 0$). Each is averaged over a correlation interval of $\tau = 2.47\ T_0$ ($17.7\ T_{min}$) and is centered at $t/T_0 = 10.49$. Note that the velocity range marked by the pairs of vertical dotted lines has shifted toward smaller $v_\parallel$, reflecting the shift toward smaller parallel phase velocities in the dispersion relation as $\beta_i$ increases, as shown in Fig.~\ref{fig:LGKDR}(a). In (c), the example correlation shows signatures of damped KAWs propagating in both directions along the background magnetic field. For $v_\parallel/v_{te} > 0$, there is evidence of two bipolar signatures, spaced far enough apart to be resolved in $v_\parallel$ space. The higher-frequency signature is centered around the upper vertical dotted line, indicating that the damped wave is at the maximum phase velocity that we expect in this simulation. This signature is also wider than the signature at smaller $v_\parallel$, consistent with our theoretical expectations illustrated in Fig.~\ref{fig:vph_fwhm}(b). Note also that the lower resonant-velocity has a significantly suppressed negative portion of its bipolar signature that is difficult to observe at this color scale, consistent with the effect of $v_\parallel^2$ weighting as illustrated in Fig.~\ref{fig:vparweight}. For $v_\parallel/v_{te} < 0$, one signature is visible just below the dotted line marking the magnitude of the most highly damped phase-velocity. The relative width of this signature is between that of the two upward-propagating wave signatures, as expected.
Panel (d) shows an example of another imbalanced signature with a clear velocity-space signature of electron Landau damping for only downward-propagating waves. In this case, there are two bipolar structures: one centered at $v_\parallel/v_{te} \simeq -1$ and another at $v_\parallel/v_{te} \simeq -0.25$. Again, the low-velocity signature is strongly affected by the $v_\parallel^2$ weighting; however, this fainter signature is persistent throughout the whole simulation. The structures in the positive velocity region of the plot do not show a clear bipolar signature indicative of electron Landau damping of upward propagating kinetic \Alfven~waves, but rather may simply be the remnants of oscillatory fluctuations that are poorly cancelled out by the average over our chosen correlation interval $\tau$.\footnote{Note that it is impossible to completely cancel out oscillatory signal from field-particle correlations in the case of broadband turbulence, since only oscillations from constant amplitude waves with periods of $\tau$ (or integer multiples of $\tau$) will exactly cancel. Furthermore, turbulence is intermittent in space and time, meaning that the magnitude of the remnant, oscillatory signal that does not cancel out over the correlation interval in broadband turbulence is expected---and generally observed---to be a highly variable value. This imperfect cancellation leads to an effective amplitude threshold for the identification of electron Landau damping.} 

The third row of Fig.~\ref{fig:gyro} shows gyrotropic correlations from the $\beta_i = 1$ simulation, from two different probe points in the midplane ($z = 0$). These correlations are averaged over an interval $\tau = 1.43\ T_0$ ($\tau = 35.0\ T_{min}$). As before, the vertical dotted lines mark out a range of resonant parallel phase velocities that are yet nearer to $v_\parallel = 0$. The largest expected velocity is now at $v_\parallel/v_{te} < 1$, meaning that the  $v_\parallel^2$-weighting will act to suppress all of the signatures relative to any incomplete cancellation that appears at $v_\parallel/ v_{te}>1$. These examples show how the non-resonant feature begins to dominate the correlations at $\beta_i \gtrsim 1$, and this feature is particularly prominent in panel (e). Despite this, a faint bipolar signature of an upward-propagating wave can still be seen in the color map of this example at the shown scale, though it has become indistinguishable in the reduced correlation due to the overwhelming magnitude of the non-resonant feature. As discussed for the damped individual waves, this feature arises due to the increased compressibility of KAWs at $\beta_i \gtrsim 1$, and due to the changing phase offset $\phi \neq \pi/2 $ between $E_\parallel$ and the compressive density fluctuation in $\partial f/\partial v_\parallel$. In many cases, this non-resonant feature hinders the observation of bipolar signatures at $\beta_i \gtrsim 1$, though at times it can be removed so that the signatures may be observed via a folding technique discussed in Section~\ref{sec:folding}. At the time of the correlation in panel (f), the non-resonant feature is smaller in amplitude relative to the resonant signatures, and a bipolar structure is visible (with a somewhat attenuated negative portion) at $v_\parallel/v_{te} \simeq 0.3$.

In summary, this section presents one of the key results of this paper, the \emph{gyrotropic velocity-space signature of electron Landau damping} $C_{E_\parallel}(v_\parallel, v_\perp)$, as exemplified for $\beta_i=0.01$ in Fig.~\ref{fig:gyro}(a) at  $v_\parallel/v_{te} \simeq -1.2$ and (b) at $v_\parallel/v_{te} \simeq +0.75$, for $\beta_i=0.1$ in Fig.~\ref{fig:gyro}(c) at  $v_\parallel/v_{te} \simeq -0.8$ and $+1.1$ and (d) at  $v_\parallel/v_{te} \simeq -1.0$, and for $\beta_i=1$ in Fig.~\ref{fig:gyro}(f) at  $v_\parallel/v_{te} \simeq +0.4$. These qualitative and quantitative features of the velocity-space signatures as a function of the ion plasma $\beta_i$ provide an essential framework for identifying electron Landau damping in both kinetic turbulence simulations and spacecraft observations. 
 
\subsection{Timestack Plots of the Field-Particle Correlation}

\begin{figure}
    \begin{center}
        \includegraphics[width=0.48\textwidth]{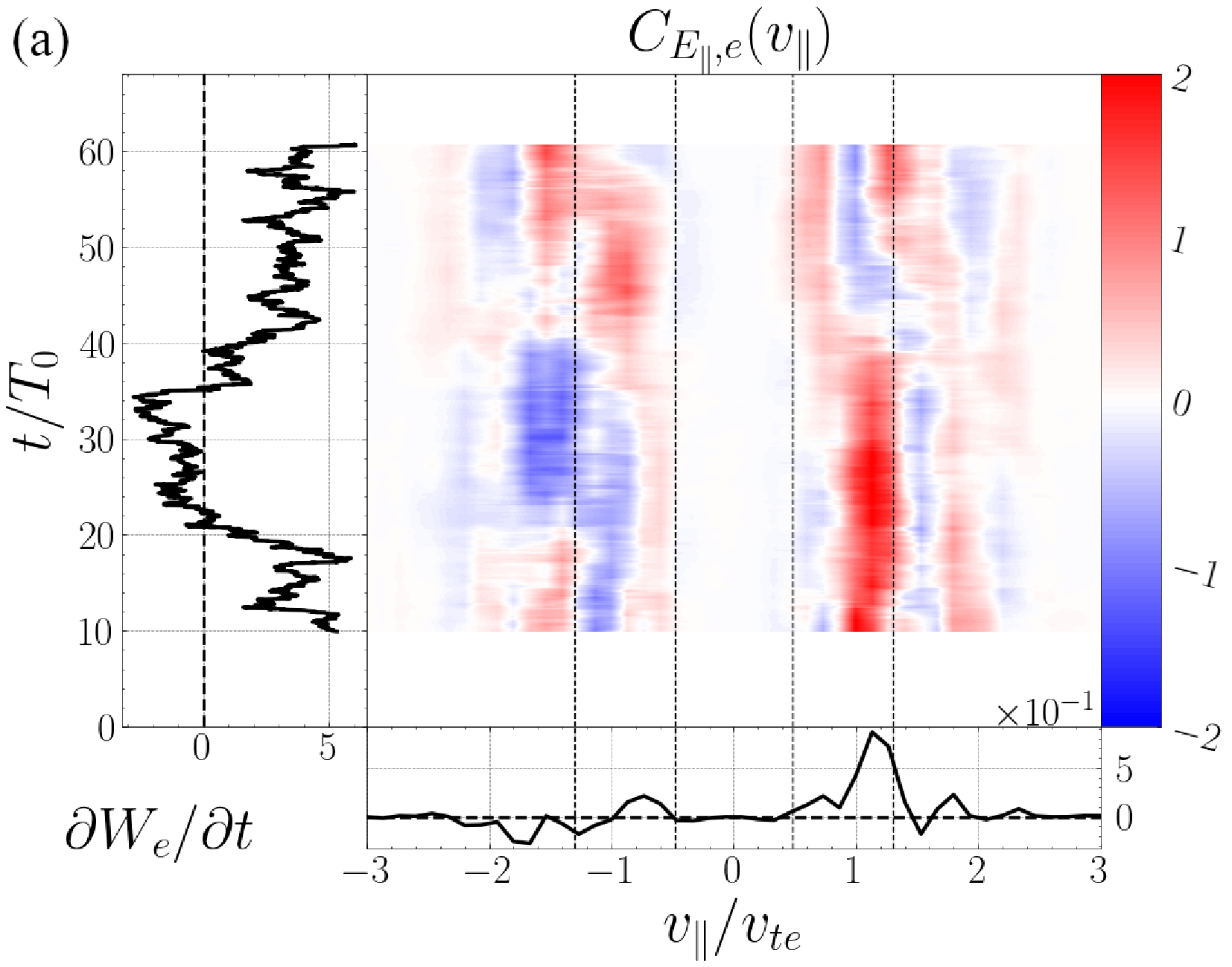}    % efpc21ao_probe020_s2_CER_nc1024_contourf_a.eps
        \includegraphics[width=0.48\textwidth]{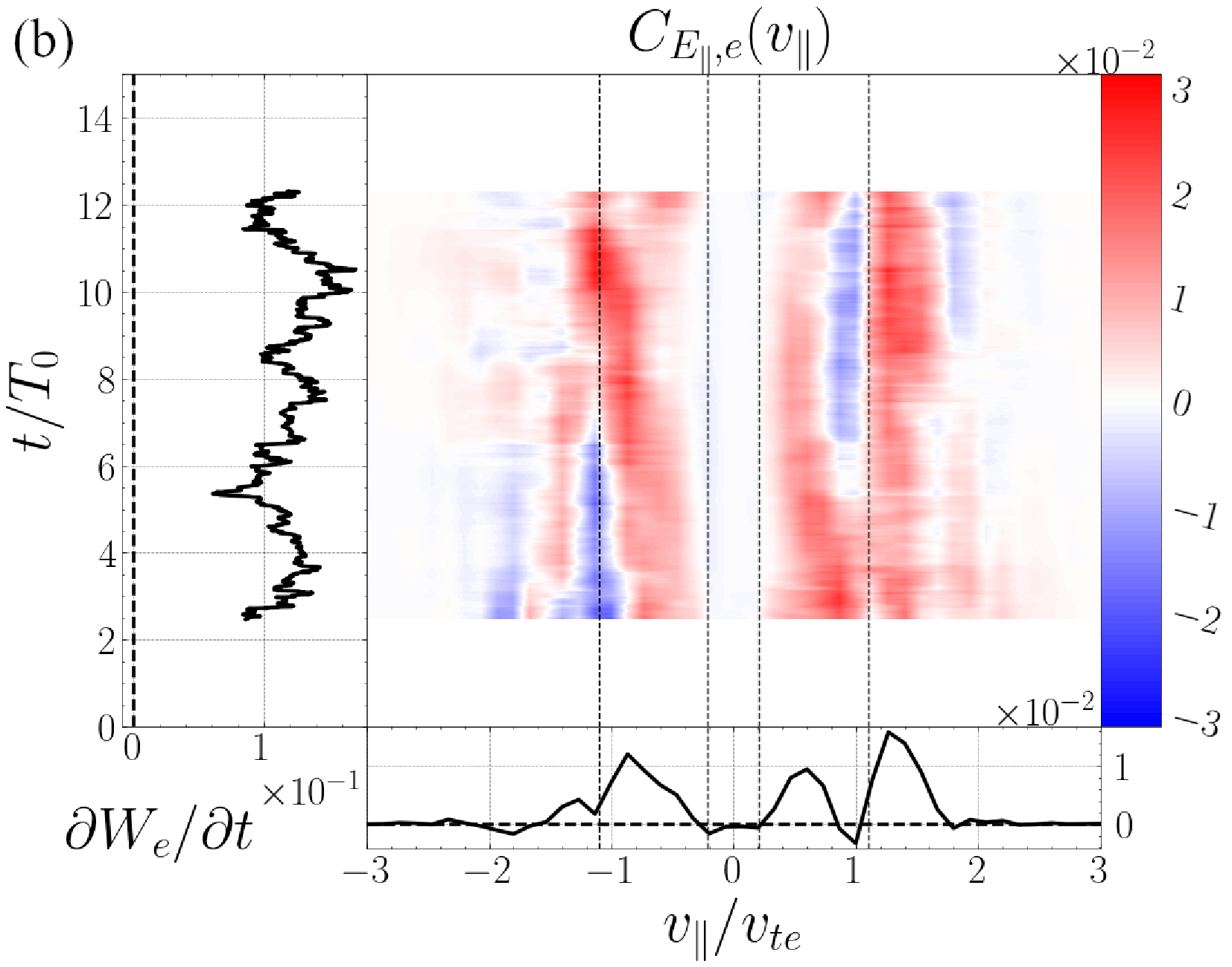}    % efpc22ak_probe021_s2_CER_nc1024_contourf_b.eps
        \includegraphics[width=0.48\textwidth]{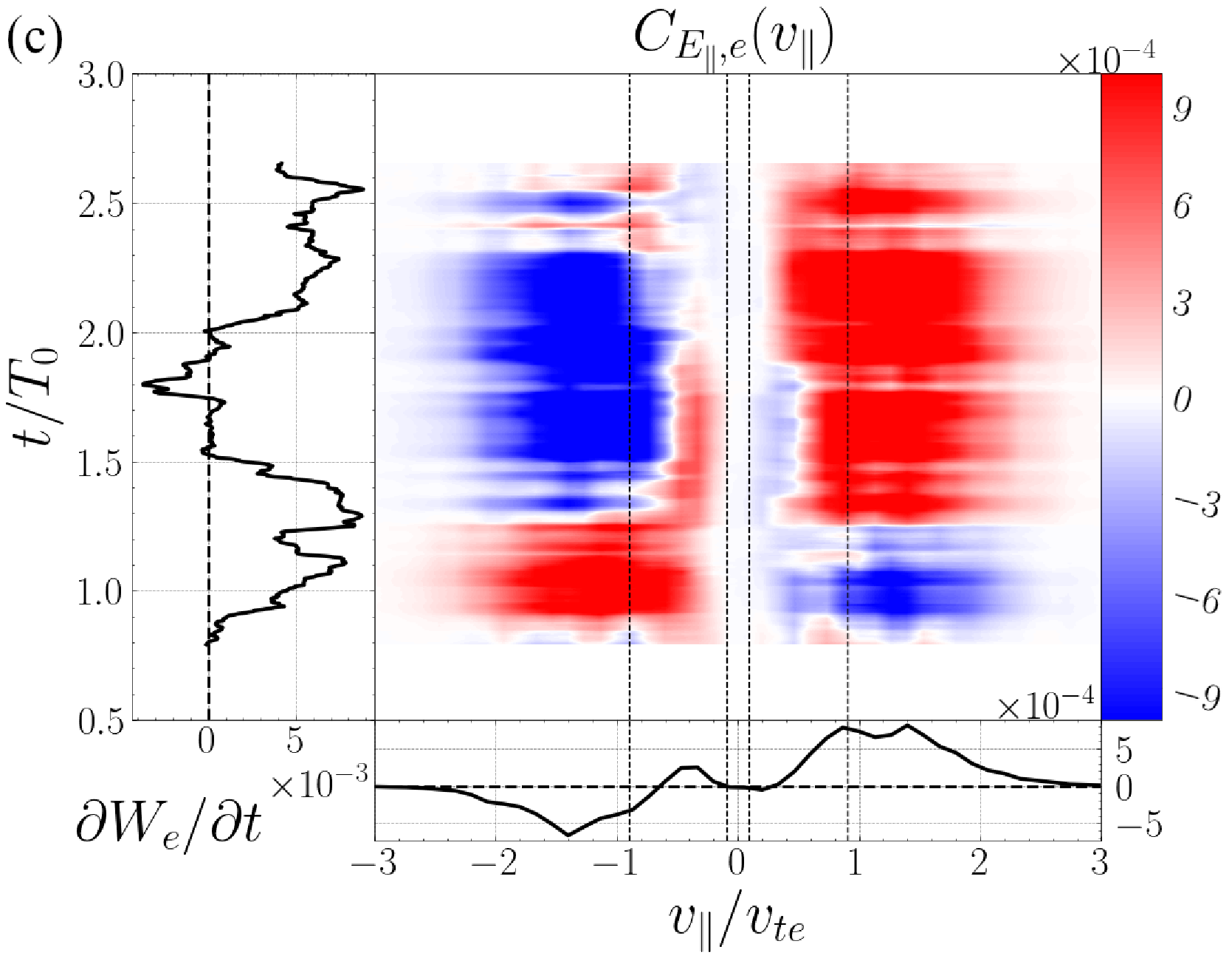}    % efpc23ai_probe016_s2_CER_nc256_contourf_c.eps
        \includegraphics[width=0.48\textwidth]{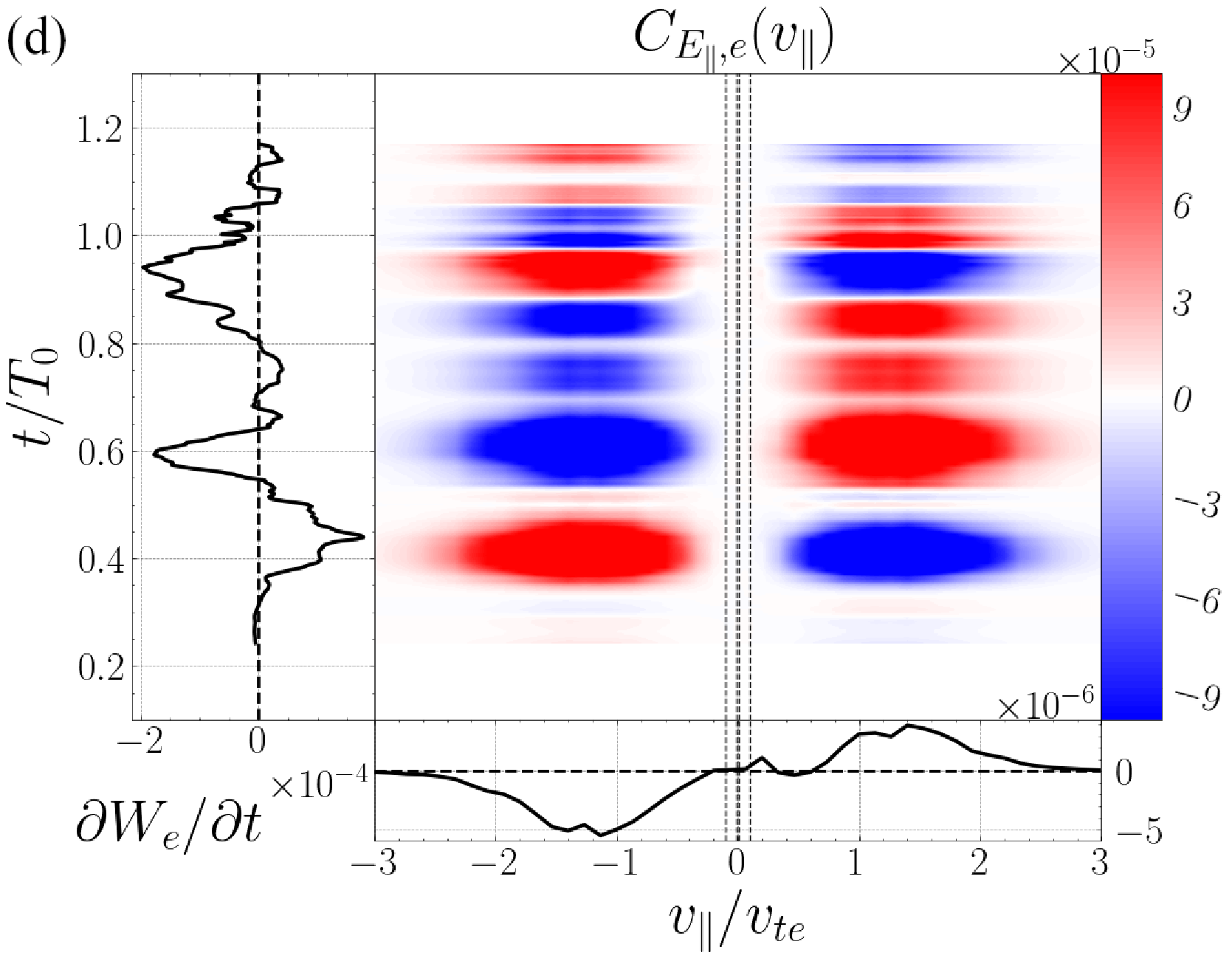}    % efpc24ap_probe003_s2_CER_nc064_contourf_d.eps
    \end{center}
    \caption{Timestacks of the reduced field-particle correlations $C_{E_\parallel, e}(v_\parallel)$ from the simulations with plasma (a) $\beta_i = 0.01$ $(\tau/T_0 \simeq 17.2,\ \tau/T_{min} \simeq 47.0)$, (b) $0.1$ $(\tau/T_0 \simeq 4.94,\ \tau/T_{min} \simeq 35.3)$, (c) $1$ $(\tau/T_0 \simeq 1.43,\ \tau/T_{min} \simeq 35.0)$, and (d) $10$ $(\tau/T_0 \simeq 0.06,\ \tau/T_{min} \simeq 1.21)$. The correlations in panels (a) and (b) are calculated from the same probe points as those in Figure~7(b) and (c), respectively.}
    \label{fig:tstacks}
\end{figure}

To observe how the gyrotropic velocity-space signatures of electron energization in Fig.~\ref{fig:gyro} evolve over time, we take advantage of the fact that these gyrotropic signatures do not strongly depend on $v_\perp$ (other than the exponential decrease with increasing $v_\perp$, a consequence of the underlying Maxwellian equilibrium velocity distribution) to integrate over $v_\perp$ and generate a timestack plot of reduced parallel correlation $C_{E_\parallel}(v_\parallel, t)$. These timestack plots are invaluable for determining the persistence of a signature over time and for establishing an overall view of the net energy transfer at a given location. In  Fig.~\ref{fig:tstacks}, we present one timestack plot from a single probe in each of the four simulations at (a) $\beta_i=0.01$, (b) $0.1$, (c) $1$, and (d) $10$.  In each, we integrate over the full duration of the simulation to obtain the time-integrated, reduced parallel correlation $C_{E_\parallel}(v_\parallel)$ at the probe position (lower panels); we also integrate over $v_\parallel$ to obtain the rate of change of the spatial energy density  $(\partial W_s/ \partial t)_{E_\parallel}=j_{e,\parallel} E_\parallel$  due to the parallel electric field.

In Fig.~\ref{fig:tstacks}(a), we show the timestack from the $\beta_i=0.01$ simulation at the same probe point as the gyrotropic correlation in Fig.~\ref{fig:gyro}(b) and averaged over the same correlation interval $\tau/T_0 \simeq 17.2,\ \tau/T_{min} \simeq 47.0$. Note that the $v_\parallel/v_{te} > 0$ signature observed at $t/T_0 = 26.18$ in the gyrotropic plot was present at the beginning of the averaging interval and persisted for over $30$ domain-scale wave periods. Other signatures (of both upward and downward propagating waves) appear later in the simulation. The proximity of these signatures in velocity-space can make it difficult to distinguish their bipolar form due to overlap with signatures at other parallel phase velocities in addition to the $v_\parallel^2$ suppression of the bipolar form. The timestack from the $\beta_i=0.1$ simulation in (b) is taken from the same probe point as the gyrotropic plot in Fig.~\ref{fig:gyro}(c) but averaged over a longer correlation interval, $\tau/T_0 \simeq 4.94,\ \tau/T_{min} \simeq 35.3$. At $t/T_0 = 10.49$---the central time of the correlation interval for the gyrotropic plot---the three signatures seen in the gyrotropic plot are clearly visible. The timestack shows that each of these signatures are present through the whole simulation, but two of the three shift somewhat in velocity, leading to the overlap of signatures at slightly different resonant parallel velocities when integrated over the full simulation (lower panel). 

In Fig.~\ref{fig:tstacks}(c), we plot a timestack from the $\beta_i=1$ simulation, taken at a midplane ($z=0$) probe that is not represented in Fig.~\ref{fig:gyro}, but averaged over the same correlation interval $\tau/T_0 \simeq 1.43,\ \tau/T_{min} \simeq 35.0$. At this value of $\beta_i$, the broad non-resonant feature has become prominent in the correlation, though the lower amplitude resonant signatures can still be seen at small parallel velocities. For instance, at $v_\parallel/v_{te} \simeq -0.2$ a signature is visible in the range $1.2 < t/T_0 < 2$, which creates an obvious positive region in the lower, time-integrated panel. Note that the negative portion of this bipolar signature is not visible in the timestack at this color scale and does not appear in the time-integrated correlation. At many probe points, such evidence of a damping signature at lower parallel velocities than the non-resonant feature is much more difficult to observe. When $\beta_i$ is increased further, these difficulties are exacerbated, as illustrated in Fig.~\ref{fig:tstacks}(d) for the $\beta_i=10$ simulation. This correlation is taken from a midplane probe, and averaged over a mere $\tau/T_0 \simeq 0.06$, or $\tau/T_{min} \simeq 1.21$. The correlation interval is a very small fraction of the the driving scale wave period $T_0$, so we would not expect to see bipolar signatures in the timestack due to these waves. The full interval of this simulation is long enough to give $\tau/T_0 \simeq 1.01$ (shown in the lower panel), so in order to display a timestack at this plasma beta we have chosen a low value of $\tau/T_0$ for the contour plot. As in the case of $\beta_i=1$, at some probe points in the $\beta_i=10$ simulation evidence of energization due to Landau damping can be found by observing positive bumps near the range of expected resonant velocities in the time-integrated correlation. However, the increased amplitude of the non-resonant feature with respect to the resonant damping signatures, coupled with the finite resolution of our parallel velocity grid, make damping signatures nearly impossible to observe clearly in correlations for $\beta_i=10$. It is worthwhile noting here that, at higher ion plasma beta values $\beta_i \gtrsim 1$, ion Landau and transit-time damping are expected to play a significant role in the dissipation of turbulent energy at scales $k_\perp \rho_i \sim 1$ \citep{Klein:2017}, so electron Landau damping of turbulent energy that reaches the kinetic \Alfven~wave cascade at $k_\perp \rho_i \gg 1$  is expected to be less significant in the overall dissipation of the turbulent cascade.

\section{Discussion: Features of the Velocity-Space Signatures of Electron Landau Damping as a Function of Plasma {\bf $\beta_i$}}\label{sec:discussion}

In the previous sections, we have demonstrated general qualitative and quantitative features that are present in the velocity-space signatures of electron Landau damping as a function of the ion plasma $\beta_i$ for linear kinetic \Alfven~waves. These indicate the general trends that should be visible in the velocity-space signatures generated by turbulence dissipation measured at different locations throughout the inner heliosphere. Next, we presented example velocity-space signatures from our suite of four kinetic plasma turbulence simulations to illustrate how these general trends appear in the context of broadband turbulent energy dissipation via the mechanism of electron Landau damping. Here, we recap these features and discuss implications for making \emph{in situ} observations of electron Landau damping. 

\subsection{Imbalanced Velocity-Space Signatures}

Since the earliest studies of turbulence in magnetohydrodynamic (MHD) plasmas, it was recognized that the turbulent cascade is mediated by nonlinear interactions between \Alfven~wavepackets propagating up and down the magnetic field \citep{Iroshnikov:1963,Kraichnan:1965,Sridhar:1994,Goldreich:1995}.  These ``collisions'' between counterpropagating \Alfven~wavepackets have been invoked as the fundamental building block of astrophysical plasma turbulence, with theoretical \citep{Howes:2013a}, numerical \citep{Nielson:2013}, and experimental \citep{Howes:2012b,Howes:2013b,Drake:2013} confirmation of the underlying physics. Further work has illustrated the role of the \Alfven~wave collisions in the self-consistent development of current sheets in kinetic plasma turbulence \citep{Howes:2016b,Verniero:2018a,Verniero:2018b} and the role of Landau damping in the dissipation of those current sheets \citep{Howes:2018}.  
When the turbulent fluctuations have more \Alfvenic~energy flux propagating in one direction along the local magnetic field than the other direction---known as \emph{imbalanced turbulence} \citep{Lithwick:2003,Lithwick:2007,Beresnyak:2008,Chandran:2008,Markovskii:2013}, or turbulence with nonzero cross helicity---it is predicted to alter the dynamics of the turbulent cascade \citep{Meyrand:2021, Squire:2022}, and therefore will likely also impact the dominant mechanisms of its dissipation.  

Even if the turbulence is balanced on longer timescales, during short periods of time local regions of imbalanced turbulence may develop. Averaged over space and time, such imbalanced regions are not thought to effect the overall energy spectra of the turbulent cascade \citep{Perez:2009}. Further, in this work we find that one may measure a local dominance of wave damping in one direction or another even if the turbulence remains relatively balanced locally. The field-particle correlation analysis of our suite of turbulence simulations frequently finds asymmetric signatures of electron Landau damping, \emph{e.g.}, Fig.~\ref{fig:gyro}(b). An inspection of the timestack plots in Fig.~\ref{fig:tstacks} clearly shows that the damping of the turbulent fluctuations varies significantly in time over the evolution of the simulations, with upward or downward propagating fluctuations dominating the energy transfer to the electrons over intervals of time from a few to tens of characteristic wave periods. This is consistent with the results of our previous simulations \citep{Horvath:2020} and with \emph{in situ} observations in the magnetosheath \citep{Afshari:2021}. Using \emph{MMS} data, \citet{Afshari:2021} connected the imbalance in the wave damping with the dominant direction of the Poynting flux in the data interval. In our suite of simulations presented here, however, we find that the overall flux of electromagnetic energy remains roughly balanced at all times. This is not surprising since the turbulence is driven by approximately equal-amplitude, oppositely-directed \Alfven~waves. Nonetheless, on shorter timescales and at individual probe positions in the simulations, local imbalances of damping arise that lead to the asymmetric signatures that are frequently observed in these simulations and in \emph{MMS} observations \citep{Afshari:2021}.

\subsection{Temporal Intermittency and the Overlap of Velocity-Space Signatures}

As was found in previous field-particle correlation studies of electron Landau damping in simulations \citep{Horvath:2020, Horvath:2022}, we find that electron damping signatures in broadband, dispersive KAW turbulence contain more structure than the simple bipolar signature of a single, damped wave. This was illustrated in Fig.~\ref{fig:gyro}(c), where the time-integrated signature at $v_\parallel > 0$ is composed of at least two overlapping bipolar signatures, creating a broadened, double-peaked structure. In the timestack plot of Fig.~\ref{fig:tstacks}(b), closely-spaced signatures overlap in the time-average to create the appearance of a single signature that `shifts' in $v_\parallel$. Note that, if the resonant velocities were spaced more closely and were unresolved along $v_\parallel$, this would result in a smoothly broadened signature. The superposition of multiple bipolar signatures creating a complicated final signature was most commonly observed in our lowest beta simulation ($\beta_i=0.01$), which has the largest range of expected resonant parallel phase velocities. Additionally, \T{AstroGK} evolves low-beta plasmas significantly faster than high-beta plasmas, resulting in a longer time interval over which temporally intermittent wave damping can be observed (see the simulation lengths, $\Delta t/T_0$, in Table~1). If we continued running the higher beta simulations to create longer time intervals, we would expect to see the effects of signature superposition ubiquitously (with the exception of the $\beta_i = 10$ simulation, in which the $v_\parallel$ resolution is insufficient to resolve multiple signatures in the region of expected damping). In observations, obtaining long time intervals---in terms of the underlying kinetic \Alfven~wave periods---is not a problem. On the other hand, finite instrumental resolution in parallel velocity 
\citep{Verniero:2021a,Verniero:2021b} is a significant limitation that must be taken into account when interpreting the velocity-space signatures from the field-particle correlation analysis of spacecraft observations. 

\subsection{Velocity-Space Resolution and Weighting by $v_\parallel^2$}

The effects of $v_\parallel^2$ weighting on the velocity-space signatures are visible in the various examples from the turbulence simulations in Figures~\ref{fig:gyro} and~\ref{fig:tstacks}. In particular, we have shown that this weighting may obscure the negative portion of a bipolar signature of electron Landau damping when the parallel resonant phase velocity falls at a sufficiently small value of $v_\parallel/v_{te}$.  From our individual wave simulations in Sec.~\ref{sec:lin}, we found that our binned parallel velocity resolution of $\Delta v_\parallel/v_{te} = 0.133$, coupled with the $v_\parallel^2$ weighting,  resulted in difficulty observing the full bipolar structure of the velocity-space signatures of electron Landau damping at parallel velocities $v_\parallel/v_{te} \le 0.3$. The entire dispersive range of parallel phase velocities
is above this threshold for the $\beta_i = 0.01$ simulations, but as $\beta_i$ is increased, more and more of the dispersive range of phase velocities falls below this threshold. For \emph{in situ} signatures, clear bipolar signatures may therefore be more easily observed for a given finite resolution in parallel velocity closer to the Sun, where $\beta_i$ may be smaller due to increased local magnetic energy density relative to the plasma thermal energy density. For plasma conditions and velocity resolutions where $v_\parallel^2$ does result in obscuring the negative portion of the characteristic bipolar structure, electron Landau damping may still be identified through the timestack correlations. If energization is persistent and localized at a given parallel velocity in a regime where the negative portion is not observed, this still indicates the action of a resonant energization mechanism that may be identified as electron Landau damping.

\subsection{Variations in Velocity-Space Signature Width}

As discussed above, in both individual kinetic \Alfven~wave and turbulence simulations, we observe that the velocity-space signature of electron Landau damping varies in width as the perpendicular wavenumber and plasma parameters change. We explain this relationship by showing that the increased width (quantified by the full-width half maximum value of the positive portion of the bipolar signature) is associated with an increased normalized damping rate of KAWs, as shown in Figure~\ref{fig:vph_fwhm}(b). Our simulations indicate that we should expect to observe widths of the velocity-space signatures of electron Landau damping that depend on the local plasma $\beta_i$ and the perpendicular scale $k_\perp \rho_i$ of the wave being damped, which is generally a monotonically increasing function of the parallel phase velocity, as shown in dispersion relation plotted in Figure~\ref{fig:LGKDR}(a). As observed in the gyrotropic plot of Fig.~\ref{fig:gyro}(a) at $v_\parallel/v_{te}>0$, if an unexpectedly wide velocity-space signature is observed, 
it may indicate the damping of multiple dispersive kinetic \Alfven~wavepackets with different characteristic perpendicular wavenumbers $k_\perp \rho_i$ (and therefore multiple resonant parallel phase velocities $\omega/k_\parallel v_{te}$).  In this case, the overlap of multiple bipolar signatures is expected to lead to a broadened velocity-space signature, as illustrated in Figure~1 of \citet{Horvath:2020}.

\subsection{Non-resonant feature}

Finally, here we show that, as the plasma parameters are varied, a non-resonant feature may arise in the field-particle correlation that obscures the signatures of secular particle energization in velocity-space. Our results show, for a constant ion-to-electron temperature ratio, that a non-resonant feature due to electron density oscillations becomes problematic for $\beta_i \geq 1$. This feature is antisymmetric in $v_\parallel$, so it cancels out when the correlation is integrated over parallel velocity, yielding zero net change to the total electron spatial energy density $W_e({\bf r}, t)$. However, it is the velocity-space signatures of the rate of change of the phase-space energy density $w_e({\bf r}, {\bf v}, t)$ that are used to distinguish the mechanisms of energy transfer using the field-particle correlation technique. Therefore, this non-resonant feature may hinder the identification of energization mechanisms in high-$\beta_i$, KAW turbulence which has non-negligible levels of compressibility and in which the phase offset between $E_\parallel$ and the compressive fluctuation of $\partial f_e/\partial v_\parallel$ has shifted away from $\phi \simeq \pi/2$. 

\subsubsection{Eliminating the Non-Resonant Feature by `Folding' Parallel Velocity-Space}
\label{sec:folding}
\begin{figure}
    \begin{center}
        \includegraphics[width=0.49\textwidth]{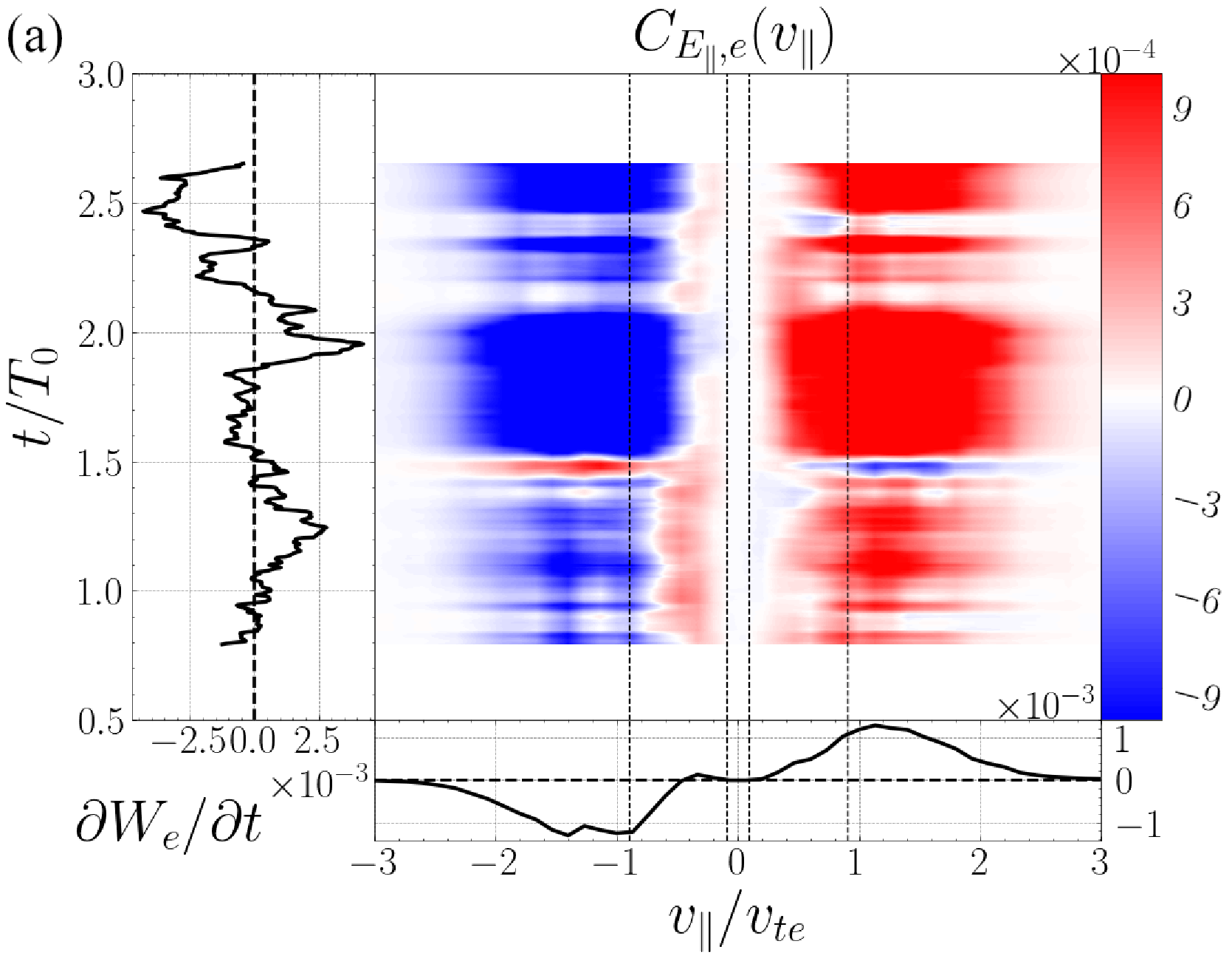}    % efpc23ai_probe007_s2_CER_nc256_contourf_a.eps
        \includegraphics[width=0.49\textwidth]{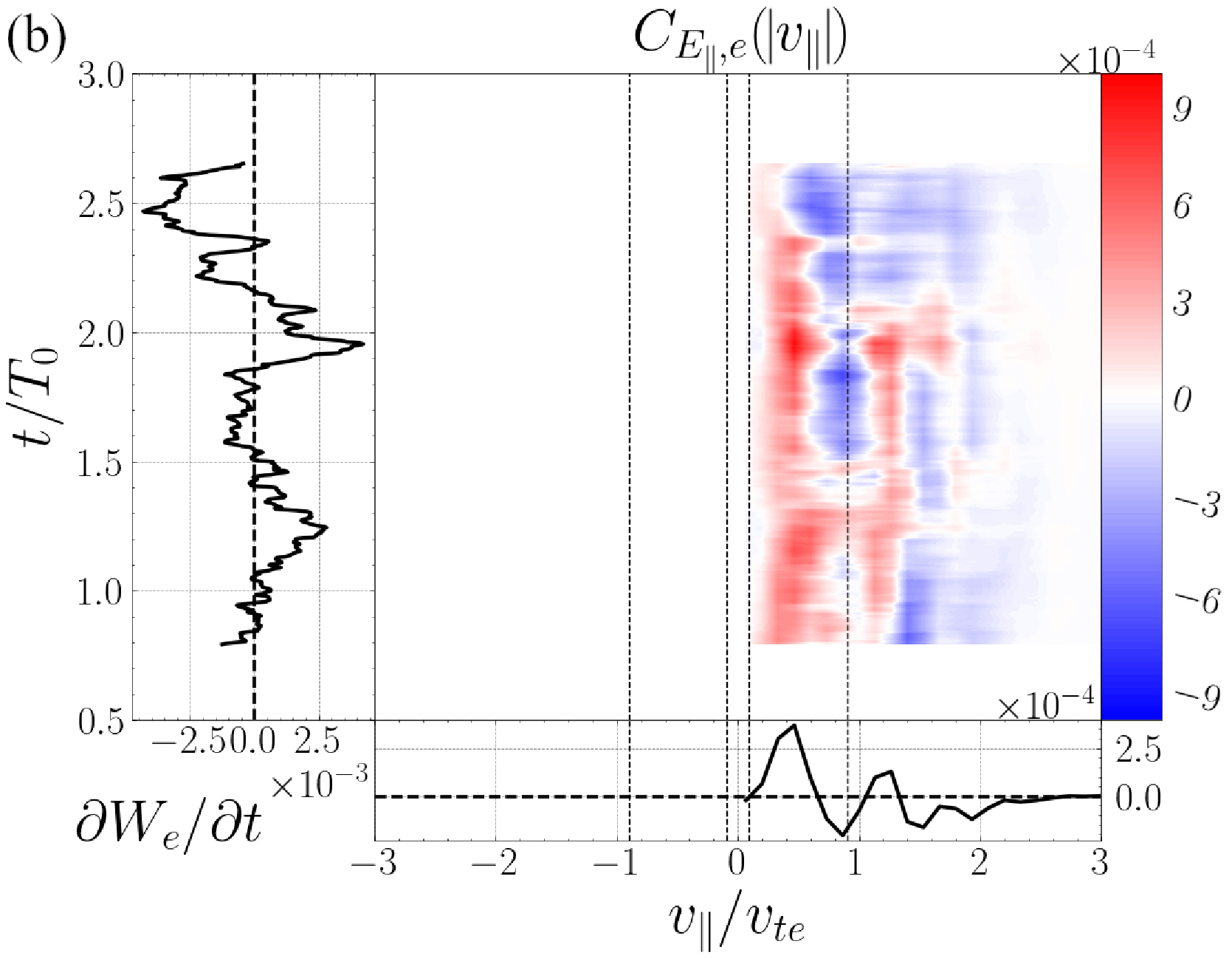}    % efpc23ai_probe007_s2_CER_nc256_contourf_folded_b.eps
    \end{center}
    \caption{(a): Reduced parallel field-particle correlation $C_{E_\parallel}(v_\parallel, t)$ timestack plot in the $\beta_i = 1$ turbulence  simulation using $\tau = 2.6\ T_{max}$. (b): The same timestack plot in (a) folded across $v_\parallel=0$ to obtain 
    $C_{E_\parallel}(|v_\parallel|,t)$, showing a clear signature of electron Landau damping at $v_\parallel/v_{te} \sim 0.1$ (albeit with the negative region suppressed by the $v_\parallel^2$ weighting) that persists throughout the simulation, with a peak amplitude about one quarter of the non-resonant feature.}
    
    \label{fig:NRF_folded}
\end{figure}

At the expense of losing the distinction between signatures of upward ($v_\parallel > 0$) and downward ($v_\parallel < 0$) propagating waves, the non-resonant feature can be removed by taking advantage of its antisymmetric nature. `Folding' the reduced parallel correlation across $v_\parallel = 0$, obtaining $C_{E_\parallel}(|v_\parallel|,t)= C_{E_\parallel}(v_\parallel > 0,t)+ C_{E_\parallel}(v_\parallel < 0,t)$, causes the large-amplitude, non-resonant feature to cancel out, often revealing an underlying bipolar velocity-space signature, as shown in Fig.~\ref{fig:NRF_folded}. In (a), we plot the timestack correlation $C_{E_\parallel}(v_\parallel, t)$ from the $\beta_i = 1$ simulation, which clearly shows an antisymmetric feature that obscures a portion of the expected velocity-range for Landau damping. When the data in the central and lower panels are folded over $v_\parallel = 0$---like closing a book with $v_\parallel = 0$ as the spine---a bipolar structure is revealed that is persistent throughout the entire simulated interval. The amplitude of the resulting signature of electron Landau damping has a peak amplitude about one-quarter of the amplitude of the broad non-resonant feature. Though by folding the correlation we have lost insight into whether the signature is due to the damping of upward or downward propagating waves, it reveals the structure in velocity-space that is necessary for identifying the mechanism of electron Landau damping. This method of `seeing beneath' the non-resonant feature was effective for our $\beta_i=1$ simulations, but we note that in our $\beta_i = 10$ simulations the Landau resonances with the electrons are so close to $v_\parallel=0$ that the velocity-space resolution of the simulation is insufficient for revealing signatures with certainty, even after folding the data.

Alternatively, high-pass filtering could be an effective way of removing a low-frequency non-resonant feature while retaining the distinction between the damping of forward and backward propagating waves. We tested this with some success in our simulations, but since the timestep in \T{AstroGK} simulations varies dynamically in order to satisfy the numerical Courant–Friedrichs–Lewy stability condition, it is  unclear if the results of the filtered \T{AstroGK} datasets are reliable. We report that applying a fast-Fourier transform, backward and forward high-pass Butterworth filters, and an inverse fast-Fourier transform did remove the non-resonant feature in tests performed on the $\beta_i=1$ simulation. Due to the variable timestep, however, we did not pursue this method of removing the non-resonant feature in depth. 

%==========================================================================

\section{Summary and Conclusions}\label{sec:conclusions}

Application of the field-particle correlation technique to measurements of weakly collisional plasma turbulence in space and astrophysical plasma environments generates velocity-space signatures of particle energization. These signatures can be used to identify the specific physical mechanisms that dominate the dissipation of the turbulence. Recent field-particle correlation analyses of \emph{MMS} observations of turbulence in Earth's magnetosheath plasma have provided the first direct evidence of electron Landau damping as a significant mechanism for the dissipation of space plasma turbulence \citep{Chen:2019,Afshari:2021}. Here, we employ linear kinetic \Alfven~wave (KAW) simulations and a suite of four gyrokinetic simulations of turbulence with $\beta_i=0.01,0.1,1,10$ and $T_i/T_e=1$ to characterize the qualitative and quantitative features of the velocity-space signature of electron Landau damping in a turbulent space plasma. Linear KAW simulations have demonstrated that the zero-crossing of the bipolar velocity-space signature of electron Landau damping falls at the resonant parallel phase velocity of the wave, and that the width of the signature increases with increasing normalized damping rate. Furthermore, unlike in the case of ion Landau damping where both the negative and positive regions of the bipolar velocity-space signature are clearly observable \citep{Klein:2017,Howes:2018,Klein:2020}, for the electron Landau damping of dispersive KAWs, the negative region of the bipolar signature can be suppressed by the $v_\parallel^2$ weighting in the field-particle correlation for waves with low parallel phase velocities compared to the electron thermal velocity, $v_{ph, \parallel}/v_{te} \le 0.3$.

A key result of this paper is the characterization of the typical gyrotropic velocity-space signatures $C_{E_\parallel}(v_\parallel,v_\perp)$ of electron Landau damping, with typical examples presented in Figure~\ref{fig:gyro} for $\beta_i=0.01,0.1,$ and $1$.  Similarly, timestack plots of the reduced parallel field-particle correlation $C_{E_\parallel}(v_\parallel,t)$, shown in Figure~\ref{fig:tstacks}, provide a clear means to determine the electron energization rates as a function of time, showing both temporal intermittency and the overlap of bipolar velocity-space signatures of KAW wavepackets with different resonant parallel phase velocities. Furthermore, even in balanced turbulence---characterized by approximately equal wave energy fluxes up and down the local mean magnetic field---over periods of a few to tens of KAW periods, one frequently observes an imbalance of the velocity-space signatures of electron Landau damping of upward and downward propagating KAWs, consistent with previous \emph{MMS} observations 
\citep{Afshari:2021}. In addition, a significant non-resonant feature that is asymmetric in $v_\parallel$ appears due to the plasma density fluctuations associated with KAWs at higher values of plasma beta, $\beta_i \gtrsim 1$. This large-amplitude, non-resonant feature can make it difficult to identify the velocity-space signatures of electron Landau damping, but a method of folding over the correlations in $v_\parallel$ can effectively eliminate the feature, revealing the generally smaller amplitude velocity-space signatures of resonant damping. All of these factors should be taken into consideration when analyzing the field-particle correlation signatures of electron Landau damping. It is our hope that these results can provide insight for interpreting future field-particle correlation analysis of \emph{in situ} spacecraft observations: helping to identify the signatures of electron Landau damping throughout the inner heliosphere and advancing the wider goal of determining the contribution of this mechanism to electron energization in the solar wind and other weakly collisional plasma environments. 

%==========================================================================
%Acknowledgements
%\section*{Acknowledgements}

%==========================================================================
%Funding
\section*{Funding}
This work was supported by NASA FINESST Grant 80NSSC20K1509 as well as NASA grants 80NSSC18K1217 and 80NSSC18K1371. Numerical simulations were performed using the Extreme Science and Engineering Discovery Environment (XSEDE), which is supported by National Science Foundation grant number ACI-1548562, through allocation TG-PHY090084.

%==========================================================================
%Declaration of interests
\section*{Declaration of interests}
The authors report no conflicts of interest.

%====================================================
\appendix
%====================================================

\section{Velocity-Space Signatures of Electron Landau Damping of Single Kinetic \Alfven~Waves} \label{sec:AppA}

Here we reproduce timestack plots of the field-particle correlations $C_{E_\parallel}(v_\parallel, t)$ in steady state for five of the twenty individual kinetic \Alfven~wave (KAW) simulations. For a $\beta_i = 1$ plasma, we plot the cases of a KAW with $k_\perp \rho_i = 2,\ 4,\ 8,\ 16,$ and $32$. Each correlation shows clear electron energization due to Landau damping of the wave by the resonant electrons, where the resonant parallel phase velocity for each wave is indicated by the vertical dotted lines. At the parallel velocity-space resolution of the \T{AstroGK} simulations, the negative portion of the bipolar damping signature is only visible in the time-integrated panels for the two smallest-scale waves (those at the largest resonant velocities), but all have a clear, localized, and persistent energization signature due to the action of Landau damping. The non-resonant feature, antisymmetric in $v_\parallel$, is visible in these correlations. Though the compressibility of the KAWs increases with $k_\perp \rho_i$ \citep{Matteini:2020}, the non-resonant feature is most clearly visible at lower $k_\perp \rho_i$ in these plots, likely due to the lower damping rate and the resonance's increased proximity to $v_\parallel = 0$ for these larger perpendicular wavelength waves.

\begin{figure}
    \begin{center}
        \includegraphics[width=0.49\textwidth]{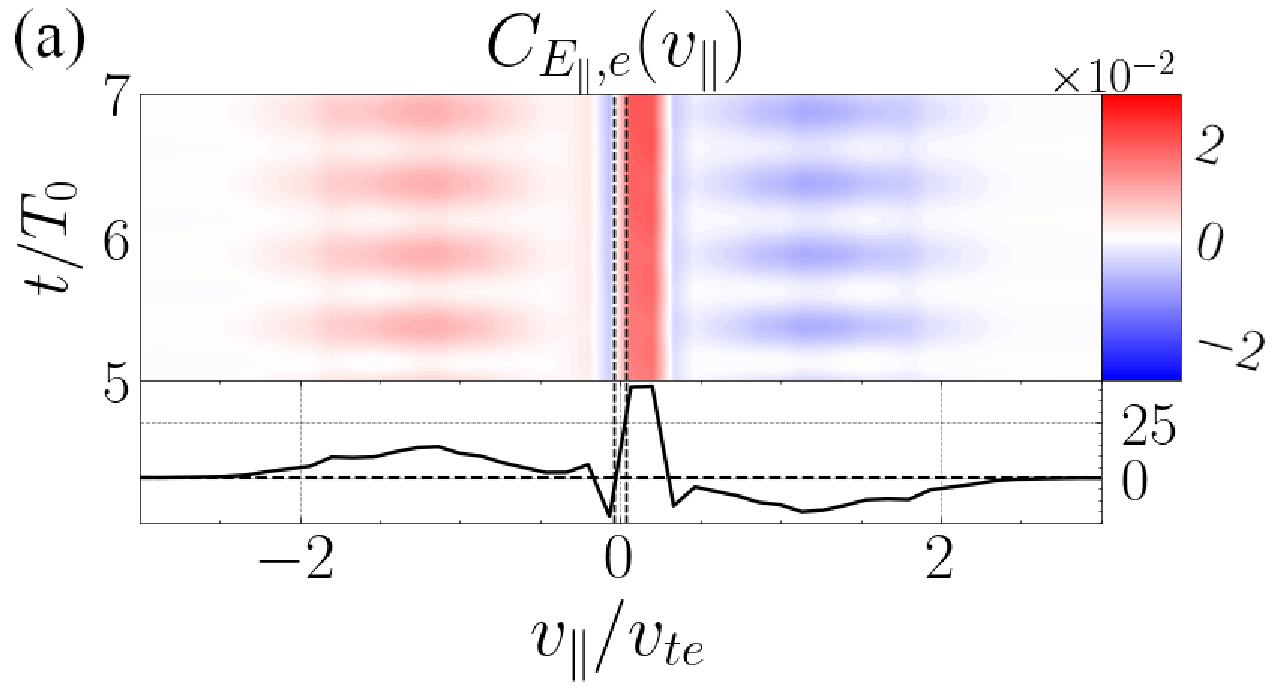}     % ekd_b1_k2_probe001_s2_CER_nc064_contourf_compressed_a.eps
        \includegraphics[width=0.49\textwidth]{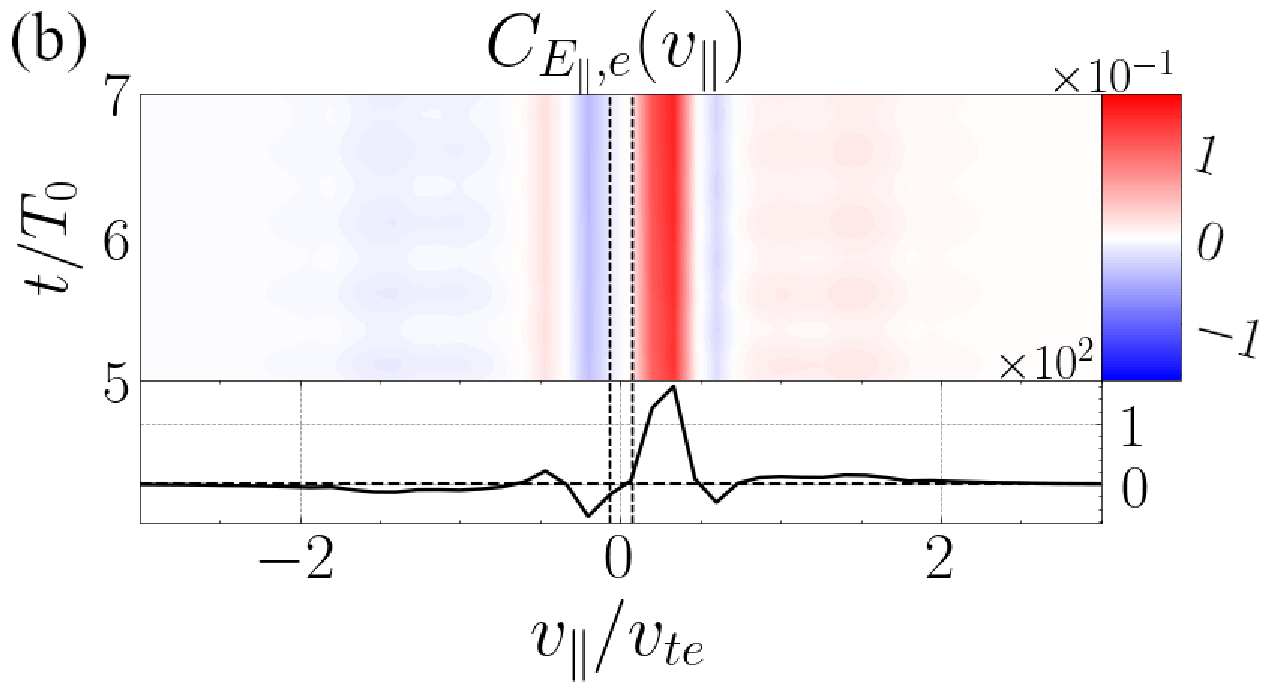}     % ekd_b1_k4_probe001_s2_CER_nc064_contourf_compressed_b.eps
        \includegraphics[width=0.49\textwidth]{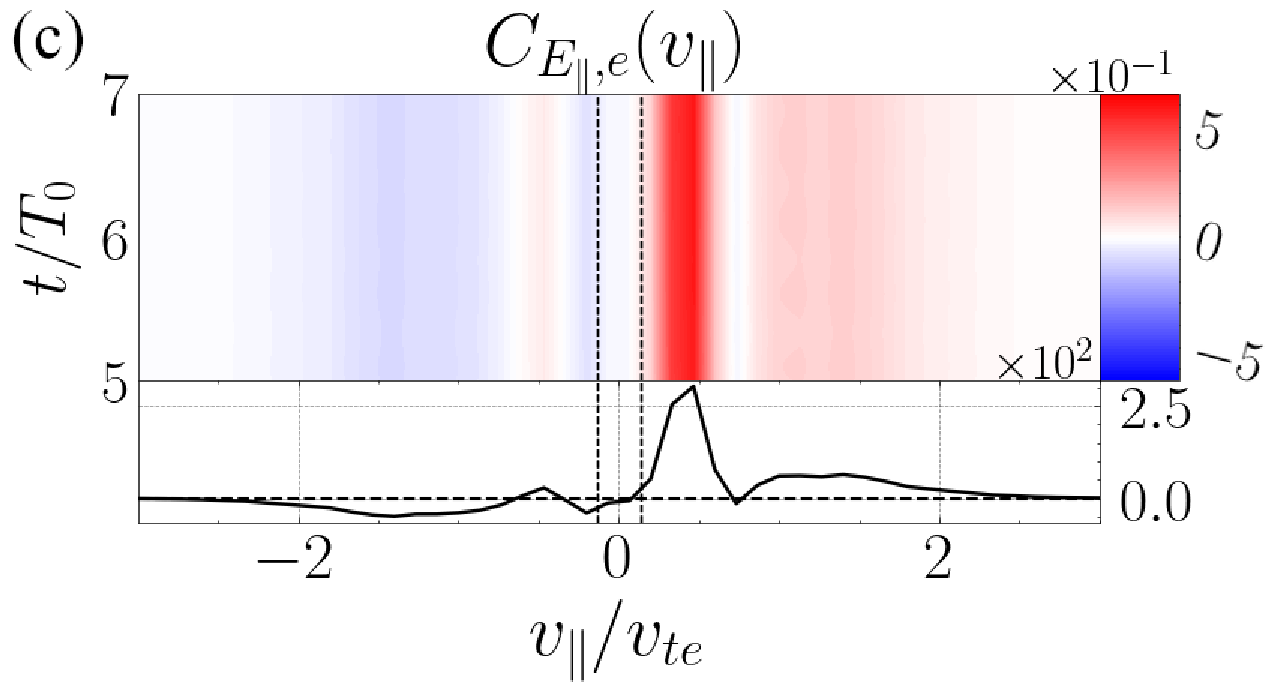}     % ekd_b1_k8_probe001_s2_CER_nc064_contourf_compressed_c.eps
        \includegraphics[width=0.49\textwidth]{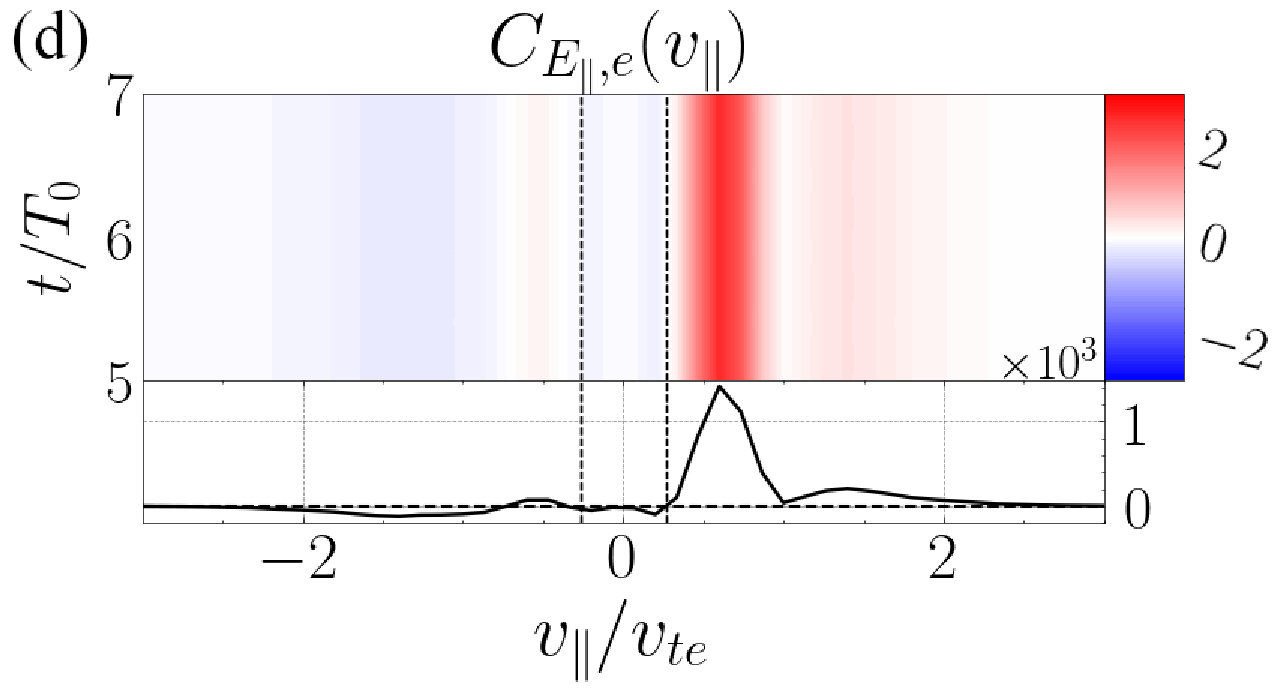}     % ekd_b1_k16_probe001_s2_CER_nc064_contourf_compressed_d.eps
        \includegraphics[width=0.49\textwidth]{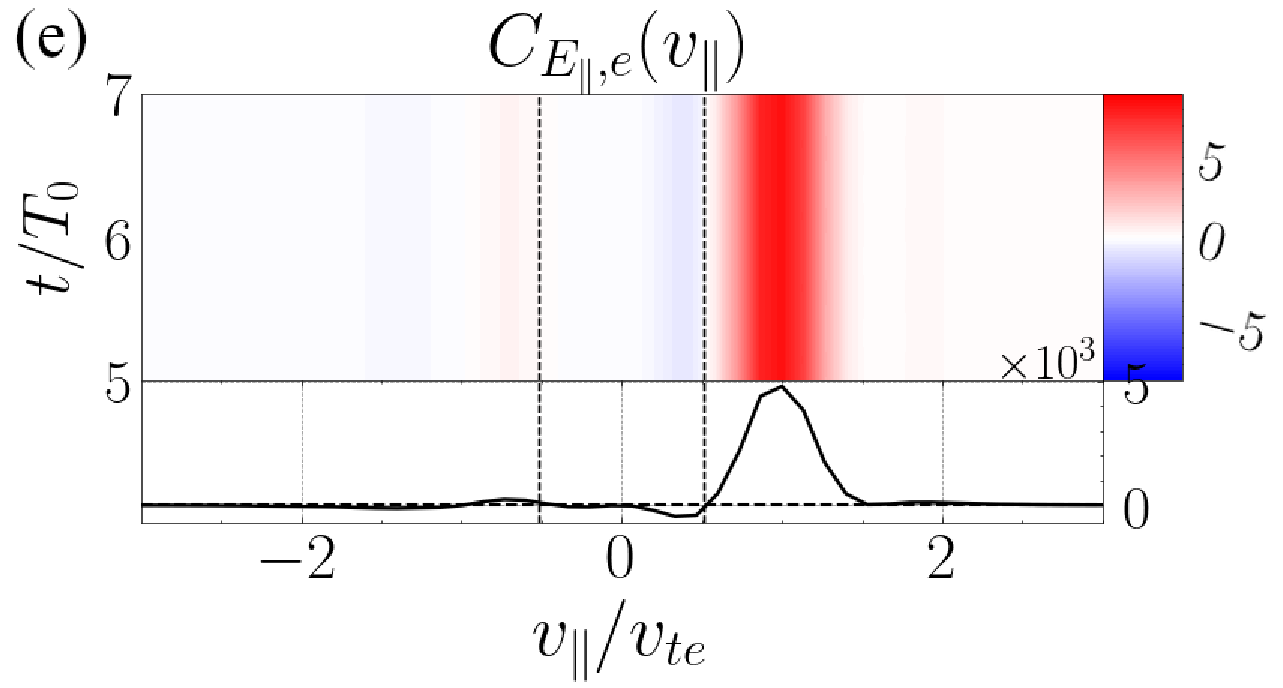}     % ekd_b1_k32_probe001_s2_CER_nc064_contourf_compressed_e.eps
    \end{center}
    \label{fig:linbeta1}
    \caption{Field-particle correlations of a set of driven, damped kinetic \Alfven~waves with $\beta_i = 1$ and  (a) 
    $k_\perp \rho_i = 2$, (b) 4, (c) 8, (d) 16, and (e) 32. The zero-crossing of the bipolar signatures correspond to the parallel wave phase velocity (vertical dotted lines), and the width of the signature increases with increasing damping rate as $k_\perp \rho_i$ increases.}
\end{figure}

%\newpage 
\section{Non-Resonant Feature Analysis}\label{sec:AppB}

\begin{figure}
    \begin{center}
        \includegraphics[width=0.48\textwidth]{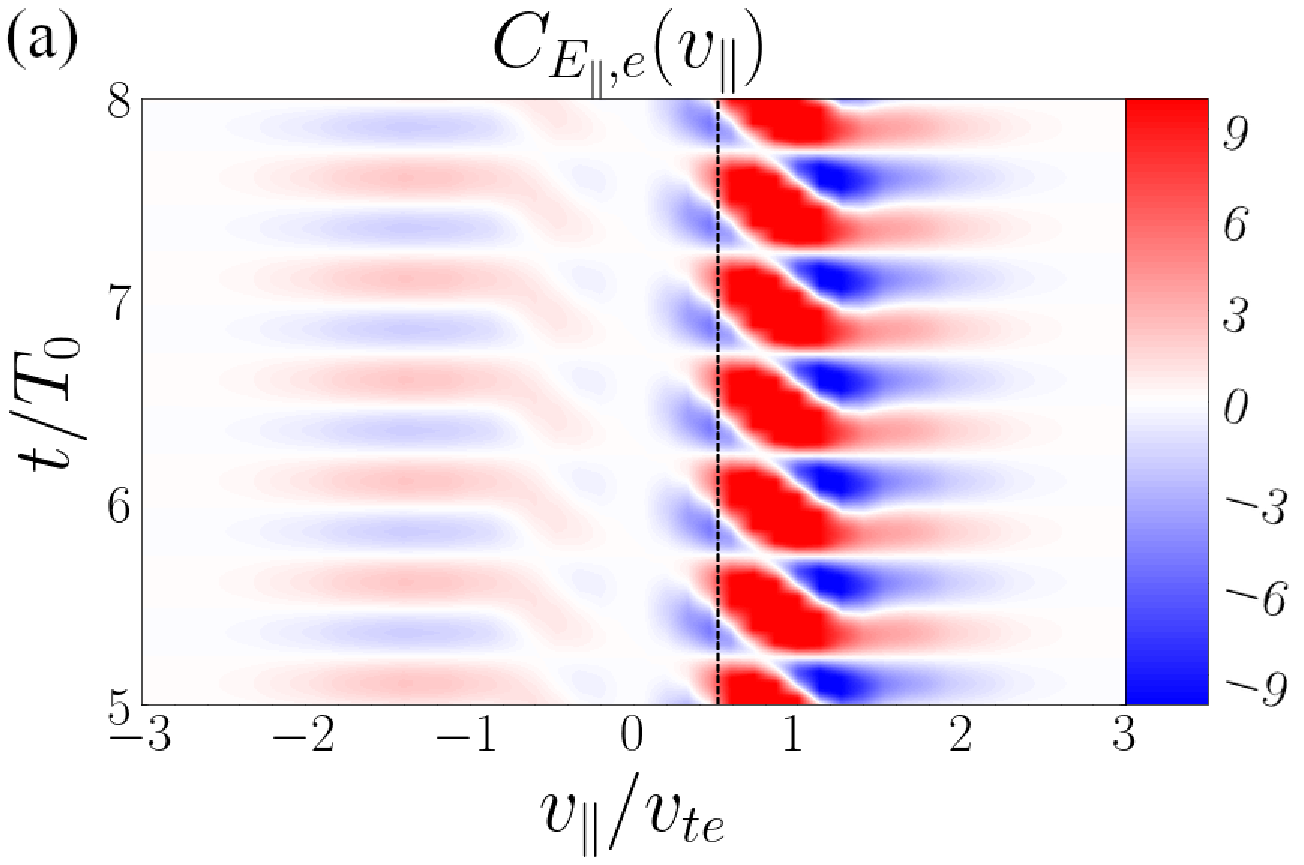}     % ekd_bp1_k8_probe001_s2_CER_nc001_contourf_tlim.eps
        \includegraphics[width=0.49\textwidth]{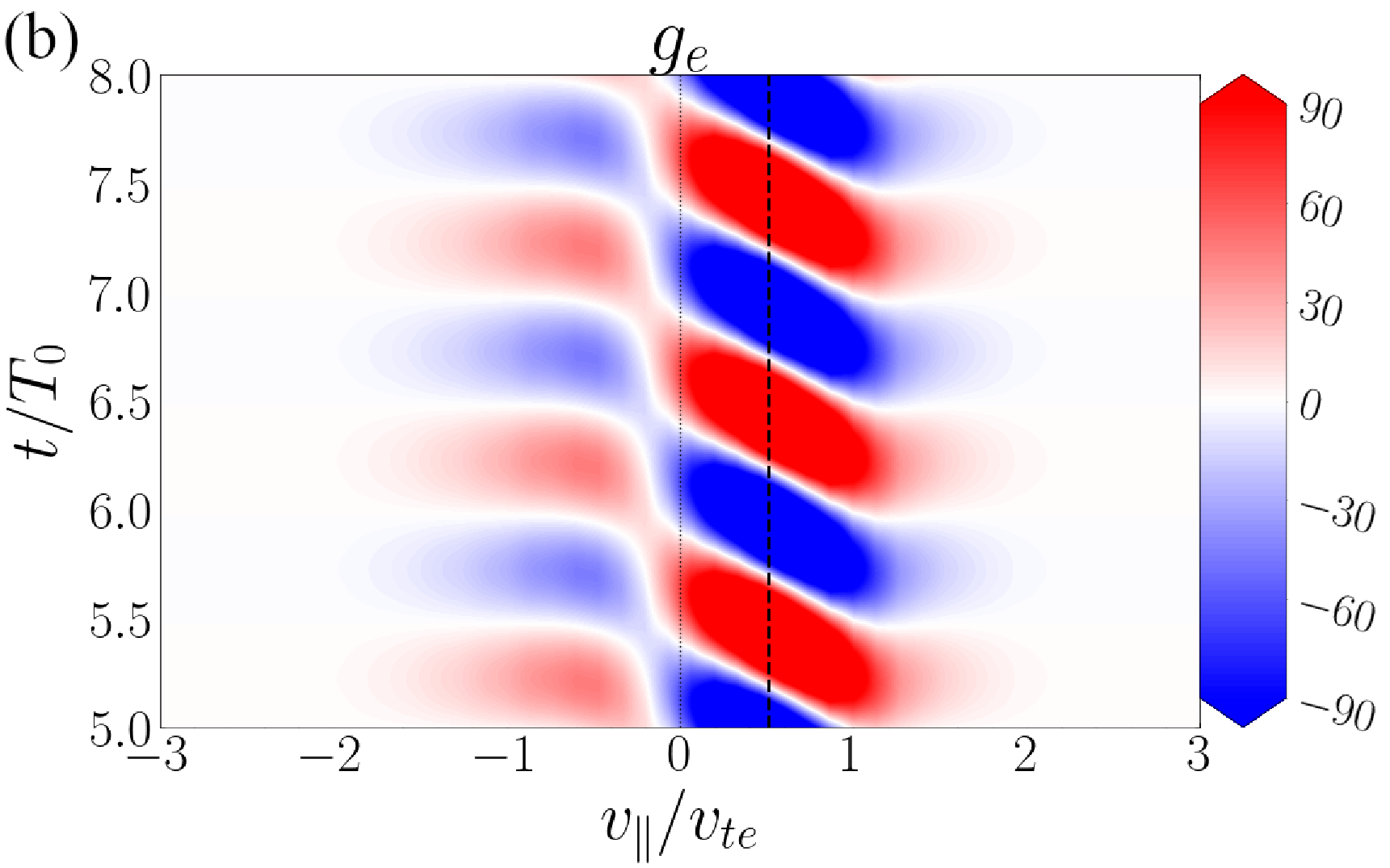}     % ekd_bp1_k8_probe001_s2_DFN_nc001_contourf_tlim.eps
        \includegraphics[width=0.48\textwidth]{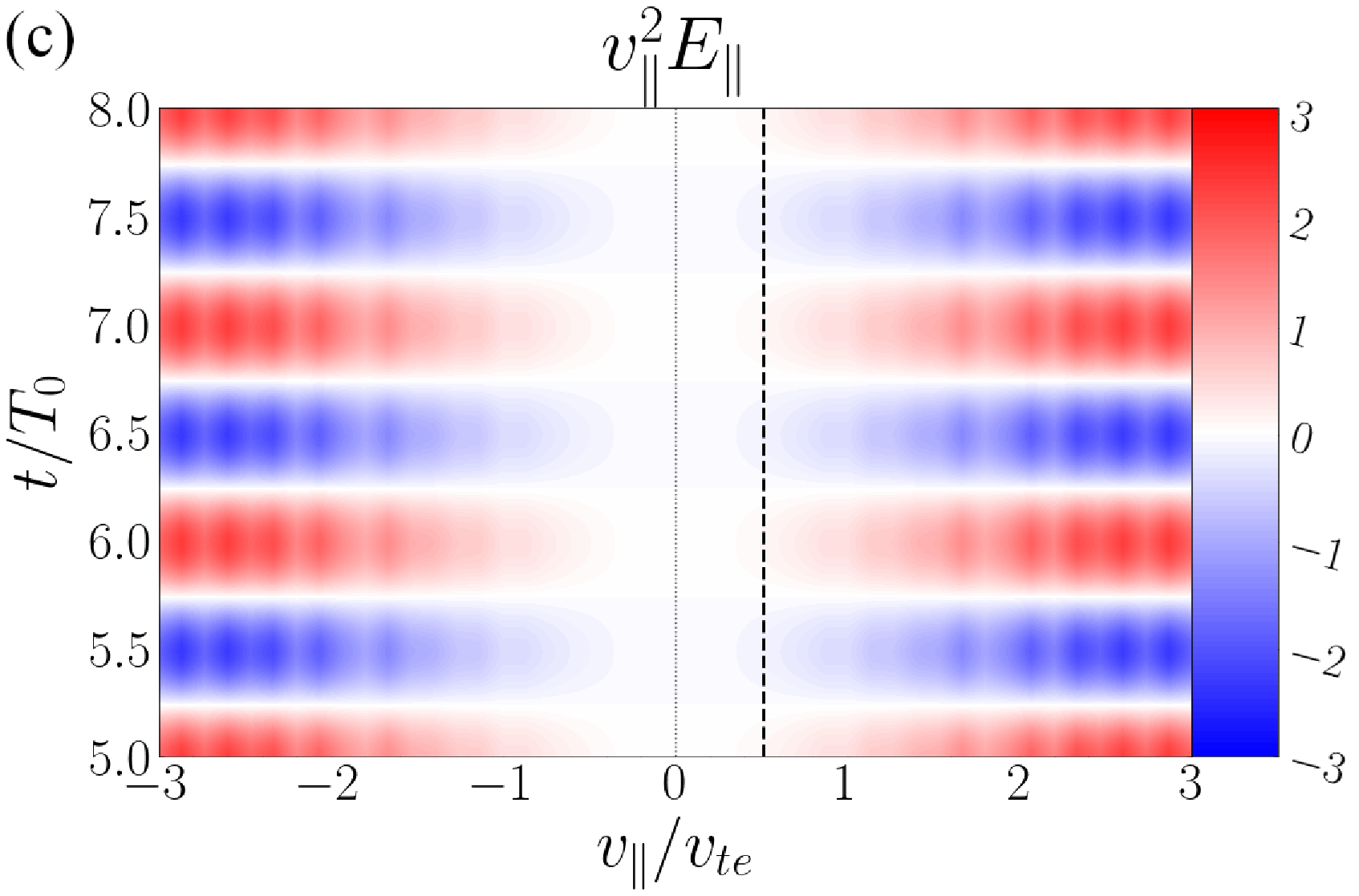}     % ekd_bp1_k8_probe001_s2_E_nc001_contourf_tlim.eps
        \includegraphics[width=0.49\textwidth]{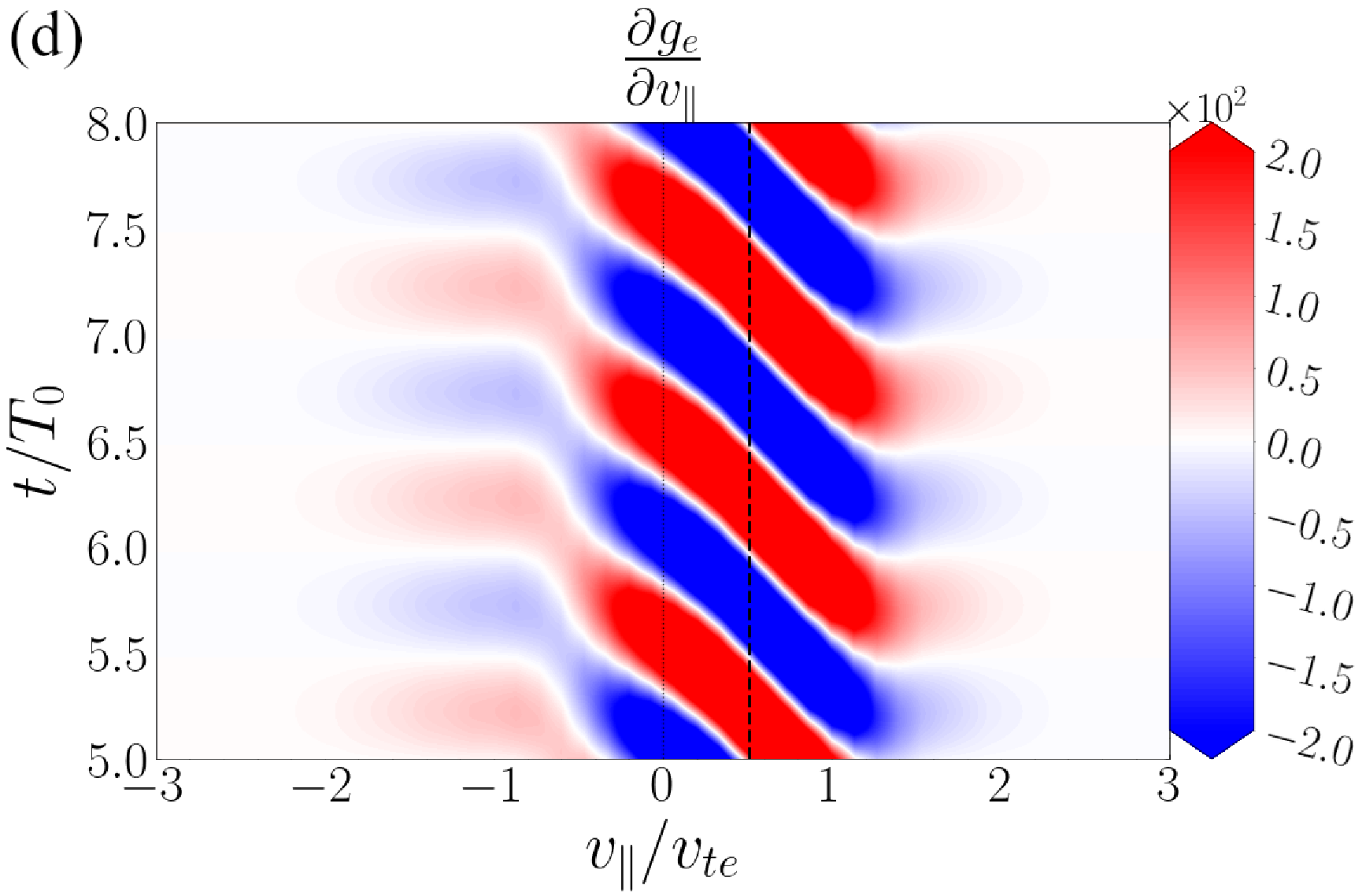}     % ekd_bp1_k8_probe001_s2_DVDFN_nc001_contourf_tlim.eps
    \end{center}
    \caption{Detail of the density perturbations giving rise to the non-resonant feature in the $E_\parallel$ field-particle correlations of high-$\beta_i$ plasmas. The parallel electric field-particle correlation $C_{E_\parallel, e}(v_\parallel)$ (a); the perturbed distribution function $g_e$ (b); the parallel electric field weighted by parallel velocity factor $E_\parallel v_\parallel^2$ (c); and the parallel velocity derivative of the perturbed distribution function $\partial g_e/\partial v_\parallel$ (d).}
    \label{fig:dn_detail}
\end{figure}

Here we explore more deeply the antisymmetric, non-resonant feature that appears in the field-particle correlation, which we attribute to the electron density perturbations, $\delta n_e$, that constitute a dynamic part of the kinetic \Alfven~wave. This non-resonant feature arises in the field-particle correlation due to (1) an increase in the amplitude of $\delta n_e$ relative to the resonant perturbations and (2) the relative phase between $E_\parallel$ and $\partial f/\partial v_\parallel$ as a function of $v_\parallel$. The compressive nature of the non-resonant feature, its amplitude, and its dependence on the phase offset $\phi$ are seen clearly when the factors in the mathematical form of the correlation are divided into separate plots, as shown in Fig.~\ref{fig:dn_detail}.

The reduced parallel field-particle correlation $C_{E_\parallel}(v_\parallel, t)$ is shown in Fig.~\ref{fig:dn_detail}(a) for the individual wave simulation at $k_\perp \rho_i = 8$, $\beta_i = 0.1$. The reduced, complementary perturbed distribution function $g_e(v_\parallel, t)$ \citep{Numata:2010} is plotted in (b). The location of the resonant KAW phase velocity is marked by the vertical dashed line at $v_\parallel/v_{te} = 0.51$, and the resonant perturbations in $g_e(v_\parallel, t)$ 
peak broadly around this resonant parallel velocity. In the background, at lower amplitude, there is a broad oscillatory signature that is an even function in $v_\parallel$ and that spans all velocities (fading to zero with the distribution function at large $v_\parallel$). This non-resonant, even perturbation  in $g_e(v_\parallel, t)$ represents compression in the electron density, $\delta n_e$, when integrated over $v_\parallel$. In (d), we show the parallel velocity derivative of the reduced perturbed distribution function, $\partial g_e/\partial v_\parallel$. Here, it can be seen that the even function in $g_e$, yielding $\delta n_e \neq 0$, turns into an odd function when the derivative with respect to $v_\parallel$ is computed. Finally, (c) contains the parallel electric field multiplied by $v_\parallel^2$; note that $E_\parallel$ itself is constant over $v_\parallel$. Upon close inspection, it can be observed that the oscillation in $E_\parallel$ and the background oscillation in $\partial g_e/\partial v_\parallel$ have the same period but are offset in phase. This phase offset $\phi$ is ultimately what causes the density oscillation to lead to the odd, non-resonant signature that appears in the field-particle correlation in panel (a). 

In the main body of this paper, we have explained how the relative amplitudes of the resonant and density perturbations to $g_e$ change due to the combined effects of increasing compressibility of KAWs and to the decreasing phase-velocity of KAWs as $\beta_i$ increases. However, we did not discuss the origin of the phase offset between the density perturbation and the parallel electric field from $\phi \simeq \pi/2$ at low $\beta_i$ to $\phi \neq \pi/2$ at higher $\beta_i$. If the electron response is approximately instantaneous, the density perturbation will be exactly $\pi/2$ out of phase with the finite $E_\parallel$ that drives it. In this case, the average acceleration of the particles by $E_\parallel$ is zero. This is approximately the background phase relationship we observe in the low $\beta_i$ ($0.01$ and $0.1$) single-wave runs. 

\begin{figure}
    \begin{center}
        \includegraphics[width=0.99\textwidth]{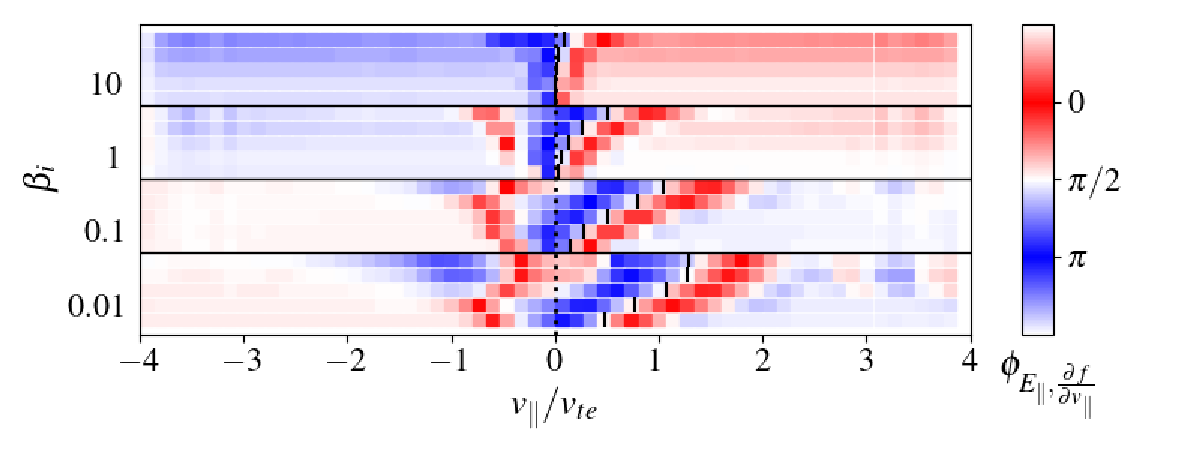}       % phi_scatter_custom.eps
    \end{center}
    \caption{The phase offset $\phi$ between $E_\parallel$ and the density oscillation in $\partial g_e/\partial v_\parallel$ as a function of $v_\parallel/v_{te}$. Horizontal sections, separated by black lines, correspond to different $\beta_i$. Within each of these sections, there are five separate strips that correspond to a different $k_\perp \rho_i$. Increasing vertically within each section, $k_\perp \rho_i$ cycles through $\{2, 4, 8, 16, 32\}$. The phase velocity of the KAW for each $\beta_i$, $k_\perp \rho_i$ pair is marked with a vertical black line.}
    \label{fig:phi}
\end{figure}

The phase offset is approximately $\phi \simeq \pi/2$ at all parallel velocities except those that resonantly interact with the $E_\parallel$ wave around $v_{ph} = \omega/k_\parallel$, as shown in Fig.~\ref{fig:phi}. For small perpendicular wavenumbers at large $\beta_i$ ($1$ and $10$) the phase offset is also $\phi \simeq \pi/2$. As $k_\perp \rho_i$ increases, however, the phase offset deviates from $\pi/2$ and moves toward $\pi$ or $0$, resulting in detectable non-resonant features in the field-particle correlation signatures. As $\beta_i$ and $k_\perp \rho_i$ increase, the compressive wave begins to lag $E_\parallel$, shifting the relative phase $\phi$ and thereby enabling $E_\parallel$ to do work on the particles, which results in a local non-zero contribution to the change in phase-space energy density. Averaged over parallel velocity, however, the contribution of the non-resonant feature goes to zero, and therefore the compressions do not change the total spatial energy density of the electrons $W_e(\V{r},t)$. 

%====================================================
\newpage
%\bibliographystyle{jpp}
%\bibliography{abbrev,conley}

\begin{thebibliography}{80}
  \expandafter\ifx\csname natexlab\endcsname\relax\def\natexlab#1{#1}\fi
  
  \bibitem[{Afshari} {\em et~al.\/}(2021){Afshari}, {Howes}, {Kletzing},
    {Hartley} \& {Boardsen}]{Afshari:2021}
  {\sc {Afshari}, A.~S., {Howes}, G.~G., {Kletzing}, C.~A., {Hartley}, D.~P. \&
    {Boardsen}, S.~A.} 2021 {The Importance of Electron Landau Damping for the
    Dissipation of Turbulent Energy in Terrestrial Magnetosheath Plasma}. {\em
    J.~Geophys.~Res.\/} {\bf 126}~(12), e29578.
  
  \bibitem[{Afshari} {\em et~al.\/}(2023){Afshari}, {Howes}, {Shuster}, {Klein},
    {McGinnis}, {Martinovic}, {Boardsen}, {Kletzing} \& {Hartley}]{Afshari:2023}
  {\sc {Afshari}, A.~S., {Howes}, G.~G., {Shuster}, J.~R., {Klein}, K.~G.,
    {McGinnis}, D., {Martinovic}, M.~M., {Boardsen}, S.~A., {Kletzing}, C.~A. \&
    {Hartley}, D.~P.} 2023 {Direct observation of ion cyclotron damping of
    turbulence in Earth’s magnetosheath plasma}. {\em Nature Comm.\/}
    Submitted.
  
  \bibitem[{Alexandrova} {\em et~al.\/}(2008){Alexandrova}, {Carbone}, {Veltri}
    \& {Sorriso-Valvo}]{Alexandrova:2008}
  {\sc {Alexandrova}, O., {Carbone}, V., {Veltri}, P. \& {Sorriso-Valvo}, L.}
    2008 {Small-Scale Energy Cascade of the Solar Wind Turbulence}. {\em
    Astrophys.~J.\/} {\bf 674}, 1153--1157.
  
  \bibitem[{Arzamasskiy} {\em et~al.\/}(2019){Arzamasskiy}, {Kunz}, {Chandran} \&
    {Quataert}]{Arzamasskiy:2019}
  {\sc {Arzamasskiy}, Lev, {Kunz}, Matthew~W., {Chandran}, Benjamin D.~G. \&
    {Quataert}, Eliot} 2019 {Hybrid-kinetic Simulations of Ion Heating in
    Alfv{\'e}nic Turbulence}. {\em Astrophys.~J.\/} {\bf 879}~(1), 53.
  
  \bibitem[{Beresnyak} \& {Lazarian}(2008)]{Beresnyak:2008}
  {\sc {Beresnyak}, A. \& {Lazarian}, A.} 2008 {Strong Imbalanced Turbulence}.
    {\em Astrophys.~J.\/} {\bf 682}, 1070--1075.
  
  \bibitem[{Brown} {\em et~al.\/}(2023){Brown}, {Juno}, {Howes}, {Haggerty} \&
    {Constantinou}]{Brown:2023}
  {\sc {Brown}, C.~R., {Juno}, J., {Howes}, G.~G., {Haggerty}, C.~C. \&
    {Constantinou}, S.} 2023 {Isolation and phase-space energization analysis of
    the instabilities in collisionless shocks}. {\em J.~Plasma Phys.\/} {\bf
    89}~(3), 905890308.
  
  \bibitem[{Bruno} \& {Carbone}(2005)]{Bruno:2005}
  {\sc {Bruno}, R. \& {Carbone}, V.} 2005 {The Solar Wind as a Turbulence
    Laboratory}. {\em Living Reviews in Solar Physics\/} {\bf 2}, 4.
  
  \bibitem[{Burch} {\em et~al.\/}(2016){Burch}, {Moore}, {Torbert} \&
    {Giles}]{Burch:2016}
  {\sc {Burch}, J.~L., {Moore}, T.~E., {Torbert}, R.~B. \& {Giles}, B.~L.} 2016
    {Magnetospheric Multiscale Overview and Science Objectives}. {\em Space
    Sci.~Rev.\/} {\bf 199}, 5--21.
  
  \bibitem[{Cerri} {\em et~al.\/}(2021){Cerri}, {Arzamasskiy} \&
    {Kunz}]{Cerri:2021}
  {\sc {Cerri}, S.~S., {Arzamasskiy}, L. \& {Kunz}, M.~W.} 2021 {On Stochastic
    Heating and Its Phase-space Signatures in Low-beta Kinetic Turbulence}. {\em
    Astrophys.~J.\/} {\bf 916}~(2), 120.
  
  \bibitem[{Chandran}(2008)]{Chandran:2008}
  {\sc {Chandran}, B.~D.~G.} 2008 {Strong Anisotropic MHD Turbulence with Cross
    Helicity}. {\em Astrophys.~J.\/} {\bf 685}, 646--658.
  
  \bibitem[{Chandran}(2010)]{Chandran:2010b}
  {\sc {Chandran}, B.~D.~G.} 2010 {Alfv{\'e}n-wave Turbulence and Perpendicular
    Ion Temperatures in Coronal Holes}. {\em Astrophys.~J.\/} {\bf 720},
    548--554.
  
  \bibitem[{Chen} {\em et~al.\/}(2019){Chen}, {Klein} \& {Howes}]{Chen:2019}
  {\sc {Chen}, C.~H.~K., {Klein}, K.~G. \& {Howes}, G.~G.} 2019 {Evidence for
    electron Landau damping in space plasma turbulence}. {\em Nature Comm.\/}
    {\bf 10}, 740.
  
  \bibitem[{Cranmer}(2009)]{Cranmer:2009b}
  {\sc {Cranmer}, S.~R.} 2009 {Coronal Holes}. {\em Living Reviews in Solar
    Physics\/} {\bf 6}, 3.
  
  \bibitem[{Drake} {\em et~al.\/}(2013){Drake}, {Schroeder}, {Howes}, {Kletzing},
    {Skiff}, {Carter} \& {Auerbach}]{Drake:2013}
  {\sc {Drake}, D.~J., {Schroeder}, J.~W.~R., {Howes}, G.~G., {Kletzing}, C.~A.,
    {Skiff}, F., {Carter}, T.~A. \& {Auerbach}, D.~W.} 2013 {Alfv{\'e}n wave
    collisions, the fundamental building block of plasma turbulence. IV.
    Laboratory experiment}. {\em Phys.~Plasmas\/} {\bf 20}~(7), 072901.
  
  \bibitem[{Edl{\'e}n}(1943)]{Edlen:1943}
  {\sc {Edl{\'e}n}, B.} 1943 {Die Deutung der Emissionslinien im Spektrum der
    Sonnenkorona. Mit 6 Abbildungen.} {\em Zeitschrift für Astrophysik\/} {\bf
    22}, 30.
  
  \bibitem[{Edl{\'e}n}(1945)]{Edlen:1945}
  {\sc {Edl{\'e}n}, B.} 1945 {The identification of the coronal lines (George
    Darwin Lecture)}. {\em Mon.~Not.~Roy.~Astron.~Soc.\/} {\bf 105}, 323.
  
  \bibitem[{Goldreich} \& {Sridhar}(1995)]{Goldreich:1995}
  {\sc {Goldreich}, P. \& {Sridhar}, S.} 1995 {Toward a Theory of Interstellar
    Turbulence. II. Strong Alfvenic Turbulence}. {\em Astrophys.~J.\/} {\bf 438},
    763.
  
  \bibitem[{Heyvaerts} \& {Priest}(1983)]{Heyvaerts:1983}
  {\sc {Heyvaerts}, J. \& {Priest}, E.~R.} 1983 {Coronal heating by phase-mixed
    shear Alfven waves}. {\em Astron.~Astrophys.\/} {\bf 117}, 220--234.
  
  \bibitem[{Horvath} {\em et~al.\/}(2020){Horvath}, {Howes} \&
    {McCubbin}]{Horvath:2020}
  {\sc {Horvath}, Sarah~A., {Howes}, Gregory~G. \& {McCubbin}, Andrew~J.} 2020
    {Electron Landau damping of kinetic Alfv{\'e}n waves in simulated
    magnetosheath turbulence}. {\em Phys.~Plasmas\/} {\bf 27}~(10), 102901.
  
  \bibitem[{Horvath} {\em et~al.\/}(2022){Horvath}, {Howes} \&
    {McCubbin}]{Horvath:2022}
  {\sc {Horvath}, Sarah~A., {Howes}, Gregory~G. \& {McCubbin}, Andrew~J.} 2022
    {Observing particle energization above the Nyquist frequency: An application
    of the field-particle correlation technique}. {\em Phys.~Plasmas\/} {\bf
    29}~(6), 062901.
  
  \bibitem[{Howes}(2008)]{Howes:2008c}
  {\sc {Howes}, Gregory~G.} 2008 {Inertial range turbulence in kinetic plasmas}.
    {\em Phys.~Plasmas\/} {\bf 15}~(5), 055904--055904.
  
  \bibitem[{Howes}(2011)]{Howes:2011c}
  {\sc {Howes}, G.~G.} 2011 {Prediction of the Proton-to-total Turbulent Heating
    in the Solar Wind}. {\em Astrophys.~J.\/} {\bf 738}, 40.
  
  \bibitem[{Howes}(2015)]{Howes:2015}
  {\sc {Howes}, G.~G.} 2015 {A dynamical model of plasma turbulence in the solar
    wind}. {\em Philosophical Transactions of the Royal Society of London Series
    A\/} {\bf 373}~(2041), 20140145--20140145.
  
  \bibitem[{Howes}(2016)]{Howes:2016b}
  {\sc {Howes}, G.~G.} 2016 {The Dynamical Generation of Current Sheets in
    Astrophysical Plasma Turbulence}. {\em Astrophys.~J.~Lett.\/} {\bf 82}, L28.
  
  \bibitem[{Howes}(2017)]{Howes:2017b}
  {\sc {Howes}, Gregory~G.} 2017 {A prospectus on kinetic heliophysics}. {\em
    Phys.~Plasmas\/} {\bf 24}~(5), 055907.
  
  \bibitem[{Howes} {\em et~al.\/}(2006){Howes}, {Cowley}, {Dorland}, {Hammett},
    {Quataert} \& {Schekochihin}]{Howes:2006}
  {\sc {Howes}, G.~G., {Cowley}, S.~C., {Dorland}, W., {Hammett}, G.~W.,
    {Quataert}, E. \& {Schekochihin}, A.~A.} 2006 {Astrophysical Gyrokinetics:
    Basic Equations and Linear Theory}. {\em Astrophys.~J.\/} {\bf 651},
    590--614.
  
  \bibitem[{Howes} {\em et~al.\/}(2008{\natexlab{{\em a\/}}}){Howes}, {Cowley},
    {Dorland}, {Hammett}, {Quataert} \& {Schekochihin}]{Howes:2008b}
  {\sc {Howes}, G.~G., {Cowley}, S.~C., {Dorland}, W., {Hammett}, G.~W.,
    {Quataert}, E. \& {Schekochihin}, A.~A.} 2008{\natexlab{{\em a\/}}} {A model
    of turbulence in magnetized plasmas: Implications for the dissipation range
    in the solar wind}. {\em J.~Geophys.~Res.\/} {\bf 113}~(A5), A05103.
  
  \bibitem[{Howes} {\em et~al.\/}(2008{\natexlab{{\em b\/}}}){Howes}, {Dorland},
    {Cowley}, {Hammett}, {Quataert}, {Schekochihin} \& {Tatsuno}]{Howes:2008}
  {\sc {Howes}, G.~G., {Dorland}, W., {Cowley}, S.~C., {Hammett}, G.~W.,
    {Quataert}, E., {Schekochihin}, A.~A. \& {Tatsuno}, T.} 2008{\natexlab{{\em
    b\/}}} {Kinetic Simulations of Magnetized Turbulence in Astrophysical
    Plasmas}. {\em Phys.~Rev.~Lett.\/} {\bf 100}~(6), 065004.
  
  \bibitem[{Howes} {\em et~al.\/}(2012){Howes}, {Drake}, {Nielson}, {Carter},
    {Kletzing} \& {Skiff}]{Howes:2012b}
  {\sc {Howes}, G.~G., {Drake}, D.~J., {Nielson}, K.~D., {Carter}, T.~A.,
    {Kletzing}, C.~A. \& {Skiff}, F.} 2012 {Toward Astrophysical Turbulence in
    the Laboratory}. {\em Phys.~Rev.~Lett.\/} {\bf 109}~(25), 255001.
  
  \bibitem[{Howes} {\em et~al.\/}(2017){Howes}, {Klein} \& {Li}]{Howes:2017}
  {\sc {Howes}, G.~G., {Klein}, K.~G. \& {Li}, T.~C.} 2017 {Diagnosing
    collisionless energy transfer using field-particle correlations:
    Vlasov-Poisson plasmas}. {\em J.~Plasma Phys.\/} {\bf 83}~(1), 705830102.
  
  \bibitem[Howes {\em et~al.\/}(2018)Howes, McCubbin \& Klein]{Howes:2018}
  {\sc Howes, Gregory~G., McCubbin, Andrew~J. \& Klein, Kristopher~G.} 2018
    Spatially localized particle energization by landau damping in current sheets
    produced by strong alfvén wave collisions. {\em J.~Plasma Phys.\/} {\bf
    84}~(1), 905840105.
  
  \bibitem[{Howes} \& {Nielson}(2013)]{Howes:2013a}
  {\sc {Howes}, G.~G. \& {Nielson}, K.~D.} 2013 {Alfv{\'e}n wave collisions, the
    fundamental building block of plasma turbulence. I. Asymptotic solution}.
    {\em Phys.~Plasmas\/} {\bf 20}~(7), 072302.
  
  \bibitem[{Howes} {\em et~al.\/}(2013){Howes}, {Nielson}, {Drake}, {Schroeder},
    {Skiff}, {Kletzing} \& {Carter}]{Howes:2013b}
  {\sc {Howes}, G.~G., {Nielson}, K.~D., {Drake}, D.~J., {Schroeder}, J.~W.~R.,
    {Skiff}, F., {Kletzing}, C.~A. \& {Carter}, T.~A.} 2013 {Alfv{\'e}n wave
    collisions, the fundamental building block of plasma turbulence. III. Theory
    for experimental design}. {\em Phys.~Plasmas\/} {\bf 20}~(7), 072304.
  
  \bibitem[{Howes} {\em et~al.\/}(2011{\natexlab{{\em a\/}}}){Howes}, {Tenbarge}
    \& {Dorland}]{Howes:2011b}
  {\sc {Howes}, G.~G., {Tenbarge}, J.~M. \& {Dorland}, W.} 2011{\natexlab{{\em
    a\/}}} {A weakened cascade model for turbulence in astrophysical plasmas}.
    {\em Phys.~Plasmas\/} {\bf 18}~(10), 102305--102305.
  
  \bibitem[{Howes} {\em et~al.\/}(2011{\natexlab{{\em b\/}}}){Howes}, {Tenbarge},
    {Dorland}, {Quataert}, {Schekochihin}, {Numata} \& {Tatsuno}]{Howes:2011}
  {\sc {Howes}, G.~G., {Tenbarge}, J.~M., {Dorland}, W., {Quataert}, E.,
    {Schekochihin}, A.~A., {Numata}, R. \& {Tatsuno}, T.} 2011{\natexlab{{\em
    b\/}}} {Gyrokinetic Simulations of Solar Wind Turbulence from Ion to Electron
    Scales}. {\em Phys.~Rev.~Lett.\/} {\bf 107}~(3), 035004.
  
  \bibitem[Iroshnikov(1963)]{Iroshnikov:1963}
  {\sc Iroshnikov, R.~S.} 1963 The turbulence of a conducting fluid in a strong
    magnetic field. {\em Astron. Zh.\/} {\bf 40}, 742, {English} Translation:
    Sov. Astron., 7 566 (1964).
  
  \bibitem[{Juno} {\em et~al.\/}(2023){Juno}, {Brown}, {Howes}, {Haggerty},
    {TenBarge}, {Wilson}, {Caprioli} \& {Klein}]{Juno:2023}
  {\sc {Juno}, James, {Brown}, Collin~R., {Howes}, Gregory~G., {Haggerty},
    Colby~C., {TenBarge}, Jason~M., {Wilson}, Lynn~B., III, {Caprioli}, Damiano
    \& {Klein}, Kristopher~G.} 2023 {Phase-space Energization of Ions in Oblique
    Shocks}. {\em Astrophys.~J.\/} {\bf 944}~(1), 15.
  
  \bibitem[{Juno} {\em et~al.\/}(2021){Juno}, {Howes}, {TenBarge}, {Wilson},
    {Spitkovsky}, {Caprioli}, {Klein} \& {Hakim}]{Juno:2021}
  {\sc {Juno}, James, {Howes}, Gregory~G., {TenBarge}, Jason~M., {Wilson},
    Lynn~B., {Spitkovsky}, Anatoly, {Caprioli}, Damiano, {Klein}, Kristopher~G.
    \& {Hakim}, Ammar} 2021 {A field-particle correlation analysis of a
    perpendicular magnetized collisionless shock}. {\em J.~Plasma Phys.\/} {\bf
    87}~(3), 905870316.
  
  \bibitem[{Kiyani} {\em et~al.\/}(2015){Kiyani}, {Osman} \&
    {Chapman}]{Kiyani:2015}
  {\sc {Kiyani}, K.~H., {Osman}, K.~T. \& {Chapman}, S.~C.} 2015 {Dissipation and
    heating in solar wind turbulence: from the macro to the micro and back
    again}. {\em Philos.~Tr.~R.~Soc.~S-A.\/} {\bf 373}~(2041),
    20140155--20140155.
  
  \bibitem[{Klein}(2017)]{Klein:2017b}
  {\sc {Klein}, K.~G.} 2017 {Characterizing fluid and kinetic instabilities using
    field-particle correlations on single-point time series}. {\em
    Phys.~Plasmas\/} {\bf 24}~(5), 055901.
  
  \bibitem[{Klein} \& {Howes}(2015)]{Klein:2015a}
  {\sc {Klein}, K.~G. \& {Howes}, G.~G.} 2015 {Predicted impacts of proton
    temperature anisotropy on solar wind turbulence}. {\em Phys.~Plasmas\/} {\bf
    22}~(3), 032903.
  
  \bibitem[{Klein} \& {Howes}(2016)]{Klein:2016}
  {\sc {Klein}, K.~G. \& {Howes}, G.~G.} 2016 {Measuring Collisionless Damping in
    Heliospheric Plasmas using Field-Particle Correlations}. {\em
    Astrophys.~J.~Lett.\/} {\bf 826}, L30.
  
  \bibitem[{Klein} {\em et~al.\/}(2017){Klein}, {Howes} \&
    {Tenbarge}]{Klein:2017}
  {\sc {Klein}, K.~G., {Howes}, G.~G. \& {Tenbarge}, J.~M.} 2017 {Diagnosing
    collisionless energy transfer using field-particle correlations: gyrokinetic
    turbulence}. {\em J.~Plasma Phys.\/} {\bf 83}~(4), 535830401.
  
  \bibitem[{Klein} {\em et~al.\/}(2020){Klein}, {Howes}, {TenBarge} \&
    {Valentini}]{Klein:2020}
  {\sc {Klein}, Kristopher~G., {Howes}, Gregory~G., {TenBarge}, Jason~M. \&
    {Valentini}, Francesco} 2020 {Diagnosing collisionless energy transfer using
    field-particle correlations: Alfv{\'e}n-ion cyclotron turbulence}. {\em
    J.~Plasma Phys.\/} {\bf 86}~(4), 905860402.
  
  \bibitem[{Klimchuk}(2006)]{Klimchuk:2006}
  {\sc {Klimchuk}, J.~A.} 2006 {On Solving the Coronal Heating Problem}. {\em
    Sol.~Phys.\/} {\bf 234}, 41--77.
  
  \bibitem[Kraichnan(1965)]{Kraichnan:1965}
  {\sc Kraichnan, R.~H.} 1965 Inertial range spectrum of hyromagnetic turbulence.
    {\em Phys.~Fluids\/} {\bf 8}, 1385--1387.
  
  \bibitem[{Landau}(1946)]{Landau:1946}
  {\sc {Landau}, L.~D.} 1946 {On the Vibrations of the Electronic Plasma}. {\em
    Journal of Physics\/} {\bf 10}, 25.
  
  \bibitem[Leamon {\em et~al.\/}(1998)Leamon, Smith, Ness, Matthaeus \&
    Wong]{Leamon:1998a}
  {\sc Leamon, R.~J., Smith, C.~W., Ness, N.~F., Matthaeus, W.~H. \& Wong, H.~K.}
    1998 Observational constraints on the dynamics of the interplanetary magnetic
    field dissipation range. {\em J.~Geophys.~Res.\/} {\bf 103}, 4775--4787.
  
  \bibitem[{Leamon} {\em et~al.\/}(1999){Leamon}, {Smith}, {Ness} \&
    {Wong}]{Leamon:1999}
  {\sc {Leamon}, Robert~J., {Smith}, Charles~W., {Ness}, Norman~F. \& {Wong},
    Hung~K.} 1999 {Dissipation range dynamics: Kinetic Alfv{\'e}n waves and the
    importance of {\ensuremath{\beta}}$_{e}$}. {\em J.~Geophys.~Res.\/} {\bf
    104}~(A10), 22331--22344.
  
  \bibitem[{Li} {\em et~al.\/}(2016){Li}, {Howes}, {Klein} \&
    {TenBarge}]{TCLi:2016}
  {\sc {Li}, T.~C., {Howes}, G.~G., {Klein}, K.~G. \& {TenBarge}, J.~M.} 2016
    {Energy Dissipation and Landau Damping in Two- and Three-dimensional Plasma
    Turbulence}. {\em Astrophys.~J.~Lett.\/} {\bf 832}, L24.
  
  \bibitem[{Lithwick} \& {Goldreich}(2003)]{Lithwick:2003}
  {\sc {Lithwick}, Yoram \& {Goldreich}, Peter} 2003 {Imbalanced Weak
    Magnetohydrodynamic Turbulence}. {\em Astrophys.~J.\/} {\bf 582}~(2),
    1220--1240.
  
  \bibitem[{Lithwick} {\em et~al.\/}(2007){Lithwick}, {Goldreich} \&
    {Sridhar}]{Lithwick:2007}
  {\sc {Lithwick}, Y., {Goldreich}, P. \& {Sridhar}, S.} 2007 {Imbalanced Strong
    MHD Turbulence}. {\em Astrophys.~J.\/} {\bf 655}~(1), 269--274.
  
  \bibitem[{Markovskii} \& {Vasquez}(2013)]{Markovskii:2013}
  {\sc {Markovskii}, S.~A. \& {Vasquez}, B.~J.} 2013 {Magnetic Helicity in the
    Dissipation Range of Strong Imbalanced Turbulence}. {\em Astrophys.~J.\/}
    {\bf 768}, 62.
  
  \bibitem[{Matteini} {\em et~al.\/}(2020){Matteini}, {Franci}, {Alexandrova},
    {Lacombe}, {Landi}, {Hellinger}, {Papini} \& {Verdini}]{Matteini:2020}
  {\sc {Matteini}, L., {Franci}, L., {Alexandrova}, O., {Lacombe}, C., {Landi},
    S., {Hellinger}, P., {Papini}, E. \& {Verdini}, A.} 2020 {Magnetic field
    turbulence in the solar wind at sub-ion scales: in situ observations and
    numerical simulations}. {\em Frontiers in Astronomy and Space Sciences\/}
    {\bf 7}, 83.
  
  \bibitem[{McCubbin} {\em et~al.\/}(2022){McCubbin}, {Howes} \&
    {TenBarge}]{McCubbin:2022}
  {\sc {McCubbin}, Andrew~J., {Howes}, Gregory~G. \& {TenBarge}, Jason~M.} 2022
    {Characterizing velocity-space signatures of electron energization in
    large-guide-field collisionless magnetic reconnection}. {\em Phys.~Plasmas\/}
    {\bf 29}~(5), 052105.
  
  \bibitem[{Meyrand} {\em et~al.\/}(2021){Meyrand}, {Squire}, {Schekochihin} \&
    {Dorland}]{Meyrand:2021}
  {\sc {Meyrand}, R., {Squire}, J., {Schekochihin}, A.~A. \& {Dorland}, W.} 2021
    {On the violation of the zeroth law of turbulence in space plasmas}. {\em
    J.~Plasma Phys.\/} {\bf 87}~(3), 535870301.
  
  \bibitem[{Montag} \& {Howes}(2022)]{Montag:2022}
  {\sc {Montag}, P. \& {Howes}, Gregory~G.} 2022 {A field-particle correlation
    analysis of magnetic pumping}. {\em Phys.~Plasmas\/} {\bf 29}~(3), 032901.
  
  \bibitem[{Nielson} {\em et~al.\/}(2013){Nielson}, {Howes} \&
    {Dorland}]{Nielson:2013}
  {\sc {Nielson}, K.~D., {Howes}, G.~G. \& {Dorland}, W.} 2013 {Alfv{\'e}n wave
    collisions, the fundamental building block of plasma turbulence. II.
    Numerical solution}. {\em Phys.~Plasmas\/} {\bf 20}~(7), 072303.
  
  \bibitem[{Numata} {\em et~al.\/}(2010){Numata}, {Howes}, {Tatsuno}, {Barnes} \&
    {Dorland}]{Numata:2010}
  {\sc {Numata}, R., {Howes}, G.~G., {Tatsuno}, T., {Barnes}, M. \& {Dorland},
    W.} 2010 {AstroGK: Astrophysical gyrokinetics code}. {\em J.~Plasma Phys.\/}
    {\bf 229}, 9347--9372.
  
  \bibitem[{Parker}(1988)]{Parker:1988}
  {\sc {Parker}, E.~N.} 1988 {Nanoflares and the solar X-ray corona}. {\em
    Astrophys.~J.\/} {\bf 330}, 474--479.
  
  \bibitem[{Perez} \& {Boldyrev}(2009)]{Perez:2009}
  {\sc {Perez}, Jean~Carlos \& {Boldyrev}, Stanislav} 2009 {Role of
    Cross-Helicity in Magnetohydrodynamic Turbulence}. {\em Phys.~Rev.~Lett.\/}
    {\bf 102}~(2), 025003.
  
  \bibitem[{Quataert}(1998)]{Quataert:1998}
  {\sc {Quataert}, E.} 1998 {Particle Heating by Alfv\'enic Turbulence in Hot
    Accretion Flows}. {\em Astrophys.~J.\/} {\bf 500}, 978--991.
  
  \bibitem[{Richardson} \& {Smith}(2003)]{Richardson:2003}
  {\sc {Richardson}, John~D. \& {Smith}, Charles~W.} 2003 {The radial temperature
    profile of the solar wind}. {\em Geophys.~Res.~Lett.\/} {\bf 30}~(5), 1206.
  
  \bibitem[{Sahraoui} {\em et~al.\/}(2013){Sahraoui}, {Huang}, {Belmont},
    {Goldstein}, {R{\'e}tino}, {Robert} \& {De Patoul}]{Sahraoui:2013b}
  {\sc {Sahraoui}, F., {Huang}, S.~Y., {Belmont}, G., {Goldstein}, M.~L.,
    {R{\'e}tino}, A., {Robert}, P. \& {De Patoul}, J.} 2013 {Scaling of the
    Electron Dissipation Range of Solar Wind Turbulence}. {\em Astrophys.~J.\/}
    {\bf 777}, 15.
  
  \bibitem[{Schekochihin} {\em et~al.\/}(2009){Schekochihin}, {Cowley},
    {Dorland}, {Hammett}, {Howes}, {Quataert} \& {Tatsuno}]{Schekochihin:2009}
  {\sc {Schekochihin}, A.~A., {Cowley}, S.~C., {Dorland}, W., {Hammett}, G.~W.,
    {Howes}, G.~G., {Quataert}, E. \& {Tatsuno}, T.} 2009 {Astrophysical
    Gyrokinetics: Kinetic and Fluid Turbulent Cascades in Magnetized Weakly
    Collisional Plasmas}. {\em Astrophys.~J.\/} {\bf 182}, 310--377.
  
  \bibitem[{Squire} {\em et~al.\/}(2022){Squire}, {Meyrand}, {Kunz},
    {Arzamasskiy}, {Schekochihin} \& {Quataert}]{Squire:2022}
  {\sc {Squire}, Jonathan, {Meyrand}, Romain, {Kunz}, Matthew~W., {Arzamasskiy},
    Lev, {Schekochihin}, Alexander~A. \& {Quataert}, Eliot} 2022 {High-frequency
    heating of the solar wind triggered by low-frequency turbulence}. {\em Nature
    Astronomy\/} {\bf 6}, 715--723.
  
  \bibitem[{Sridhar} \& {Goldreich}(1994)]{Sridhar:1994}
  {\sc {Sridhar}, S. \& {Goldreich}, P.} 1994 {Toward a theory of interstellar
    turbulence. 1: Weak Alfvenic turbulence}. {\em Astrophys.~J.\/} {\bf 432},
    612--621.
  
  \bibitem[{Stix}(1992)]{Stix:1992}
  {\sc {Stix}, T.~H.} 1992 {\em {Waves in Plasmas}\/}. New York: American
    Institute of Physics.
  
  \bibitem[{TenBarge} \& {Howes}(2013)]{TenBarge:2013a}
  {\sc {TenBarge}, J.~M. \& {Howes}, G.~G.} 2013 {Current Sheets and
    Collisionless Damping in Kinetic Plasma Turbulence}. {\em
    Astrophys.~J.~Lett.\/} {\bf 771}~(2), L27.
  
  \bibitem[{TenBarge} {\em et~al.\/}(2014){TenBarge}, {Howes}, {Dorland} \&
    {Hammett}]{TenBarge:2014}
  {\sc {TenBarge}, J.~M., {Howes}, G.~G., {Dorland}, W. \& {Hammett}, G.~W.} 2014
    {An oscillating Langevin antenna for driving plasma turbulence simulations}.
    {\em Comp.~Phys.~Comm.\/} {\bf 185}~(2), 578--589.
  
  \bibitem[{TenBarge} {\em et~al.\/}(2012){TenBarge}, {Podesta}, {Klein} \&
    {Howes}]{TenBarge:2012b}
  {\sc {TenBarge}, J.~M., {Podesta}, J.~J., {Klein}, K.~G. \& {Howes}, G.~G.}
    2012 {Interpreting Magnetic Variance Anisotropy Measurements in the Solar
    Wind}. {\em Astrophys.~J.\/} {\bf 753}~(2), 107.
  
  \bibitem[{Told} {\em et~al.\/}(2015){Told}, {Jenko}, {TenBarge}, {Howes} \&
    {Hammett}]{Told:2015}
  {\sc {Told}, D., {Jenko}, F., {TenBarge}, J.~M., {Howes}, G.~G. \& {Hammett},
    G.~W.} 2015 {Multiscale Nature of the Dissipation Range in Gyrokinetic
    Simulations of Alfv{\'e}nic Turbulence}. {\em Phys.~Rev.~Lett.\/} {\bf
    115}~(2), 025003.
  
  \bibitem[{Verniero} \& {Howes}(2018)]{Verniero:2018b}
  {\sc {Verniero}, J.~L. \& {Howes}, G.~G.} 2018 {The Alfv{\'e}nic nature of
    energy transfer mediation in localized, strongly nonlinear Alfv{\'e}n
    wavepacket collisions}. {\em J.~Plasma Phys.\/} {\bf 84}~(1), 905840109.
  
  \bibitem[{Verniero} {\em et~al.\/}(2018){Verniero}, {Howes} \&
    {Klein}]{Verniero:2018a}
  {\sc {Verniero}, J.~L., {Howes}, G.~G. \& {Klein}, K.~G.} 2018 {Nonlinear
    energy transfer and current sheet development in localized Alfv{\'e}n
    wavepacket collisions in the strong turbulence limit}. {\em J.~Plasma
    Phys.\/} {\bf 84}~(1), 905840103.
  
  \bibitem[{Verniero} {\em et~al.\/}(2021{\natexlab{{\em a\/}}}){Verniero},
    {Howes}, {Stewart} \& {Klein}]{Verniero:2021a}
  {\sc {Verniero}, J.~L., {Howes}, G.~G., {Stewart}, D.~E. \& {Klein}, K.~G.}
    2021{\natexlab{{\em a\/}}} {Determining Threshold Instrumental Resolutions
    for Resolving the Velocity Space Signature of Ion Landau Damping}. {\em
    Journal of Geophysical Research (Space Physics)\/} {\bf 126}~(5), e28361.
  
  \bibitem[{Verniero} {\em et~al.\/}(2021{\natexlab{{\em b\/}}}){Verniero},
    {Howes}, {Stewart} \& {Klein}]{Verniero:2021b}
  {\sc {Verniero}, J.~L., {Howes}, G.~G., {Stewart}, D.~E. \& {Klein}, K.~G.}
    2021{\natexlab{{\em b\/}}} {PATCH: Particle Arrival Time Correlation for
    Heliophysics}. {\em Journal of Geophysical Research (Space Physics)\/} {\bf
    126}~(5), e28940.
  
  \bibitem[{Verscharen} {\em et~al.\/}(2019){Verscharen}, {Klein} \&
    {Maruca}]{Verscharen:2019}
  {\sc {Verscharen}, Daniel, {Klein}, Kristopher~G. \& {Maruca}, Bennett~A.} 2019
    {The multi-scale nature of the solar wind}. {\em Living Rev.~Solar Phys.\/}
    {\bf 16}~(1), 5.
  
  \bibitem[{Wilson} {\em et~al.\/}(2018){Wilson}, {Stevens}, {Kasper}, {Klein},
    {Maruca}, {Bale}, {Bowen}, {Pulupa} \& {Salem}]{Wilson:2018}
  {\sc {Wilson}, Lynn~B., III, {Stevens}, Michael~L., {Kasper}, Justin~C.,
    {Klein}, Kristopher~G., {Maruca}, Bennett~A., {Bale}, Stuart~D., {Bowen},
    Trevor~A., {Pulupa}, Marc~P. \& {Salem}, Chadi~S.} 2018 {The Statistical
    Properties of Solar Wind Temperature Parameters Near 1 au}. {\em
    Astrophys.~J.~Supp.\/} {\bf 236}~(2), 41.
  
  \bibitem[{Withbroe} \& {Noyes}(1977)]{Withbroe:1977}
  {\sc {Withbroe}, G.~L. \& {Noyes}, R.~W.} 1977 {Mass and energy flow in the
    solar chromosphere and corona}. {\em Ann.~Rev.~Astron.~Astrophys.\/} {\bf
    15}, 363--387.
  
  \bibitem[{Zhou} {\em et~al.\/}(2023){Zhou}, {Liu} \& {Loureiro}]{Zhou:2023}
  {\sc {Zhou}, Muni, {Liu}, Zhuo \& {Loureiro}, Nuno~F.} 2023 {Electron heating
    in kinetic-Alfv{\'e}n-wave turbulence}. {\em Proc.~Nat.~Acad.~Sci.\/} {\bf
    120}~(23), e2220927120.
  
  \end{thebibliography}

\end{document}